\documentclass[prl,superscriptaddress,twocolumn,notitlepage,longbibliography]{revtex4-1}

\usepackage{amsfonts}
\usepackage{subfigure}
\usepackage{amsmath}
\usepackage{amssymb}
\usepackage{amsbsy} 
\usepackage{epsfig}
\usepackage{graphicx}
\usepackage{epstopdf}
\usepackage{bbm}
\usepackage{xcolor}

\usepackage[unicode=true,
 bookmarks=true,bookmarksnumbered=false,bookmarksopen=false,
 breaklinks=false,pdfborder={0 0 1},backref=false,colorlinks=true]
 {hyperref}
\hypersetup{
 linkcolor=magenta,urlcolor=blue,citecolor=blue,pdfstartview={FitH},urlcolor=blue}

\def\be{\begin{equation}} \def\ee{\end{equation}}
\def\bea{\begin{eqnarray}} \def\eea{\end{eqnarray}}

\begin{document}

\title{Floquet Spin Splitting and Spin Generation in Antiferromagnets}

\author{Bo Li}
\email{libphysics@xjtu.edu.cn}
\affiliation{MOE Key Laboratory for Nonequilibrium Synthesis and Modulation of Condensed Matter,\\
Shaanxi Province Key Laboratory of Quantum Information and Quantum Optoelectronic Devices,\\
School of Physics, Xi’an Jiaotong University, Xi’an 710049, China}

\author{Ding-Fu Shao}
\email{dfshao@issp.ac.cn}
\affiliation{Key Laboratory of Materials Physics, Institute of Solid State Physics, HFIPS, Chinese Academy of Sciences, Hefei 230031, China}

\author{Alexey A. Kovalev}
\email{alexey.kovalev@unl.edu}
\affiliation{Department of Physics and Astronomy and Nebraska Center for Materials and Nanoscience,
University of Nebraska, Lincoln, Nebraska 68588, USA}

\date{\today}

\begin{abstract}
In antiferromagnetic spintronics, accessing the spin degree of freedom is essential for generating spin currents and manipulating magnetic order, which generally requires lifting spin degeneracy. This is typically achieved through relativistic spin-orbit coupling or non-relativistic spin splitting in altermagnets. Here, we propose an alternative approach: a dynamical spin splitting induced by an optical field in antiferromagnets. By coupling the driven system to a thermal bath, we demonstrate the emergence of steady-state pure spin currents, as well as linear-response longitudinal and transverse spin currents. Crucially, thermal bath engineering enables a nonrelativistic Edelstein effect— the generation of a net spin accumulation—without relying on spin–orbit coupling.
Our results provide a broadly applicable and experimentally tunable route to control spins in antiferromagnets, offering new opportunities for spin generation and manipulation in antiferromagnetic spintronics.

\end{abstract}

\maketitle
The frontier of present spintronics research largely focuses on antiferromagnetic systems, due to their ultra-fast dynamics and free of magnetic stray field~\cite{Baltz2018AFspintronics,Jungwirth2016AFspintronics,Han2023RevcoherentAFM,Jungwirth2018RevDiretionAFM,Šmejkal2018RevTopoAFM}. However, in many collinear antiferromagnets, the electronic bands remain spin degenerate due to a preserved antiunitary effective time-reversal symmetry, typically realized as the combination of time reversal with spatial inversion or a sublattice translation. This symmetry protection imposes a fundamental constraint on antiferromagnetic spintronics, where direct manipulation of real spin degrees of freedom is essential. To access the spin degree of freedoms, spin-orbit coupling (SOC) is typically required, enabling a variety of spin-related phenomena, such as spin generation, spin Hall effect, and spin-orbit torques~\cite{Jungwirth2016afm, Baltz2018RMPafmSpintronics}.

Recently, a new class of magnetic materials with non-relativistic spin splitting, dubbed altermagnets, gathered extensive attention~\cite{Hayami2019spinSplitAFM,Yuan2020spinSplitting,Yuan2021spiltNoSoc,Libor2020CrystalHall,libor2022prxAltermagnet,Mazin2020PRXaltermagnet,GUO2023spinsplit,Hayami2020DesignSplit,Rafael2021SpinSplitter,Liu2022spinSymmetry,Smejkal2022landscapeAltermagnet,Bai2022splittingtorque,Karube2022splitterTorque,Feng2022anormalousHallAltermagnet,Smejkal2023ChiralMagnon,Sato2024PAlterAnormalousHall,Das2025honeycombaltermagnet}. In altermagnets, the effective time reversal symmetry is absent because the involved lattice symmetry is intrinsically broken by the lattice structure. More importantly, the non-relativistic spin splitting is typically larger than that arising from SOC, which is favorable for utilizing the spin degree of freedom.

Another route to lifting spin degeneracy is by explicitly breaking time-reversal symmetry, as in the case of Zeeman splitting induced by a magnetic field. However, the resulting energy scales are typically negligible compared to the electronic band structure ($\sim 1$ eV). For example, a 1 T magnetic field yields a splitting on the order of $10^{-4}\sim 10^{-5}$ eV. As an alternative, time-reversal symmetry can be broken dynamically using an optical field, offering a more efficient means to lift spin degeneracy. In this work, we investigate light-induced spin splitting in antiferromagnets possessing effective time-reversal symmetry. Using Floquet theory~\cite{Takashi2019FloquetEigineering,Rudner2020floquet}, we demonstrate that an optical field with suitable intensity and frequency can induce substantial spin splitting in the electronic quasienergy bands via non-equilibrium effects. With appropriate thermal bath engineering, the driven system can support SOC-independent steady-state spin and charge currents, as well as spin accumulation.
Our findings uncover an experimentally tunable mechanism for controlling the spin degree of freedom in spin-degenerate antiferromagnets, opening new avenues for spin generation and manipulation in antiferromagnetic spintronics

\begin{figure}
\centering
\includegraphics[width=1\linewidth]{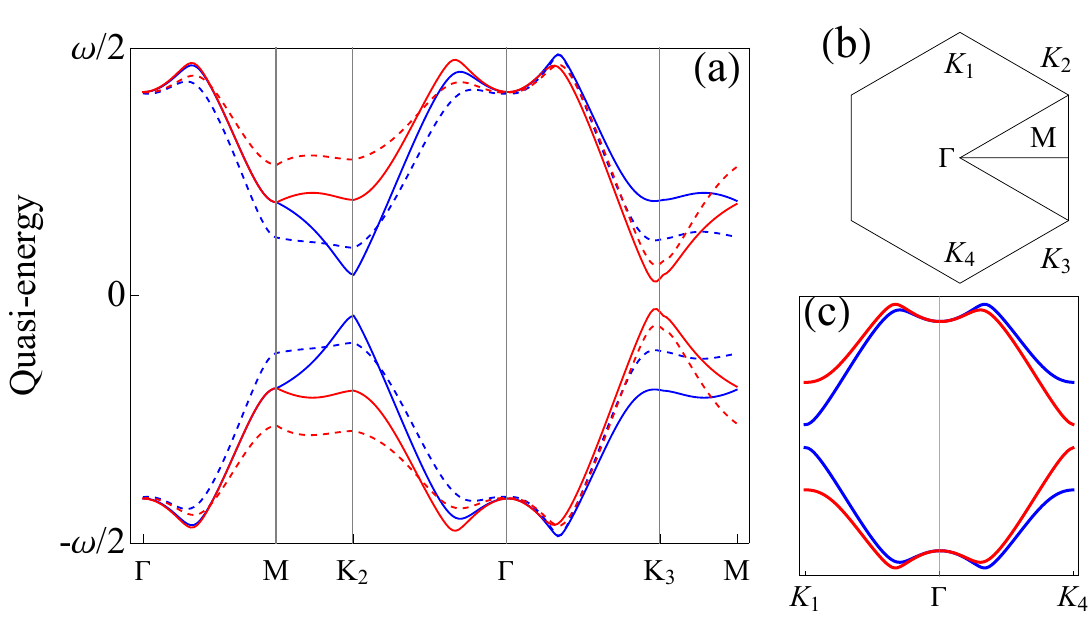}
\caption{(a) Quasi-energy band structure in honeycomb-lattice AFM with N\'eel order, where blue (red) color stands for spin up (down) and solid (dashed) line represents the case without (with) SOC ( with $\lambda_{\text{SO}}= 0.2$). (b) The Brillouin zone. (c) Quasi-energy band structure along $K_1-\Gamma-K_4$, where SOC is zero. In (a), (c), $\varphi=\pi/3, A_0a=1,\omega=4$ and $t=1,\lambda=0.5$.} \label{fig:honeycombAFM}
\end{figure}

\textcolor{blue}{\textit{Floquet spin splitting in AFM}.} We consider an electronic system with collinear antiferromagnetic order on honeycomb lattice, as realized in materials such as MnPX$_3$ (X$=$S, Se)~\cite{Long2020MnPS3,LIU2020MnPSe3}. The momentum-space Hamiltonian matrix under basis $(c_{\mathbf kA,\uparrow},c_{\mathbf kA,\downarrow},c_{\mathbf kB,\uparrow},c_{\mathbf kB,\downarrow})^T$ is given by
\begin{eqnarray}\label{eq:Hamilontian}
H(\mathbf k)=\left(
\begin{array}{cc}  \lambda_{\text{SO}}\xi_{\mathbf k}\sigma^z+\lambda\mathbf n\cdot\boldsymbol{\sigma} & t\gamma_{\mathbf k} \\
t\gamma^\ast_{\mathbf k} & -\lambda_{\text{SO}}\xi_{\mathbf k}\sigma^z-\lambda\mathbf n\cdot\boldsymbol{\sigma}
\end{array}\right)
\end{eqnarray}
where $\gamma_{\mathbf k}=\sum_{i=1}^3 e^{-i\mathbf k\cdot\boldsymbol{\delta}_i}$ with $\boldsymbol{\delta}_1=a(1,0), \boldsymbol{\delta}_{2,3}=a(-1/2,\pm\sqrt{3}/2)$ (here, $a$ is lattice constant), and $\xi_{\mathbf k}=\sum_{i=1}^32\sin(\mathbf{k}\cdot\boldsymbol{\kappa}_i)$ with $\boldsymbol{\kappa}_1=a(0,\sqrt{3})$, $\boldsymbol{\kappa}_{2,3}=a(\mp3/2,-\sqrt{3}/2)$. Here, $\lambda_{\text{SO}}$ is the intrinsic SOC, $\lambda$ represents the exchange coupling, $\sigma^i$ is the Pauli matrix acting in spin space, and $\mathbf n$ is the N\'eel vector. We assume $\mathbf n \parallel \hat z$ to preserve spin conservation.
The system preserves combined parity-time ($\mathcal P \mathcal T$) symmetry, ensuring spin degeneracy throughout the band structure.

To break the $\mathcal P \mathcal T$ symmetry, we apply polarized light described by a time-dependent vector potential
$\boldsymbol{\mathcal A}=A_0(\sin\omega t, \sin(\omega t+\varphi),0)$, where $A_0=E_0/\omega$ (set $e=\hbar=1$) with $E_0$ being the electric field amplitude. Under inversion $\mathcal P$ and time reversal $\mathcal T$, the vector potential transforms as $\mathcal P$: $\boldsymbol{\mathcal A}(t)\rightarrow -\boldsymbol{\mathcal A}(t)$, and $\mathcal T$: $\boldsymbol{\mathcal A}(t)\rightarrow -\boldsymbol{\mathcal A}(-t)$, which together yield $\mathcal P\mathcal T: \boldsymbol{\mathcal A}(t)\rightarrow \boldsymbol{\mathcal A}(-t)$. Elliptical or circular polarization ($\varphi\neq 0,\pi$) explicitly breaks the $\mathcal P\mathcal T$ symmetry. In contrast, for linear polarization ($\varphi=0,\pi$), $\mathcal P\mathcal T$ enforces $\boldsymbol{\mathcal A}(-t)=-\boldsymbol{\mathcal A}(t)$, which is equivalent to a time translation by $\pi/\omega$. As a result, all physical observables remain invariant over one driving period, and the $\mathcal P\mathcal T$ symmetry is effectively preserved.  The light couples to the system via the Peierls substitution $H(\mathbf k)\rightarrow H(t)= H[\mathbf k+\boldsymbol{\mathcal A}(t)]$. To preserve antiferromagnetic order, we focus on a subgap driving regime $J_{ex}\ll\hbar\omega\ll\Delta_{c}$, where $J_{ex}$ is exchange energy between local spins and $\Delta_{c}$ the charge gap associated with their formation~\cite{Batignani2015ultrafastAFM,Takashi2019FloquetEigineering, Walldorf2019drivenHubbard,Torre2021RMPnonthermal}. Typically, $J_{ex}\sim t_h^2/U$ and $\Delta_c\sim U$ where $t_h$ is the electron hopping energy and $U$ refers to local repulsive Coulomb energy, satisfying $t_h\ll U$. Taking $t_h\lesssim t$, a drive with $\hbar\omega\sim t$ lies within this subgap window. In this regime, magnon absorption is suppressed and heating via charge excitations occurs only on exponentially long time scales ($\sim e^{U/\hbar\omega}$), which can be further mitigated by coupling to a thermal bath~\cite{Oka2012Mott,Abanin2015slowheating,Sensarma2010Hubbard}. By contrast, antiferromagnetic order may also survive in the ultrafast regime $\hbar\omega\gg U\gg t$, but the resulting spin splitting is perturbatively small. 

The resulting periodically driven system is naturally analyzed within the framework of Floquet theory~\cite{rudner2020floquetengineershandbook,Rudner2020floquet}. 
The eigenstate is represented as a Floquet state: $ |\psi_n(t)\rangle=e^{-i\varepsilon_n t/\hbar}|\phi_n(t)\rangle$ where $\varepsilon_n$ is the quasi-energy, and  $|\phi_n(t+T)\rangle=|\phi_n(t)\rangle$ with $T=2\pi/\omega$. The periodic part of Floquet state respects $(\varepsilon_n+i\partial_t)|\phi_n(t)\rangle=H(t)|\phi_n(t)\rangle$. This equation can be further translated to an equation of associated Fourier components: $(\varepsilon_n+m\omega)|\phi_n^{(m)}\rangle= \sum_{m^\prime}H^{(m-m^\prime)}|\phi_n^{(m^\prime)}\rangle$, where $|\phi_n(t)\rangle=\sum_me^{-im\omega t}|\phi_n^{(m)}\rangle$ and $H^{(m)}=\frac{1}{T}\int_0^T dt e^{im\omega t}H(t)$. The quasi-energy $\varepsilon_n$ is well-defined up to $m\omega$ ($m\in$ Integers). Therefore, it is enough to confine the quasi-energy to the first ``Floquet-Brillouin Zone"(FBZ): $-\omega/2\leq \varepsilon_n<\omega/2$.

It is straightforward to obtain the Fourier component of  Hamiltonian~\eqref{eq:Hamilontian}:
\begin{eqnarray}\label{eq:FourierHamiltonian}
H^{(m)}=\left(\begin{array}{cc}
    0 & h^{(m)} \\
   (h^{(-m)})^\ast & 0
\end{array}\right)\sigma^0+h^{(m)}_{\text{SO}}\tau^z\sigma^z+\lambda\tau^z\sigma^z\delta_{m,0},    
\end{eqnarray}
where $\tau^i$ is the Pauli matrix in the sublattice space.
Here, $h^{(m)}=\sum_{i=1}^3 e^{-i\mathbf k\cdot\boldsymbol{\delta}_i}e^{-im\theta_i}J_m(\zeta_i A_0a)$ and
$h^{(m)}_{\text{SO}}=-i\lambda_{\text{SO}}\sum_{i=1}^3[e^{i\mathbf k\cdot\boldsymbol{\kappa}_i}(-1)^m-e^{-i\mathbf k\cdot\boldsymbol{\kappa}_i}]e^{im\tilde{\theta}_i}J_m(\tilde{\zeta}_iA_0a)$, where $J_m(\cdots)$ is the $m$-th Bessel function and all involved parameters are listed in table.~\ref{table_transform}. The quasi-energy band structure can be obtained by applying Eq.~\eqref{eq:FourierHamiltonian} to the Fourier-transformed eigen equation. In this system, spin remains a good quantum number, allowing the band structure and corresponding transport properties to be analyzed within spin-resolved subspaces. 

\begin{table}[ht]
	\centering
	\begin{tabular}{|c| c| c| c|}
		\hline\hline 
		$\theta_1$& $\theta_{2,3}$ & $\zeta_1$ & $\zeta_{2,3}$ \\ [0.5ex] 
		\hline 
		0& $\mp\text{sign}(\pi-\varphi)\cos^{-1}\Big[\frac{\frac{1}{2}\mp\frac{\sqrt{3}}{2}\cos\varphi}{\sqrt{\mathcal N_{\mp}}}\Big]$ & $1$ & $-\sqrt{\mathcal N_{\mp}}$ \\
		\hline\hline
	    $\tilde{\theta}_1$& $\tilde{\theta}_{2,3}$ & $\tilde{\zeta}_1$ & $\tilde{\zeta}_{2,3}$
		\\[1ex]
		\hline
		$-\varphi$& $\text{sign}(\pi-\varphi)\cos^{-1}\Big[\frac{\mp\frac{3}{2}-\frac{\sqrt{3}}{2}\cos\varphi}{\sqrt{3\mathcal N_{\pm}}}\Big]$ & $\sqrt{3}$ & $\sqrt{3\mathcal N_{\pm}}$\\
		\hline 
	\end{tabular}
	\caption{Expression of parameters in the Fourier transformed Hamiltonian, where $\mathcal N_\pm=1\pm \frac{\sqrt{3}}{2}\cos\varphi$.} \label{table_transform}
\end{table}

The application of light is expected to lift the spin degeneracy of the quasi-energy bands. Indeed, the band structure in Fig.~\ref{fig:honeycombAFM} is spin non-degenerate, even in the absence of SOC. Importantly, the spin-split energy is surprisingly large,  comparable to the scale of the original band structure and exceeding the typical SOC. This offers an efficient way to approach the spin degrees of freedom. It is worth noting that if $\lambda_{\text{SO}}=0$, the Hamiltonian Eq.~\eqref{eq:Hamilontian} has a dual symmetry between the two spin sectors:
\begin{eqnarray}\label{eq:dual}
\tau^x H_{\uparrow}[\mathbf k+\boldsymbol{\mathcal A}(t)]\tau^x=H_{\downarrow}[-\mathbf k-\boldsymbol{\mathcal A}(t)].   
\end{eqnarray}
This symmetry originates from the combined action of spatial inversion ($\mathcal P$) and a twofold spin rotation about the $x$ axis $C_{2x}^{\mathrm{spin}}$, such that $\mathcal P C_{2x}^{spin} H[\mathbf k+\boldsymbol{\mathcal A}(t)](\mathcal P C_{2x}^{spin})^{-1}=H[\mathbf k+\boldsymbol{\mathcal A}(t)]$, where the inversion exchanges the sublattices (represented by $\tau^x$) and transforms $\mathbf k+\boldsymbol{\mathcal A}$ to $-\mathbf k-\boldsymbol{\mathcal A}$, and the spin rotation flip spin $\sigma^z$ to $-\sigma^z$.
The dual symmetry leads to a dual relation between quasi-energies: $\varepsilon_{u,d}^{\uparrow}(\mathbf k)=\varepsilon_{u,d}^{\downarrow}(-\mathbf k)$ [e.g., see Fig.~\ref{fig:honeycombAFM} (c)], where $u,d$ refer to the up and down bands in the FBZ. Notice that in Fig.~\ref{fig:honeycombAFM} the equivalence among valleys is absent because the three-fold rotation symmetry of the original model is broken by the optical field. As a result, the dual relation between quasi-energies does not apply to the path $\mathbf K_3-\boldsymbol{\Gamma}-\mathbf K_2$. On the other hand, a nonzero intrinsic SOC can explicitly break the dual symmetry in Eq.~\eqref{eq:dual}.

The spin splitting is a cooperative effect of exchange coupling and the light. To see this in a clean way, we investigate the special case of off-resonance ($\omega\gg t$) and weak driving field ($A_0a\ll 1$), for which the original bands are dressed by the light to yield a spin-split term.  We focus on the vicinity near the valley $\mathbf K_4(\mathbf K_1) =(0,\mp\frac{4\pi}{3\sqrt{3}})$ point, where $\mathcal H_v(t)=v_F[\tau^x\eta^z(q_y+\mathcal A_y)+\tau^y(q_x+\mathcal A_x)]+s\lambda\tau^z$ and $\lambda_{\text{SO}}$ is set to zero for simplicity. Here, $v_F=3at/2$, $\mathbf q=\mathbf k-\mathbf K_4$ (or $\mathbf q=\mathbf k-\mathbf K_1$), $\eta^i$ denote the Pauli matrix for valley freedoms,  and $s=\pm 1$ for up and down spins. The effective Floquet Hamiltonian~\cite{Kitagawa2011photoQuantumHall,Goldman2014effectiveFloquet} up to $O[(aA_0)^4]$ is 
\begin{eqnarray}
H^F_{eff}&&\approx\mathcal H_v^{(0)}+\frac{[\mathcal H_v^{(-1)},\mathcal H_v^{(+1)}]}{\omega}\nonumber\\
&&=v_F(\tau^x\eta^zq_y+\tau^yq_x)+s\lambda\tau^z- M\tau^z\eta^z
\end{eqnarray}
where $M=\frac{(v_F A_0)^2}{\omega}\sin\varphi$, and $\mathcal H^{(\pm)}_v$ is the Fourier component of $\mathcal H_v(t)$. It is readily to obtain $\varepsilon^s_{\mathbf{K}_4(\mathbf{K}_1)}(\mathbf q)=\pm\sqrt{v_F^2q^2+(s\lambda-\chi M)^2}$ with $\chi=\pm 1$ being valley index. It is clear that the quasi-energy at each valley becomes spin-dependent, driven by the combined effects of exchange coupling and optical driving. When the intrinsic SOC is considered, $M$ is shifted to $M-s3\sqrt{3}\lambda_{\text{SO}}$, thus breaking the dual relation $\varepsilon^s_{\mathbf{K}_1}=\varepsilon^{-s}_{\mathbf{K}_4}$.


In general, a periodically driven isolated system tends toward an infinite-temperature state due to energy absorption from the drive. To avoid this heating problem and reach a steady state, the system must be coupled to a heat bath. Interestingly, we find that the nature of spin transport or accumulation depends sensitively on the type of bath: distinct spin-related behaviors emerge when coupling to bosonic versus fermionic reservoirs.

\textcolor{blue}{\textit{Steady state and spin currents}.} 
We first consider electron coupling to a bosonic phonon bath, described by the electron-phonon Hamiltonian
\begin{eqnarray}
H_{ep}=\sum_{\mathbf k,\mathbf q}\sum_{\nu,\nu^\prime}\sum_{\lambda}g^\lambda_{\nu^\prime\nu}(\mathbf k-\mathbf q,\mathbf k) c^\dagger_{\mathbf k-\mathbf q,\nu^\prime}c_{\mathbf k,\nu}(b^\dagger_{\mathbf q\lambda}+b_{-\mathbf q\lambda}).   
\end{eqnarray}
Here, $\nu,\nu^\prime$ label the sublattices, $b^\dagger_{\mathbf q\lambda} (b_{-\mathbf q\lambda})$ is the creation (anihillation) operator for $\lambda$ phonon modes, and $g^\lambda_{\nu^\prime\nu}(\mathbf k-\mathbf q,\mathbf k)$ is the electron-phonon coupling matrix~\cite{Thingstad2020phononSC, Yu2024geomentryEPC}. The phonon modes are described by $H_{ph}=\sum_{\lambda,\mathbf k}\hbar\omega_{\lambda\mathbf k}b^\dagger_{\mathbf q\lambda}b_{\mathbf q\lambda}$.  The steady-state occupation can be obtained by numerically solving the kinetic equation
\begin{eqnarray}\label{eq:kinetic}
\partial_t\rho_{\mathbf k\alpha}=&&\sum_{\mathbf k^\prime\alpha^\prime}   W_{\mathbf k\alpha,\mathbf k^\prime\alpha^\prime}(1-\rho_{\mathbf k,\alpha})\rho_{\mathbf k^\prime,\alpha^\prime}-W_{\mathbf k^\prime\alpha^\prime,\mathbf k\alpha}\nonumber\\
&&\qquad\times(1-\rho_{\mathbf k^\prime\alpha^\prime})\rho_{\mathbf k\alpha},
\end{eqnarray}
where $\rho_{\mathbf k,\alpha}$ denotes the occupation number for $\alpha$ quasi-band at momentum $\mathbf k$, and $W_{\mathbf k\alpha,\mathbf k^\prime\alpha^\prime}$ is the scattering rate determined by the electron–phonon coupling and the Floquet eigenstates, see details in Ref.~\cite{Seetharam2015population, Supp}.

\begin{figure}
\centering
\includegraphics[width=1\linewidth]{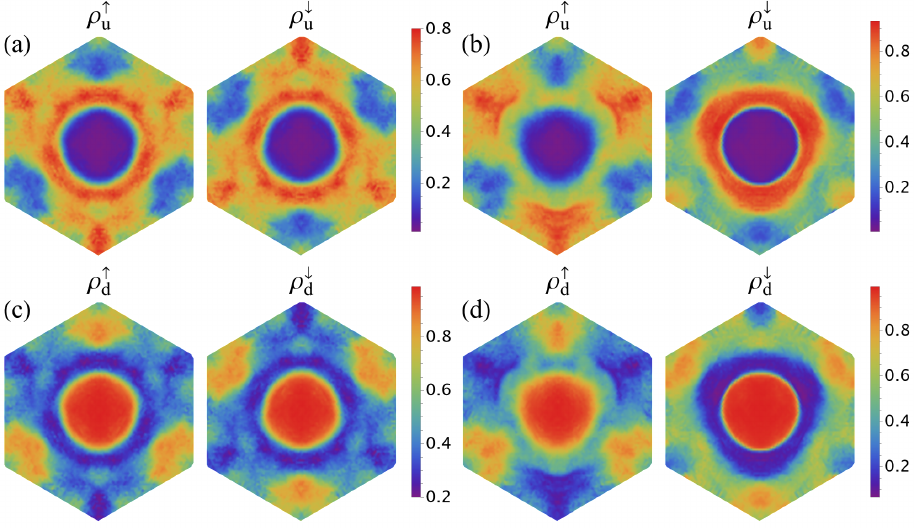}
\caption{Steady-state population for the quasi-energy band (in the first Floquet BZ) in each spin sector. The first and second rows display the population for upper and lower bands, respectively. In (a,c), $\lambda_{\text{SO}}=0$, in (b,d) $\lambda_{\text{SO}}=0.1$. Other parameters are  $\varphi=\pi/2, A_0a=1,\omega=1$, and $t=1$.}
\label{fig:population}
\end{figure}

\begin{figure}
\centering
\begin{tabular}{cc}
\includegraphics[width=1\linewidth]{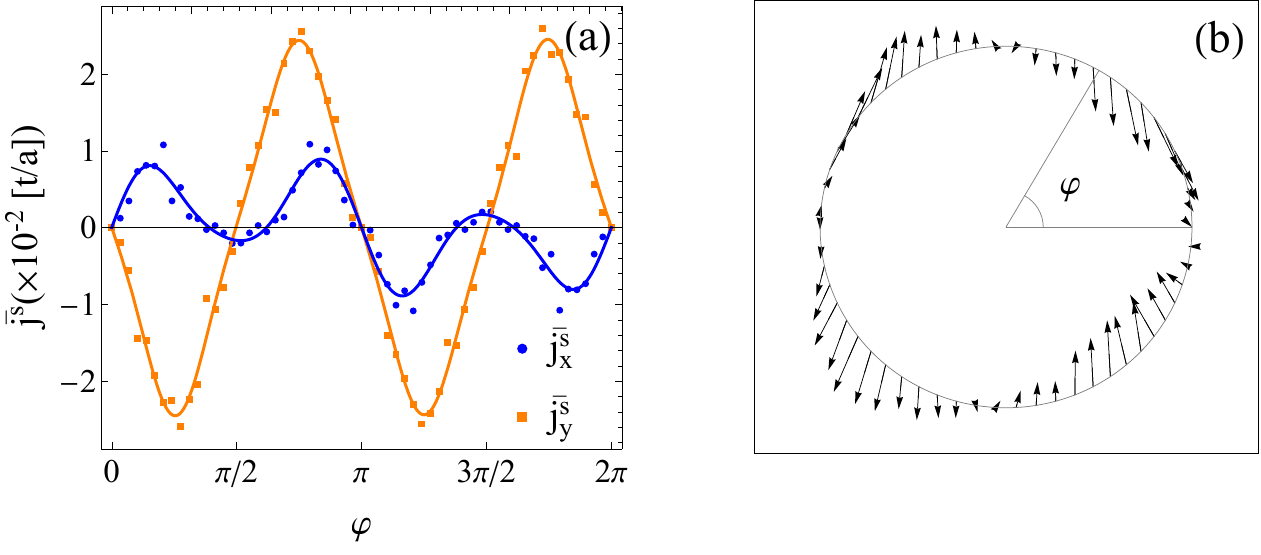}\\
\includegraphics[width=1\linewidth]{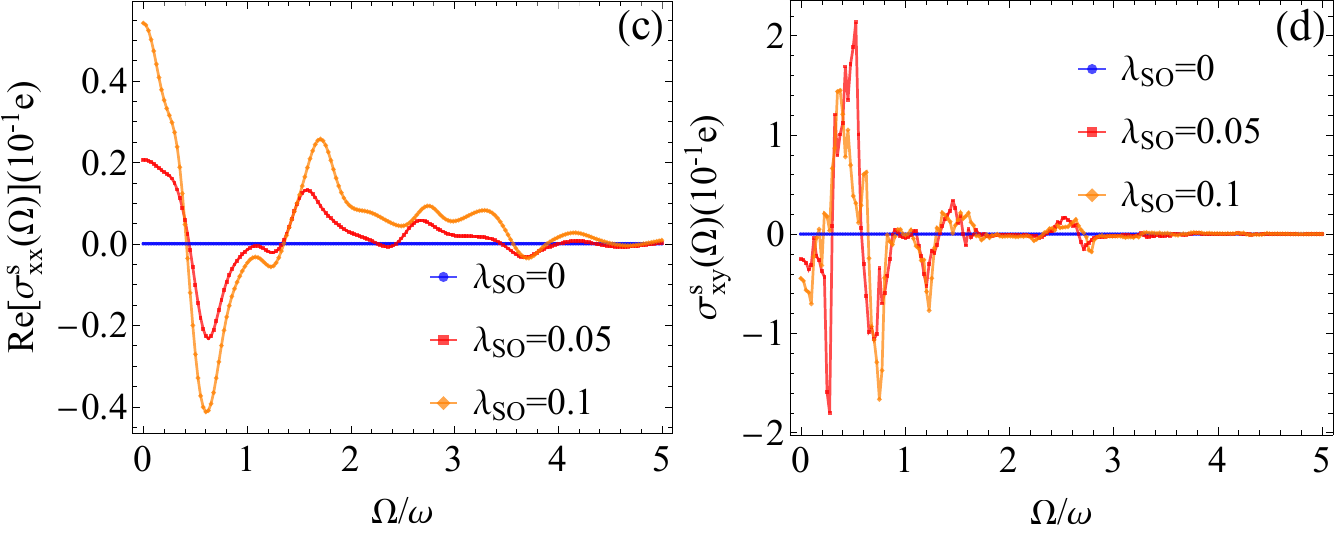}
\end{tabular}
\caption{(a) Steady state spin current with $\lambda_{\text{SO}}=0$, where the unit $t/a\sim 10^9$ eV/m. (b) The diagram for the current direction and magnitude. (c,d) Longitudinal and transverse optical conductivity in honeycomb AFM with $\varphi=\pi/2$. In the plots other parameters are $A_0a=1,\omega=1$, $t=1,\lambda=0.5$, and $T_{\text{ph}}=0.01 t$.}\label{fig:spinCurrent_honeycomb}
\end{figure}

The steady-state distribution obtained from Eq.~\eqref{eq:kinetic} deviates markedly from the conventional Fermi-Dirac form. In particular, the initially fully occupied lower Floquet bands become partially depleted, while the empty upper bands acquire finite occupation, as shown in Fig.\ref{fig:population}. This redistribution leads to an effectively metallic steady state, despite the presence of an initial band gap, as evident in the finite conductivity shown in Fig.~\ref{fig:spinCurrent_honeycomb}(c). Notably, the dual symmetry in Eq.~\eqref{eq:dual} imposes a corresponding symmetry in the momentum-resolved occupations: $\rho^{\uparrow}_{u,d}(\mathbf k)=\rho^{\downarrow}_{u,d}(-\mathbf{k})$, see Fig.~\ref{fig:population} (a),{\color{blue}(c)}. This duality in population plays a key role in enabling spin transport, as we discuss below.


In the non-equilibrium steady state, a nonvanishing current is allowed to exist.
The averaged spin current density over one period is given by $\bar{\mathbf j}^{s}=\bar{\mathbf j}^{\uparrow}-\bar{\mathbf j}^{\downarrow}$ with
\begin{eqnarray}\label{eq:steadycurrent}
\bar{\mathbf j}^{\sigma}=
\sum_n\int\frac{d^2\mathbf k}{(2\pi)^2}\rho^{\sigma}_n(\mathbf k)\partial_{\mathbf k}\varepsilon^{\sigma}_{n\mathbf k},
\end{eqnarray}
where $\sigma=\uparrow,\downarrow$~\cite{Supp}. The inversion symmetry in each spin sector is broken, see Fig.~\ref{fig:honeycombAFM} (c) and Fig.~\ref{fig:population}.
In the absence of intrinsic SOC, the dual symmetry [Eq.~\eqref{eq:dual}] ensures that the spin-resolved currents are equal in magnitude and opposite in direction, resulting in a pure spin current with vanishing net charge current. When intrinsic SOC is introduced, this dual symmetry is lifted, allowing both spin and charge currents to coexist. The magnitude and direction of the steady-state spin current in the SOC-free case are shown in Figs.~\ref{fig:spinCurrent_honeycomb}(a),(b) as functions of the light polarization angle, revealing a high degree of tunability via the optical field. 

In Fig.~\ref{fig:spinCurrent_honeycomb}(a), the oscillating spin current exhibits two characteristic symmetry properties.
\textbf{(i)} The current is antisymmetric about $\varphi=\pi$, i.e.,
$\bar{\mathbf j}^s(2\pi-\varphi)=-\bar{\mathbf j}^s(\varphi)$. \textbf{(ii)} In the interval $[0,\pi]$, the components $\bar j_x^s$ and $\bar j_y^s$ are respectively symmetric and antisymmetric about $\pi/2$, satisfying
$\bar j_x^s(\tfrac{\pi}{2}-\varphi)=\bar j_x^s(\tfrac{\pi}{2}+\varphi)$ and
$\bar j_y^s(\tfrac{\pi}{2}-\varphi)=-\bar j_y^s(\tfrac{\pi}{2}+\varphi)$. Property \textbf{(i)} follows from the relation
$H_{\sigma}[\mathbf k+\boldsymbol{\mathcal A}(2\pi-\varphi,t)]
=\mathcal T H^{\ast}_{-\sigma}[-\mathbf k+\boldsymbol{\mathcal A}(\varphi,t)]\mathcal T^{-1}$,
which implies
$\varepsilon^{\sigma}_{n\mathbf k}(2\pi-\varphi)=\varepsilon^{-\sigma}_{n,-\mathbf k}(\varphi)$ and
$\rho^{\sigma}_n(\mathbf k,2\pi-\varphi)=\rho^{-\sigma}_n(-\mathbf k,\varphi)$; inserting these into Eq.~\eqref{eq:steadycurrent} directly yields the antisymmetry.
Property \textbf{(ii)} originates from the $\mathcal M_x\mathcal T$ symmetry of the lattice, where $\mathcal M_x$ denotes mirror reflection about the $x$ axis. Under $\mathcal M_x\mathcal T$, the gauge field transforms as
$\boldsymbol{\mathcal A}(\tfrac{\pi}{2}+\varphi,t)\to\boldsymbol{\mathcal A}(\tfrac{\pi}{2}-\varphi,t)$,
while the spin current transforms as $(j_x^s,j_y^s)\to(j_x^s,-j_y^s)$, enforcing the symmetry relations in \textbf{(ii)}.


Beyond the steady-state spin current, linear-response spin transport induced by a weak external electric field provides an additional degree of control over spin manipulation~\cite{Dehghani2015OpticalCond,Chen2018FloquetSemimetal,rudner2020floquetengineershandbook}, particularly in systems where inversion symmetry prohibits a steady-state current. Unlike the steady-state response, the linear optical spin conductivity can be strongly constrained by the dual symmetry in Eq.~\eqref{eq:dual}, which enforces cancellation between contributions from the two spin sectors. Consequently, a finite response requires the inclusion of intrinsic spin-orbit coupling. Figures~\ref{fig:spinCurrent_honeycomb}(c),(d) show the computed longitudinal and transverse spin conductivities, respectively, where the oscillatory behavior originates from resonances between the external ac field frequency $\Omega$ and the Floquet quasi-energy gaps. Notably, both components are finite in the DC limit, despite the fact that the undriven, half-filled system is a band insulator. This indicates that the interplay between periodic driving and phonon-mediated relaxation leads to an effectively metallic steady state, enabling finite spin transport under small electric fields.

{\color{blue}\textit{Nonrelativistic Edelstein effect.}} The spin-conserving nature of electron-phonon coupling prohibits net spin accumulation, as particle numbers in each spin sector are conserved. This constraint can be lifted by coupling the system to a fermionic reservoir through tunneling terms of the form $c_i^\dagger d_j$, which enables exchange of particles between the system ($c_i^\dagger$) and the leads ($d_j$). In the presence of a spin-split Floquet band structure, effective spin nonconservation then emerges as an imbalance between spin-up and spin-down particle inflow into the driven system, giving rise to a finite spin accumulation~\cite{note1}. Importantly, spin remains locally conserved and spin current is well defined in this mechanism, in sharp contrast to scenarios where spin–orbit coupling explicitly breaks local spin conservation~\cite{Shitade2022spinAccumulation}.

To model the bath engineering effect, we consider a system coupled to two fermionic electrodes on its left and right sides, characterized by chemical potentials $\mu_L$ and $\mu_R$, respectively. The coupling to each lead is described by the parameters $\Gamma_{L}$ and $\Gamma_{R}$. The resulting spin-related phenomena are analyzed using Floquet-Keldysh theory~\cite{Oka2009photoGraphen,Liu2025FloquetWeyl, Kitagawa2011photoQuantumHall,Vahid2024FloquetGreen, Supp}.

\begin{figure}
    \centering
    \includegraphics[width=1\linewidth]{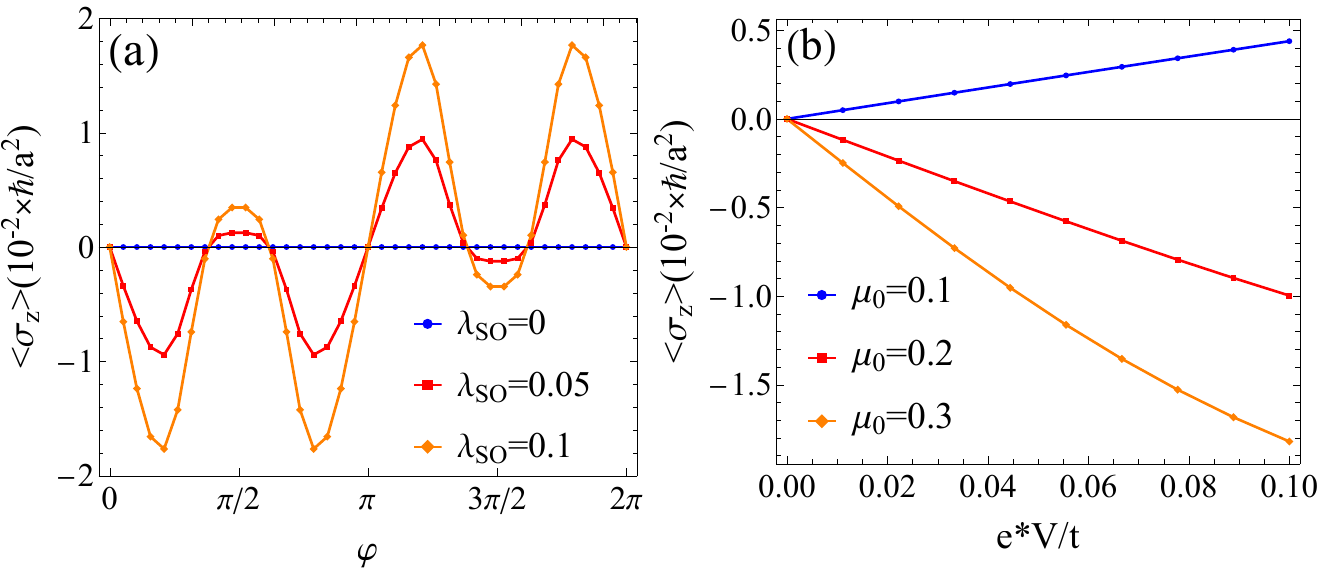}    \caption{Spin accumulation by contacting the system to electrodes with chemical potential $\mu_L, \mu_R$ on its left and right. (a) two leads are symmetric with $\mu_L=\mu_R = 0$. (b) Voltage induced spin accumulation, where $\mu_{L/R}=\mu_0\pm V/2$ and $\lambda_{\text{SO}}=0$, $\varphi=\pi/2$. Other parameters used in the calculations are $t=1, A_0a=1, \lambda = 0.5$, $\Gamma_L=\Gamma_R = 0.1t$, and $\omega =1$; the system contains 10 unit cells along the longitudinal direction.}
    \label{fig:spinAccumulation}
\end{figure}

When analyzing spin accumulation, the dual symmetry expressed in Eq.~\eqref{eq:dual} remains crucial. If the two leads are symmetric, i.e., $\mu_L=\mu_R$, the dual symmetry enforces vanishing spin accumulation even through the spin degeneracy and particle conservation are broken, necessitating finite SOC to induce spin accumulation [ Fig.~\ref{fig:spinAccumulation} (a)], where the oscillatory pattern is dictated by the same symmetry that governs the steady-state spin current discussed before~\cite{Supp}. Remarkably, the dual symmetry can also be broken by introducing asymmetry between the leads. For example, in Fig.~\ref{fig:spinAccumulation}(b), we consider leads with chemical potentials 
$\mu_{L/R}=\mu_0\pm V/2$. This asymmetry lifts the dual symmetry and leads to a sizable spin accumulation, which emerges as a linear response to the applied voltage $V$. The sign reversal of the spin signal serves as a hallmark of bath engineering. This SOC-independent effect is known in the literature as the nonrelativistic Edelstein effect~\cite{Aronov1989spin,EDELSTEIN1990spin,Pari2025NRedelstein,Chakraborty2025NRedelstein}.

\textcolor{blue}{\textit{Experimental feasibility}}. To assess experimental feasibility, we consider the following parameters:  $\omega=1$eV (set $t=1$eV), $A_0a=1$, and $a=5\mathring{A}$. These correspond to an electric field strength  $E_0=A_0\omega/e=2\times 10$MV/cm, resulting in a laser intensity $I\simeq 5\times 10^{11}$W/cm$^2$, which lies well within experimentally accessible regimes~\cite{Kitagawa2011photoQuantumHall,Schuler2022FloquetWse2}.
The input power delivered to the system can be estimated as
$P_{in}= \frac{1}{T}\int_0^T dt\mathbf E(t)\cdot \mathbf j^{e}(t)\sim E_0|\bar{\mathbf j}^e|\sim 10^8-10^9$W/cm$^2$, where $\mathbf E(t)=-\partial_t \boldsymbol{\mathcal A}$ is the electric field, $\mathbf j^e$ is the charge current. The current can be estimated as $\mathbf j^e\sim\frac{e}{\hbar}\mathbf j^{\uparrow/\downarrow}\approx 10^{-5} \sim 10^{-6}\frac{e}{\hbar}\frac{eV}{m}$ [see Fig.~\ref{fig:spinCurrent_honeycomb} (a)] in each spin sector. In the steady state, this input power must be balanced by dissipation. Taking phonons as a representative bath, the scattering rate is estimated as $R\sim 2\pi g_0^2 D_{ph}/\hbar\sim 10^{13}s^{-1}$, with electron-phonon coupling strength $g_0\approx 0.05 eV$ and phonon density of state $D_{ph}\sim 1 \text{eV}^{-1}$~\cite{Supp}. The corresponding dissipated power is roughly $P_{diss}\sim R \frac{(D_{ph}\Delta E) n_{ph} }{S_{u.c.}}\varepsilon_{ph}\sim 10^7 - 10^{9}$W/cm$^2$ where $\Delta E\approx 0.1\sim 1\hbar\omega$ is the phonon energy window, $n_{ph}\sim 1$ is the number of relevant phonon modes, $S_{u.c.}\sim a^2$ is the area of unit cell, and $\varepsilon_{ph}\approx 0.1\sim 1\hbar\omega$ is the phonon energy. From these estimates, the power balance condition 
$P_{in}=P_{diss}$ appears feasible. In practice, additional dissipation channels—such as external reservoirs, substrates, or magnons—can further help mitigate heating in driven systems.

\textcolor{blue}{\textit{Conclusion}.} In recent years, there has been a great interest in magnetic materials with compensated magnetic order but without spin degeneracy. We propose a new non-relativistic way to lift the Kramers spin degeneracy in antiferromagnetic systems by using light. The optically induced spin splitting is useful for generating both non-perturbative (steady-state) and perturbative (linear-response) spin currents, as well as spin accumulation. Our proposal uncovers the great potential of optical method for spin generation in antiferromagnetic spintronics. Experimental confirmation and application of our prediction are highly feasible, given that the proposal is generically valid for various antiferromagnetic materials. In the future, optical engineering for magnetic torque will be an intriguing direction. Moreover, controlling spin-related transport via tuning the thermal bath is also a very attractive idea.

\textcolor{blue}{\textit{Acknowledgements}.} 
This work was supported by the National Key R\&D Program of China (Grant No. 2024YFB3614101), the National Natural Science Foundation of China (Grants Nos. 12404185, 12274411, 12241405, and 52250418), the Basic Research Program of the Chinese Academy of Sciences Based on Major Scientific Infrastructures (Grant No. JZHKYPT-2021-08), and the CAS Project for Young Scientists in Basic Research (Grant No. YSBR-084). AAK acknowledges the support
by the U.S. Department of Energy, Office of Science, Basic Energy Sciences, under Award No. DE-SC0021019.


\textcolor{blue}{\textit{Note added}:} Recently, we became
aware of two relevant studies by S.
Huang et al.~\cite{Huang2025spinsplitting} and T. Zhu et al.~\cite{Zhu2025splitting}, which report findings partially overlapping with ours.

\bibliographystyle{apsrev4-1-title}
\bibliography{splitting}

\begin{thebibliography}{63}
\bibitem [{not()}]{note1}%
  \BibitemOpen
  \href@noop {} {}\bibinfo {note} {While the total spin of the combined system and leads is conserved, spin nonconservation within the driven region implies a net transfer of spin angular momentum from the leads into the system.}\BibitemShut {Stop}%
\end{thebibliography}

\appendix
\section{End Matter}

\textcolor{blue}{\textit{Other examples}.} We have illustrated the optically induced spin splitting and associated spin generation phenomena using a prototypical honeycomb antiferromagnet, but the underlying mechanism applies broadly to antiferromagnetic systems with appropriate symmetries, as demonstrated below.

\begin{figure}
\centering
\begin{tabular}{cc}
\includegraphics[width=0.5\linewidth]{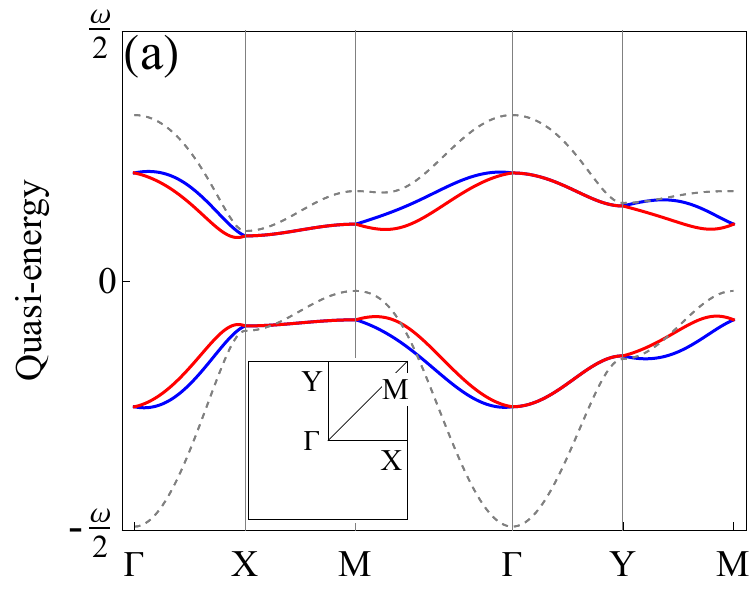}&
\includegraphics[width=0.5\linewidth]{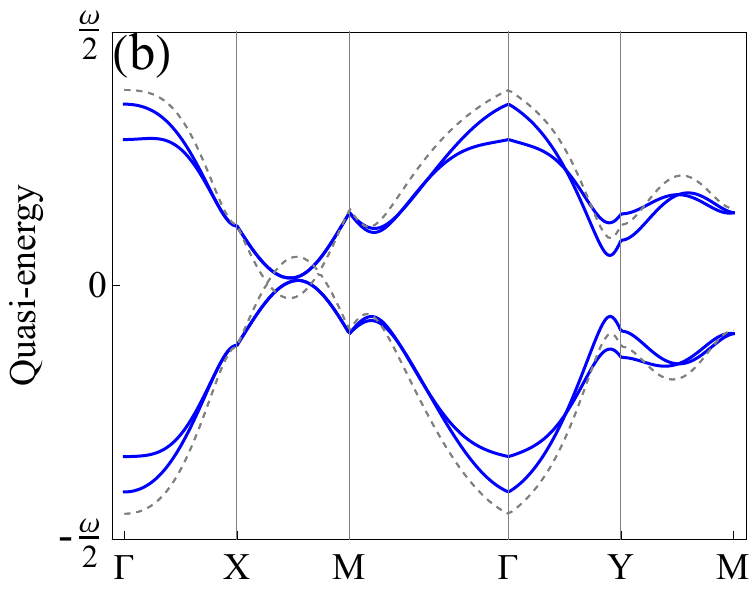}
\end{tabular}
\caption{Quasi-band structure in the first Floquet Brillouin zone. (a) The quasi-energy band of a nonsymmorphic AFM model, where blue (red) color denotes spin-up (down) bands, and parameters are $\varphi=\pi/2, A_0a=2,\omega=5$, $t=1$, $t^\prime=0.2$, $w=0.6$, and $\lambda=0.5$.
(b) Quasi-energy bands of the minimal model of tetragonal CuMnAs (each band is not spin-resolved), where $\varphi=\pi/2, A_0a= 1,\omega=5$, $t=1, t^\prime=0.08,\lambda=0.6, \alpha_R=0.8$, and $\mathbf n=(1,0,0)$. In both (a) and (b), the dashed gray lines represent the original degenerate bands, and they share the same Brillouin zone, see the inset of (a).}
\label{fig:examples}
\end{figure}

First, we consider an AFM model on square lattice with nonsymmorphic symmetry (possible materials include SrMnPb, SrMnSn, etc.~\cite{Niu2020AFMnonsymmorphicTI}), which is discussed in the context of antiferromagnetic Dirac semi-metal~\cite{Young2015DiracSM,Wang2017AFMdirac,Niu2020AFMnonsymmorphicTI}. The Hamiltonian reads $H(\mathbf k)=H_0(\mathbf k)+H^\prime(\mathbf k)$ where 
\begin{eqnarray}
H_0(\mathbf k)&=&-2t\cos\frac{k_x}{2}\cos\frac{k_y}{2}\tau^x-t^\prime(\cos k_x+\cos k_y)\nonumber\\
&&+\lambda\tau^z\boldsymbol{\sigma}\cdot\mathbf n, \label{eq:squarelattice}\\
H^\prime(\mathbf k)&=&w\sin\frac{k_y}{2}\cos\frac{k_x}{2}\tau^y.   
\end{eqnarray}
Here, $H_0$ describes the square-lattice model with AFM order, $H^\prime$ reflects the anisotropic nearest hopping strength~\cite{Niu2020AFMnonsymmorphicTI}.
As shown in Fig.~\ref{fig:examples} (a), the light-induced quasi-energy bands are spin-split even without involving SOC. It is worth noting that $H^\prime$ breaks the inversion symmetry, which is important for the spin splitting~\cite{Supp}.


Another example with PT symmetry is the minimal model for tetragonal CuMnAs ~\cite{Smejkal2017DiracAFM, Watanabe2021photocurrent}
\begin{eqnarray}
H(\mathbf k)&=&H_0(\mathbf k)+\alpha_R\tau^z(\sigma^y\sin k_x-\sigma^x\sin k_y)
\end{eqnarray}
where $H_0$ is given by Eq.~\eqref{eq:squarelattice}, $\alpha_R$ is the Rashba spin orbit coupling parameter. In contrast to the above examples, even with Kramers degeneracy, spin is not conserved here. Therefore, optical driving not only removes the spin degeneracy [see Fig.~\ref{fig:examples} (b)] but also affects the spin texture in each quasi-energy band, see Ref.~\cite{Supp,Ghorashi2025FloquetAltermagnet}. This offers an opportunity for engineering N\'eel torque, which is an interesting direction for future study.

\end{document}


\title{Supplemental Material for ``Floquet Spin Splitting and Spin Generation in Antiferromagnets"}

\author{Bo Li}
\email{libphysics@xjtu.edu.cn}
\affiliation{MOE Key Laboratory for Nonequilibrium Synthesis and Modulation of Condensed Matter,\\
Shaanxi Province Key Laboratory of Quantum Information and Quantum Optoelectronic Devices,\\
School of Physics, Xi’an Jiaotong University, Xi’an 710049, China}

\author{Ding-Fu Shao}
\email{dfshao@issp.ac.cn}
\affiliation{Key Laboratory of Materials Physics, Institute of Solid State Physics, HFIPS, Chinese Academy of Sciences, Hefei 230031, China}

\author{Alexey A. Kovalev}
\email{alexey.kovalev@unl.edu}
\affiliation{Department of Physics and Astronomy and Nebraska Center for Materials and Nanoscience,
University of Nebraska, Lincoln, Nebraska 68588, USA}

\maketitle

\onecolumngrid

\tableofcontents

\section{Details of the honeycomb model}

\begin{figure}
\centering
\includegraphics[width=0.6\linewidth]{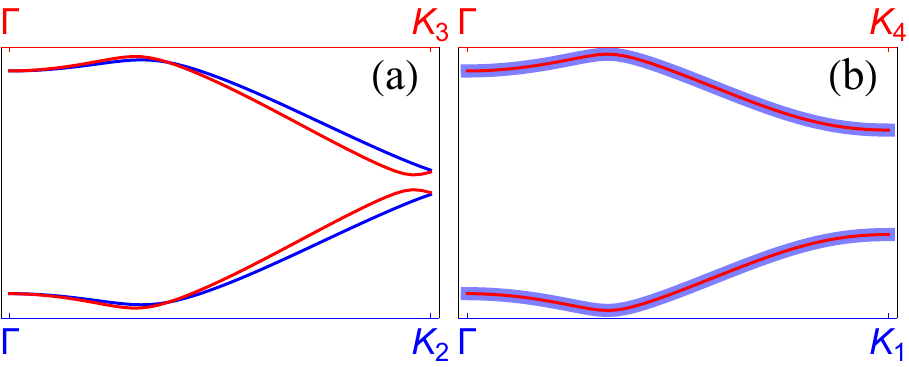}
\caption{Comparison between spin-up and down bands along different paths, showing that the three-fold rotation of the honeycomb lattice is broken by light. Here, $\varphi=\pi/3, A_0a=1,\omega=4$ and $t=1,\lambda=0.5$.}
\label{fig:comparison}
\end{figure}

Here, we elaborate on the three-fold rotation symmetry breaking by light in the honeycomb model. In Fig.~\eqref{fig:comparison}, we compare the quasi-energy dispersion among different paths. If the three-fold rotation symmetry still holds, the path $\Gamma-K_1$ and $\Gamma-K_3$ are supposed to be equivalent, so do the path $\Gamma-K_2$ and $\Gamma-K_4$. According to the dual relation $\varepsilon_{u,d}^\uparrow(\mathbf k)=\varepsilon_{u,d}^{\downarrow}(-\mathbf k)$, the quasi-energy dispersions for opposite spins along $\Gamma-K_1$ and $\Gamma-K_3$ are degenerate, as plotted in Fig.~\ref{fig:comparison} (b). However, such a degeneracy is not transmitted to the path $\Gamma-K_1$ and $\Gamma-K_3$, as shown in Fig.~\ref{fig:comparison} (a). This shows the three-fold rotation symmetry is broken by light.

\section{Steady state and transport-- the case for phonon bath}

In this section, we first derive a kinetic equation for the steady-state quasienergy-band occupations of a Floquet electronic system coupled to a bosonic bath, following Ref.~\cite{Seetharam2015population}. We then specialize to electron–phonon coupling in a hexagonal lattice, which is used to determine the steady state in the main text.

\subsection{Kinetic theory for the steady state}

We focus on the time scale much longer than the driving period, the occupation function for a given Floquet state $|\psi_{\mathbf k,n}\rangle$ respects the following kinetic equation
\begin{eqnarray}
\partial_t \rho_{\mathbf k,n}=I_{coll}[\rho_{\mathbf k,n}]    
\end{eqnarray}
where the r.h.s. denotes the collision integral due to coupling to different thermal baths.

When we consider the system is coupled to bosonic bath, there are two process contribute dominantly and compete with each other. (i) The photon-assisted radiative recombination process $I^{rec}$, during which electrons jump from nondriven conduction band to valence band by emitting a photon; however, this process increase the excitation density of the Floquet band by promoting lower-band electrons to upper band electrons due to the external driving force. (ii) The electron-phonon scattering process $I^{ph}$, which relaxes the excited electrons to the bottom of the upper Floquet band or relaxes the excitated electrons in the upper band to the lower band. 

Let us consider a bosonic bath described by $H_b=\sum_{\mathbf q,\lambda}E_{\lambda,\mathbf q} b^\dagger_{\mathbf q,\lambda}b_{\mathbf q,\lambda}$ and the electronic system is coupled to the bath via
\begin{eqnarray}
H_{int}=\sum_{\mathbf k,\mathbf q}\sum_{\nu,\nu^\prime}\sum_{\lambda}g^\lambda_{\nu^\prime\nu}(\mathbf k-\mathbf q,\mathbf k) c^\dagger_{\mathbf k-\mathbf q,\nu^\prime}c_{\mathbf k,\nu}(b^\dagger_{\mathbf q,\lambda}+b_{-\mathbf q,\lambda}),   
\end{eqnarray}
where $g^\lambda_{\nu^\prime\nu}(\mathbf k-\mathbf q,\mathbf k) $ is the coupling matrix element.
Because of the Hermiticity of $H_{int}$, the coupling matrix element satisfies the following relation
\begin{eqnarray}\label{eq:Hermitian_cond}
[g^\lambda_{\nu^\prime\nu}(\mathbf k-\mathbf q,\mathbf k)]^\ast=g^\lambda_{\nu\nu^\prime}(\mathbf k,\mathbf k-\mathbf q).    
\end{eqnarray}
A state in the Floquet basis is given as  follows: 
\begin{eqnarray}
|\psi_{\mathbf k,\alpha}(t)\rangle=\sum_n e^{-i(\varepsilon_{\mathbf k,\alpha}+n\omega)t}|\phi^n_{\mathbf k,\alpha}\rangle.    
\end{eqnarray}
Given the completeness of the basis $\mathbbm 1=\sum_{\alpha}|\psi_{\mathbf k,\alpha}(t)\rangle\langle \psi_{\mathbf k,\alpha}(t)|$, the creation and annihilation operator in original basis can be represented in the Floquet basis 
\begin{eqnarray}
c_{\mathbf k\nu}^\dagger=\sum_{\alpha,n}e^{i(\varepsilon_{\mathbf k\alpha}+n\omega)t} \langle\phi^n_{\mathbf k,\alpha}|\nu\mathbf k\rangle f^\dagger_{\mathbf k\alpha}(t),\nonumber\\
c_{\mathbf k\nu}=\sum_{\beta,m}e^{-i(\varepsilon_{\mathbf k\beta}+m\omega)t} \langle\nu\mathbf k|\phi^m_{\mathbf k,\beta}\rangle f_{\mathbf k\beta}(t),
\end{eqnarray}
where $f^\dagger_{\mathbf k\alpha}\leftrightarrow|\psi_{\mathbf k,\alpha}(t)\rangle$. Then, the electron-boson interaction is turned to Floquet basis
\begin{eqnarray}
H_{int}=\sum_{\mathbf k,\mathbf q}\sum_{\alpha,\beta}\sum_{\lambda}\sum_n e^{i(\varepsilon_{\mathbf k-\mathbf q,\alpha}-\varepsilon_{\mathbf k,\beta})t+in\omega t}\mathcal G_{\lambda,\alpha\beta}^{(n)}(\mathbf k-\mathbf q,\mathbf k)f^\dagger_{\mathbf k-\mathbf q,\alpha}(t)f_{\mathbf k,\beta}(t)(b^\dagger_{\mathbf q,\lambda}+b_{-\mathbf q,\lambda}).    
\end{eqnarray}
Here, 
\begin{eqnarray}
\mathcal G_{\lambda,\alpha\beta}^{(n)}(\mathbf k-\mathbf q,\mathbf k)&&=\sum_{\nu,\nu^\prime}\sum_m g^\lambda_{\nu^\prime\nu}(\mathbf k-\mathbf q,\mathbf k) \langle\phi^{m+n}_{\mathbf k-\mathbf q,\alpha}|\nu^\prime\mathbf k-\mathbf q\rangle \langle\nu\mathbf k|\phi^m_{\mathbf k,\beta}\rangle\nonumber\\
&&=\sum_m \langle\phi^{m+n}_{\mathbf k-\mathbf q,\alpha}|V^\lambda(\mathbf k-\mathbf q,\mathbf k)|\phi^m_{\mathbf k,\beta}\rangle
\end{eqnarray}
where $V^\lambda(\mathbf k-\mathbf q,\mathbf k)=g^\lambda_{\nu^\prime\nu}(\mathbf k-\mathbf q,\mathbf k)|\nu^\prime\mathbf k-\mathbf q\rangle \langle\nu\mathbf k|$.
Following Eq.~\eqref{eq:Hermitian_cond}, it is readily to check 
\begin{eqnarray}\label{eq:hermitian_cond1}
[\mathcal G_{\lambda,\alpha\beta}^{(n)}(\mathbf k-\mathbf q,\mathbf k)]^\ast= \mathcal G_{\lambda,\beta\alpha}^{(-n)}(\mathbf k,\mathbf k-\mathbf q).   
\end{eqnarray}

Our goal is to determine the equation of motion of $\rho_{\mathbf{k}\alpha}=\langle f^\dagger_{\mathbf k,\alpha}(t)f_{\mathbf k,\alpha}(t)\rangle$, where $\langle\cdots\rangle=\langle\Psi|\cdots|\Psi\rangle$ with $|\Psi\rangle$ being the state of whole system, including the electronic system, baths and their coupling, i.e., $i\partial_t|\Psi\rangle=H_{tot}|\Psi\rangle$. Here and after we will use the compact subscript $a=(\mathbf k,\alpha)$. The time evolution of operator is given by $f^\dagger_a(t)=U(t,t^\prime)f_a(t^\prime)U^\dagger(t,t^\prime)$ where $U(t,t^\prime)=\overrightarrow{\mathcal T}\exp[-i\int_{t^\prime}^t d\tau H(\tau)]$, $\overrightarrow{\mathcal T}$ is the time ordering operator. Therefore, the EOM for $f^\dagger_a$ reads
\begin{eqnarray}
i\partial_t f^\dagger_a(t)=[H(t),f^\dagger_a(t)]. 
\end{eqnarray}
For more general tensor $P_{ab}=\langle f^\dagger_a(t) f_b(t)\rangle$, its EOM is given as below
\begin{eqnarray}\label{eq:tensordyn1}
i\partial_t P_{ab}&&= i\partial_t\left(\langle\Psi|\right)f^\dagger_a f_b|\Psi\rangle+\langle\Psi|f^\dagger_a f_b\left(i\partial_t|\Psi\rangle\right)+\langle (i\partial_t f^\dagger_a)f_b\rangle+\langle f^\dagger_a(i\partial_t f_b)\rangle\nonumber\\
&&=\langle [f_a^\dagger f_b,H_{tot}-H(t)]\rangle.
\end{eqnarray}
More explicitly, 
\begin{eqnarray}\label{eq:tensordyn2}
[f^\dagger_a f_b, H_{int}]=&& \sum_{a^\prime,b^\prime}\sum_\lambda\sum_{n}e^{i(\varepsilon_{a^\prime}-\varepsilon_{b^\prime})t+in\omega t}\mathcal G_{\lambda,a^\prime b^\prime}^{(n)}(\mathbf k_{b^\prime}-\mathbf k_{a^\prime})\langle [f^\dagger_a f_b, f^\dagger_{a^\prime}f_{b^\prime}]A_{\lambda,\mathbf k_{b^\prime}-\mathbf k_{a^\prime}}\nonumber\\
=&&\sum_{b^\prime}\sum_\lambda\sum_{n} e^{i(\varepsilon_b-\varepsilon_{b^\prime})t+in\omega t}\mathcal G^{(n)}_{\lambda,bb^\prime}(\mathbf k_{b^\prime}-\mathbf k_b)\langle f^\dagger_a f_{b^\prime}A_{\lambda,\mathbf k_{b^\prime}-\mathbf k_b}\rangle \nonumber\\
&&-\sum_{a^\prime}\sum_\lambda\sum_n e^{i(\varepsilon_{a^\prime}-\varepsilon_{a})t+in\omega t}\mathcal G^{(n)}_{\lambda,a^\prime a}(\mathbf k_{a}-\mathbf k_{a^\prime})\langle f^\dagger_{a^\prime} f_{b}A_{\lambda,\mathbf k_{a}-\mathbf k_{a^\prime}}\rangle\nonumber\\
=&&\sum_{b^\prime}\sum_\lambda\sum_{n} e^{i(\varepsilon_b-\varepsilon_{b^\prime})t+in\omega t}\mathcal G^{(n)}_{\lambda,bb^\prime}(\mathbf q)\langle f^\dagger_a f_{b^\prime}A_{\lambda,\mathbf q}\rangle - \sum_{a^\prime}\sum_{\lambda}\sum_n e^{i(\varepsilon_{a^\prime}-\varepsilon_{a})t+in\omega t}\mathcal G^{(n)}_{\lambda,a^\prime a}(\mathbf q)\langle f^\dagger_{a^\prime} f_{b}A_{\lambda,\mathbf q}\rangle,\nonumber\\
\end{eqnarray}
where $A_{\lambda,\mathbf q}=b^\dagger_{\mathbf q,\lambda}+b_{-\mathbf q,\lambda}$, $\mathcal G^{(n)}_{\lambda,a^\prime b^\prime}(\mathbf q)=\mathcal G^{(n)}_{\lambda,\alpha\beta}(\mathbf k-\mathbf q,\mathbf k)$, and $\sum_{a^\prime}=\sum_{\alpha^\prime}\sum_{\mathbf k^\prime}\rightarrow\sum_{\alpha^\prime}\sum_{\mathbf q}$ (similar for $\sum_{b^\prime}$). Here, we used the relation $[f^\dagger_a f_b, f^\dagger_{a^\prime}f_{b^\prime}]=f^\dagger_af_{b^\prime}\delta_{ba^\prime}-f^\dagger_{a^\prime}f_b\delta_{ab^\prime}$. For simplicity, we focus on the diagonal coherence regime $P_{ab}\propto\delta_{ab}$. It is straightfoward to show that
\begin{eqnarray}\label{eq:fbinteraction}
i\partial_t\langle f^\dagger_{a} f_{b}b_{-\mathbf q,\lambda}\rangle&&=\langle[f^\dagger_a f_bb_{-\mathbf{q},\lambda},H_{int}+H_b]\rangle\nonumber\\
&&=E_{\lambda,-\mathbf q}\langle f^\dagger_{a} f_{b}b_{-\mathbf q,\lambda}\rangle +\sum_{a^\prime,b^\prime}\sum_n e^{i(\varepsilon_{a^\prime}-\varepsilon_{b^\prime})t+in\omega t}\mathcal G^{(n)}_{\lambda,a^\prime b^\prime}(\mathbf q^\prime)\langle[f_a^\dagger f_b b_{-\mathbf q,\lambda},f^\dagger_{a^\prime}f_{b^\prime}A_{\lambda,\mathbf q^\prime}]\rangle\nonumber\\
&&= E_{\lambda,-\mathbf q}\langle f^\dagger_{a} f_{b}b_{-\mathbf q,\lambda}\rangle+
\sum_{a^\prime}\sum_n e^{in\omega t}\mathcal G^{(n)}_{\lambda,a^\prime a^\prime}(0)P_{aa}P_{a^\prime a^\prime}\delta_{ab}[\Tilde{n}_B(E_{\lambda,-\mathbf q})-n_B(E_{\lambda,-\mathbf q})]\delta_{\mathbf q^\prime,-\mathbf q}\nonumber\\
&&-
\sum_{n}e^{i(\varepsilon_b-\varepsilon_a)t+in\omega t}\mathcal G^{(n)}_{\lambda,ba}(-\mathbf q)[P_{aa}(1-P_{bb})\Tilde{n}_B(E_{-\mathbf q})-P_{bb}(1-P_{aa})n_B(E_{-\mathbf q})]
\end{eqnarray}
where $\mathcal G^{(n)}_{\lambda,a^\prime a^\prime}(0)=0$, i.e., there is no scattering between the same state, and $\Tilde{n}_B(x)=1+n_B(x)$ with $n_B(x)$ being the Bose-Einstein distribution $n_B(x)=1/(e^{\beta x}-1)$. In the derivation, we used 
\begin{eqnarray}
\langle[f_a^\dagger f_b b_{-\mathbf q},f^\dagger_{a^\prime}f_{b^\prime}A_{\mathbf q^\prime}]\rangle=&&\langle f_a^\dagger f_b f^\dagger_{a^\prime} f_{b^\prime}\rangle\langle b_{-\mathbf q}A_{\mathbf q^\prime}\rangle -\langle f^\dagger_{a^\prime} f_{b^\prime}f_a^\dagger f_b \rangle\langle A_{\mathbf q^\prime} b_{-\mathbf q}\rangle\nonumber\\
=&&\delta_{\mathbf{q}^{\prime},-\mathbf q}[P_{aa}P_{a^\prime a^\prime}\delta_{ab}\delta_{a^\prime b^\prime}-P_{aa}(1-P_{bb})\delta_{a b^\prime}\delta_{a^\prime b}][1-n_B(E_{-\mathbf q})]\nonumber\\
&&- \delta_{\mathbf{q}^{\prime},-\mathbf q}[P_{aa}P_{a^\prime a^\prime}\delta_{ab}\delta_{a^\prime b^\prime}-P_{bb}(1-P_{aa})\delta_{a b^\prime}\delta_{a^\prime b}]n_B(E_{-\mathbf q})
\end{eqnarray}
where $\langle b_{-\mathbf q}A_{\mathbf q^\prime}\rangle=\delta_{\mathbf{q}^{\prime},-\mathbf q}[1+n_B(E_{-\mathbf q})]$ and $\langle A_{\mathbf q^\prime} b_{-\mathbf q}\rangle=\delta_{\mathbf{q}^{\prime},-\mathbf q}n_B(E_{-\mathbf q})$. Here,
\begin{eqnarray}
\langle f_a^\dagger f_b f^\dagger_{a^\prime} f_{b^\prime}\rangle=&&\delta_{ab}\delta_{a^\prime b^\prime}(1-\delta_{aa^\prime})\langle f_a^\dagger f_b\rangle\langle f^\dagger_{a^\prime}f_{b^\prime}\rangle\nonumber\\
&&+\delta_{ab^\prime}\delta_{ba^\prime}(1-\delta_{aa^\prime})\langle f_a^\dagger f_{b^\prime}\rangle\langle 1-f^\dagger_{a^\prime} f_{b}\rangle\nonumber\\
&&+\delta_{ab}\delta_{a^\prime b^\prime}\delta_{aa^\prime}\langle f^\dagger_a f_a f^\dagger_a f_a\rangle\nonumber\\
=&&\delta_{ab}\delta_{a^\prime b^\prime}\langle f_a^\dagger f_a\rangle\langle f^\dagger_{a^\prime}f_{a^\prime}\rangle+\delta_{ab^\prime}\delta_{ba^\prime}\langle f_a^\dagger f_{a}\rangle(1-\langle f^\dagger_{b} f_{b}\rangle)\nonumber.
\end{eqnarray}
Given that the differential equation of form 
$i\partial_t\mathcal O=\lambda\mathcal O+\xi(t)$ can be rewritten as $i\partial_t(e^{i\lambda t}\mathcal O)=e^{i\lambda t}\xi(t)$, Eq.~\eqref{eq:fbinteraction} is solved by
\begin{eqnarray}
\langle f^\dagger_{a} f_{b}b_{-\mathbf q,\lambda}\rangle=ie^{-iE_{\lambda,-\mathbf q}t}\sum_n \int_0^t d\tau e^{i(\varepsilon_b-\varepsilon_a)\tau+in\omega\tau}e^{i E_{\lambda,-\mathbf q}\tau} \mathcal G^{(n)}_{\lambda,ba}(-\mathbf q)[P_{aa}(1-P_{bb})\Tilde{n}_B(E_{-\mathbf q})-P_{bb}(1-P_{aa})n_B(E_{-\mathbf q})].\nonumber\\
\end{eqnarray}
Considering that  $\langle b^\dagger_{\mathbf q,\lambda}A_{\lambda,\mathbf q^\prime}\rangle=\delta_{\mathbf{q}^{\prime},-\mathbf q}n_B(E_{\lambda,\mathbf q})$, $\langle A_{\lambda,\mathbf q^\prime} b^\dagger_{\mathbf q,\lambda}\rangle=\delta_{\mathbf{q}^{\prime},-\mathbf q}\Tilde{n}_B(E_{\lambda,\mathbf q})$ and $[b_{\mathbf q,\lambda}^\dagger, H_b]=-E_{\lambda,\mathbf q}b^\dagger_{\mathbf q,\lambda}$, we obtain 
\begin{eqnarray}
\langle f^\dagger_{a} f_{b}b^\dagger_{\mathbf q,\lambda}\rangle=ie^{iE_{\lambda,\mathbf q}t}\sum_n \int_0^t d\tau e^{i(\varepsilon_b-\varepsilon_a)\tau+in\omega\tau}e^{-iE_{\lambda,\mathbf q}\tau}\mathcal G^{(n)}_{\lambda,ba}(-\mathbf q)[P_{aa}(1-P_{bb})n_B(E_{\lambda,\mathbf q})-P_{bb}(1-P_{aa})\Tilde{n}_B(E_{\lambda,\mathbf q})].\nonumber\\    
\end{eqnarray}

From Eq.~\eqref{eq:tensordyn1},\eqref{eq:tensordyn2} we obtain the dynamic equation for occupation density as below
\begin{eqnarray}
\partial_t\rho_a&&=\sum_{a^\prime}\sum_{\lambda}\sum_{n}\Big( \frac{1}{i}e^{i(\varepsilon_a-\varepsilon_{a^\prime})t+in\omega t}\mathcal G^{(n)}_{\lambda,aa^\prime}(\mathbf q)\langle f^\dagger_a f_{a^\prime}A_{\lambda,\mathbf q}\rangle -\frac{1}{i} e^{i(\varepsilon_{a^\prime}-\varepsilon_{a})t-in\omega t}\mathcal G^{(-n)}_{\lambda,a^\prime a}(-\mathbf q)\langle f^\dagger_{a^\prime} f_{a}A_{\lambda,-\mathbf q}\rangle\Big)\nonumber\\
&&=2\sum_{a^\prime}\sum_\lambda\sum_{n}\text{Re}\left( \frac{1}{i}e^{i(\varepsilon_a-\varepsilon_{a^\prime})t+in\omega t}\mathcal G^{(n)}_{\lambda,aa^\prime}(\mathbf q)\langle f^\dagger_a f_{a^\prime}A_{\lambda,\mathbf q}\rangle\right)\nonumber\\
&&=2\sum_{a^\prime}\sum_\lambda\sum_{n,m}\text{Re}\left(e^{i(\varepsilon_a-\varepsilon_{a^\prime})t+in\omega t}e^{-iE_{\lambda,-\mathbf q}t}\frac{e^{i(\varepsilon_{a^\prime}-\varepsilon_a+m\omega+E_{\lambda,-\mathbf q})t}-1}{i(\varepsilon_{a^\prime}-\varepsilon_a+m\omega+E_{\lambda,-\mathbf q}+i0^+)}[\mathcal G^{(n)}_{\lambda,aa^\prime}(\mathbf q)\mathcal G^{(m)}_{\lambda,a^\prime a}(-\mathbf q)]\right)\nonumber\\
&&\qquad\qquad\quad\times[\rho_a\bar{\rho}_{a^\prime}\Tilde{n}_B(E_{\lambda,-\mathbf q})-\rho_{a^\prime}\Bar{\rho}_an_B(E_{\lambda,-\mathbf q})]\nonumber\\ 
&&+2\sum_{a^\prime}\sum_\lambda\sum_{n,m}\text{Re}\left(e^{i(\varepsilon_a-\varepsilon_{a^\prime})t+in\omega t}e^{iE_{\lambda,\mathbf q}t}\frac{e^{i(\varepsilon_{a^\prime}-\varepsilon_a+m\omega-E_{\lambda,\mathbf q})t}-1}{i(\varepsilon_{a^\prime}-\varepsilon_a+m\omega-E_{\lambda,\mathbf q}+i0^+)}[\mathcal G^{(n)}_{\lambda,aa^\prime}(\mathbf q)\mathcal G^{(m)}_{\lambda,a^\prime a}(-\mathbf q)]\right)\nonumber\\
&&\qquad\qquad\quad\times[\rho_a\bar{\rho}_{a^\prime}n_B(E_{\lambda,\mathbf q})-\rho_{a^\prime}\Bar{\rho}_a\Tilde{n}_B(E_{\lambda,-\mathbf q})]\nonumber\\ 
&&\approx2\pi\sum_{a^\prime}\sum_{\lambda}\sum_{n}\Big(\delta(\varepsilon_{a^\prime}-\varepsilon_a+n\omega+E_{\lambda,-\mathbf q})|\mathcal G^{(n)}_{\lambda,a^\prime a}(-\mathbf q)\mathcal|^2[\rho_{a^\prime}\Bar{\rho}_an_B(E_{\lambda,-\mathbf q})-\rho_a\bar{\rho}_{a^\prime}\Tilde{n}_B(E_{\lambda,\mathbf q})]\nonumber\\
&&\qquad\qquad\quad +\delta(\varepsilon_{a^\prime}-\varepsilon_a-n\omega-E_{\lambda,\mathbf q})|\mathcal G^{(n)}_{\lambda,a^\prime a}(-\mathbf q)\mathcal|^2 [\rho_{a^\prime}\Bar{\rho}_a\Tilde{n}_B(E_{\lambda,\mathbf q})-\rho_a\bar{\rho}_{a^\prime}n_B(E_{\lambda,\mathbf q})]\Big).
\end{eqnarray}
In the last step, only the term with $m=-n$ is preserved and other fast oscillating terms are omitted. Now, we recover the abbreviated indices $a=(\mathbf k,\alpha)$ and  $a^\prime=(\alpha^\prime,\mathbf{k}-\mathbf{q})$
\begin{eqnarray}
\partial_t\rho_{\mathbf k\alpha}=&&2\pi\sum_{\mathbf q,\alpha^\prime}\sum_\lambda\sum_{n}\delta(\varepsilon_{\mathbf k-\mathbf{q},\alpha^\prime}-\varepsilon_{\mathbf k,\alpha}+n\omega+E_{\lambda,-\mathbf q})|\mathcal G^{(n)}_{\lambda,\alpha^\prime\alpha}(\mathbf k-\mathbf q,\mathbf k)|^2[\rho_{\mathbf k-\mathbf q,\alpha^\prime}\Bar{\rho}_{\mathbf k,\alpha}n_B(E_{\lambda,-\mathbf q})\nonumber\\
&&\qquad\qquad-\rho_{\mathbf k,\alpha}\bar{\rho}_{\mathbf k-\mathbf q,\alpha^\prime}\Tilde{n}_B(E_{\lambda,-\mathbf q})] +\delta(\varepsilon_{\mathbf k-\mathbf q,\alpha^\prime}-\varepsilon_{\mathbf k,\alpha}+n\omega-E_{\lambda,\mathbf q})|\mathcal G^{(n)}_{\lambda,\alpha^\prime\alpha}(\mathbf k-\mathbf q,\mathbf k)|^2\nonumber\\
&&\qquad\qquad\times[\rho_{\mathbf k-\mathbf q,\alpha^\prime}\Bar{\rho}_{\mathbf k,\alpha}\Tilde{n}_B(E_{\lambda,\mathbf q})-\rho_{\mathbf k,\alpha}\bar{\rho}_{\mathbf k-\mathbf q,\alpha^\prime}n_B(E_{\lambda,\mathbf q})].   
\end{eqnarray}
To simplify the equation above, we identify the scattering matrix as following:
\begin{eqnarray}\label{eq:scatteringmatrix}
W_{\mathbf k\alpha,\mathbf k^\prime\alpha^\prime}=&&2\pi\sum_\lambda\sum_{n} |\mathcal G^{(n)}_{\lambda,\alpha^\prime\alpha}(\mathbf k^\prime,\mathbf k)|^2 \Big[\delta(\varepsilon_{\mathbf k-\mathbf{q},\alpha^\prime}-\varepsilon_{\mathbf k,\alpha}+n\omega+E_{\lambda,-\mathbf q})n_B(E_{\lambda,-\mathbf q})\nonumber\\
&&\qquad\qquad+\delta(\varepsilon_{\mathbf k-\mathbf q,\alpha^\prime}-\varepsilon_{\mathbf k,\alpha}+n\omega-E_{\lambda,\mathbf q})\Tilde{n}_B(E_{\lambda,\mathbf q})\Big]\nonumber\\
=&&2\pi\sum_\lambda\sum_{n} |\mathcal G^{(n)}_{\lambda,\alpha\alpha^\prime}(\mathbf k,\mathbf k^\prime)|^2 \Big[\delta(\varepsilon_{\mathbf k,\alpha}-\varepsilon_{\mathbf k^\prime,\alpha^\prime}+n\omega-E_{\lambda,\mathbf k^\prime-\mathbf k})-\delta(\varepsilon_{\mathbf k,\alpha}-\varepsilon_{\mathbf k^\prime,\alpha^\prime}+n\omega+E_{\lambda,\mathbf k-\mathbf k^\prime})\Big]\nonumber\\
&&\qquad\qquad\quad\times n_B(\varepsilon_{\mathbf k,\alpha}-\varepsilon_{\mathbf k^\prime,\alpha^\prime}+n\omega)\nonumber\\
=&&2\pi\sum_\lambda\sum_{n} \Big|\sum_m\langle\phi^{m+n}_{\mathbf k,\alpha}|V^\lambda(\mathbf k,\mathbf k^\prime)|\phi^m_{\mathbf k^\prime,\alpha^\prime}\rangle\Big|^2 \Big[\delta(\varepsilon_{\mathbf k,\alpha}-\varepsilon_{\mathbf k^\prime,\alpha^\prime}+n\omega-E_{\lambda,\mathbf k^\prime-\mathbf k})-\nonumber\\
&&\qquad\qquad
\delta(\varepsilon_{\mathbf k,\alpha}-\varepsilon_{\mathbf k^\prime,\alpha^\prime}+n\omega+E_{\lambda,\mathbf k-\mathbf k^\prime})\Big] n_B(\varepsilon_{\mathbf k,\alpha}-\varepsilon_{\mathbf k^\prime,\alpha^\prime}+n\omega)
\end{eqnarray}
where $\mathbf k^\prime =\mathbf k-\mathbf q$. Applying the scattering matrix notation, the dynamical equation of occupation density becomes
\begin{eqnarray}\label{eq:kineticeq}
\partial_t\rho_{\mathbf k\alpha}=\sum_{\mathbf k^\prime\alpha^\prime}   W_{\mathbf k\alpha,\mathbf k^\prime\alpha^\prime}(1-\rho_{\mathbf k,\alpha})\rho_{\mathbf k^\prime,\alpha^\prime}-W_{\mathbf k^\prime\alpha^\prime,\mathbf k\alpha}(1-\rho_{\mathbf k^\prime\alpha^\prime})\rho_{\mathbf k\alpha}. 
\end{eqnarray}
The equation above can be written in a compact form:
\begin{eqnarray}\label{eq:dynamics}
\partial_t|\rho\rangle=(1-P)W|\rho\rangle -PW^T|1-\rho\rangle,  
\end{eqnarray}
where
$|\rho(t)\rangle =(\rho_1,\rho_2,\cdots, \rho_M)^T$ with $\rho_\nu=(\rho_{\nu\mathbf k_1},\rho_{\nu\mathbf k_2},\cdots,\rho_{\nu\mathbf k_N})$, where $M$ denotes the total number of quasi-bands in the first Floquet-Brillouin Zone, and $N$ stands for the number of discrete momentum values (equal to the number of lattice sites); $P=\text{Diag}(P_1,P_2,\cdots,P_M)$, where $P_\nu=\text{Diag}(\rho_{\nu\mathbf k_1},\rho_{\nu\mathbf k_2},\cdots,\rho_{\nu\mathbf k_N})$. Moreover, the electron number is conserved in the electron-phonon scattering. Suppose the initial state has a filling factor $n_f$, the total electron number is given by $n_fMN$. We have the normalization condition:
\begin{eqnarray}
\frac{1}{MN}\sum_{\mu,\mathbf k_j}\rho_{\mu,\mathbf k_j}=n_f.    
\end{eqnarray}
The electron number conservation can be directly seen by multiplying Eq.~\eqref{eq:dynamics} by $\langle 1|=\langle 1,1,\cdots,1 |$ ( $2MN$ components), leading to $\partial_t\sum_{\alpha,\mathbf k}\rho_{\alpha\mathbf k}=0$.\\

\textbf{Estimation of the energy dissipation}. In the main text, we discussed the balance between the energy dissipation to the thermal bath and the input power by the light. Here, we estimate the energy dissipation based on the scattering matrix Eq.~\eqref{eq:scatteringmatrix} and the kinetic equation~\eqref{eq:kineticeq}. The scattering rate can be approximated to
\begin{eqnarray}
R\sim&&\sum_{\mathbf k^\prime}   W_{\mathbf k\alpha,\mathbf k^\prime\alpha^\prime}\cdots \nonumber\\
=&& \sum_{\mathbf k^\prime-\mathbf k} \frac{2\pi}{\hbar}\sum_\lambda\sum_{n} \Big|\sum_m\langle\phi^{m+n}_{\mathbf k,\alpha}|V^\lambda(\mathbf k,\mathbf k^\prime)|\phi^m_{\mathbf k^\prime,\alpha^\prime}\rangle\Big|^2 \Big[\delta(\varepsilon_{\mathbf k,\alpha}-\varepsilon_{\mathbf k^\prime,\alpha^\prime}+n\omega-E_{\lambda,\mathbf k^\prime-\mathbf k})n_B(E_{\lambda,\mathbf k^\prime-\mathbf k})\nonumber\\
&&\qquad\qquad+
\delta(\varepsilon_{\mathbf k,\alpha}-\varepsilon_{\mathbf k^\prime,\alpha^\prime}+n\omega+E_{\lambda,\mathbf k-\mathbf k^\prime})\Big(1+n_B(E_{\lambda,\mathbf k^\prime-\mathbf k})\Big)\Big]\cdots\nonumber\\
=&&\frac{2\pi}{\hbar}\sum_\lambda\sum_{n} \int dE_{\lambda,\mathbf k^\prime-\mathbf k}D_{ph}(E_{\lambda,\mathbf k^\prime-\mathbf k}) \Big|\sum_m\langle\phi^{m+n}_{\mathbf k,\alpha}|V^\lambda(\mathbf k,\mathbf k^\prime)|\phi^m_{\mathbf k^\prime,\alpha^\prime}\rangle\Big|^2 \times(\cdots)\cdots\nonumber\\
\sim &&\frac{2\pi}{\hbar} D_{ph}g_0^2
\end{eqnarray}
where $D_{ph}$ is the phonon density of state (per unit energy) that is assumed uniform in the involved energy range, and $V^\lambda(\mathbf k,\mathbf k^\prime)\sim g_0$ with $g_0$ being the strength of electron-phonon coupling, see Eq.~\eqref{eq:couplingconst} below. The energy dissipation (per unit area)  is estimated as
\begin{eqnarray}
P_{diss}\sim R\frac{D_{ph}\Delta E}{S_{u.c.}}E_{ph},    
\end{eqnarray}
which approximately counts the energy relaxed to phonon modes per unit time and per unit area. Plugging parameters as in the main text can give the energy dissipation power.


\subsection{Electron-phonon interaction in a honeycomb lattice}

In this part, we specify the coupling matrix between electrons and phonons in a honeycomb lattice structure, following Ref.~\cite{Thingstad2020phononSC, Yu2024geomentryEPC}. In a lattice model, the phonons come from the variation of hopping energy due to the vibration of the ions' position. We only consider the contribution from nearest-neighbor hopping. Specifically, we consider the following Hamiltonian:
\begin{eqnarray}
H=\sum_{\mathbf r,\mu}t(\mathbf r+\mathbf u_{A,\mathbf r},\mathbf r+\boldsymbol{\delta}_{\mu}+\mathbf u_{B,\mathbf r+\boldsymbol{\delta}_\mu})(c^\dagger_{\mathbf r}c_{\mathbf r+\boldsymbol{\delta}_\mu}+h.c.)   
\end{eqnarray}
where $t(\cdots)$ is the hopping energy as a function of the position of two neighboring ions, $\mathbf u_{A,\mathbf r}$ and $\mathbf u_{B,\mathbf r+\boldsymbol{\delta}_\mu}$ are the position deviation of type-$A$ and $B$ ions at the corresponding position. We assume the hopping energy only depends on the distance between two ions, i.e., $t(\mathbf x_1,\mathbf x_2)=t(|\mathbf x_1-\mathbf x_2|)$. We can obtain the electron-phonon coupling by expanding the Hamiltonian to the leading order of the field $\mathbf u_{A,\mathbf r}$ and $\mathbf u_{B,\mathbf r+\boldsymbol{\delta}_\mu}$:\begin{eqnarray}
H_{el-ph}&&=\sum_{\mathbf r,\mu}(\mathbf u_{A,\mathbf r}-\mathbf u_{B,\mathbf r+\boldsymbol{\delta}_\mu})\cdot\boldsymbol{\nabla} t(\mathbf r)|_{\mathbf r=\boldsymbol{\delta}_{\mu}}(c^\dagger_{\mathbf r}c_{\mathbf r+\boldsymbol{\delta}_\mu}+h.c.)\nonumber\\
&&=\sum_{\mathbf r,\mu}\sum_{i=x,y}(u^i_{A,\mathbf r}-u^i_{B,\mathbf r+\boldsymbol{\delta}_\mu})f_i(\delta_\mu)(c^\dagger_{\mathbf r}c_{\mathbf r+\boldsymbol{\delta}_\mu}+h.c.),
\end{eqnarray} 
where $f_i(\delta_\mu)=\partial_i t(r)|_{r=\delta_\mu}$ with $\delta_\mu=|\boldsymbol{\delta}_\mu|$. Note that the $z$ component contribution of phonon modes vanishes because of the mirror symmetry with respect to the 2D plane. Next, we perform a Fourier transformation to convert the Hamiltonian to momentum space:
\begin{eqnarray}\label{eq:epcoupling}
H_{el-ph}&&=\sum_{\mathbf r,\mu}\sum_{i=x,y}\frac{1}{\sqrt{N}}\sum_{\mathbf q}\Big(e^{-i\mathbf q\cdot\mathbf r}u^i_{A,\mathbf q}-e^{-i\mathbf q\cdot(\mathbf r+\boldsymbol{\delta}_{\mu})}u^i_{B,\mathbf q}\Big)f_i(\delta_\mu)
\Big(\frac{1}{N}\sum_{\mathbf k_1,\mathbf k_2}e^{i\mathbf k\cdot\mathbf r}e^{i\mathbf k\cdot(\mathbf r+\boldsymbol{\delta}_\mu)}c^\dagger_{A\mathbf k_1}c_{B\mathbf k_2}\nonumber\\
&&\qquad\qquad\qquad+
\frac{1}{N}\sum_{\mathbf k_1,\mathbf k_2}e^{-i\mathbf k\cdot\mathbf r}e^{-i\mathbf k\cdot(\mathbf r+\boldsymbol{\delta}_\mu)}c^\dagger_{B\mathbf k_2}c_{A\mathbf k_1}
\Big)\nonumber\\
&&=\frac{1}{\sqrt{N}}\sum_{\mathbf k_1,\mathbf k_2}\sum_\mu \sum_{i=x,y}c^\dagger_{A\mathbf k_1}c_{B\mathbf k_2}f_i(\delta_\mu)\Big(u^i_{A,\mathbf k_1-\mathbf k_2}e^{-i\mathbf k_2\cdot\boldsymbol{\delta}_\mu}-u^i_{B,\mathbf k_1-\mathbf k_2}e^{-i\mathbf k_1\cdot\boldsymbol{\delta}_\mu}\Big)\nonumber\\
&&\qquad\qquad\qquad\qquad\quad +
c^\dagger_{B\mathbf k_1}c_{A\mathbf k_2}f_i(\delta_\mu)\Big(u^i_{A,\mathbf k_1-\mathbf k_2}e^{i\mathbf k_2\cdot\boldsymbol{\delta}_\mu}-u^i_{B,\mathbf k_1-\mathbf k_2}e^{-i\mathbf k_2\cdot\boldsymbol{\delta}_\mu}\Big)\nonumber\\
&&=\frac{1}{\sqrt{N}}\sum_{\mathbf k_1,\mathbf k_2}\sum_{i=x,y} c^\dagger_{A\mathbf k_1}c_{B\mathbf k_2}\Big(\sum_\mu f_i(\delta_\mu)e^{-i\mathbf k_2\cdot\boldsymbol{\delta}_\mu}\delta_{\tau,A}-\sum_\mu f_i(\delta_\mu)e^{-i\mathbf k_1\cdot\boldsymbol{\delta}_\mu}\delta_{\tau,B}\Big)u^i_{\tau,\mathbf k_1-\mathbf k_2}\nonumber\\
&&\qquad\qquad\qquad\quad +
c^\dagger_{B\mathbf k_1}c_{A\mathbf k_2}\Big(\sum_\mu f_i(\delta_\mu)e^{i\mathbf k_1\cdot\boldsymbol{\delta}_\mu}\delta_{\tau,A}-\sum_\mu f_i(\delta_\mu)e^{i\mathbf k_2\cdot\boldsymbol{\delta}_\mu}\delta_{\tau,B}\Big)u^i_{\tau,\mathbf k_1-\mathbf k_2}.
\end{eqnarray}
Note that
\begin{eqnarray}\label{eq:factor}
f_i(\delta_\mu)=\partial_i r\partial_rt(r)|_{r=\delta_\mu}= \frac{\delta_{\mu,i}}{r}t(r)\partial_r\log t(r) |_{r=\delta_\mu=a} =\bar{\gamma}\delta_{\mu,i} 
\end{eqnarray}
where $\bar{\gamma}=\gamma t$ with
\begin{eqnarray}
 \gamma=\frac{1}{r}\partial_r\log t(r)|_{r=a}.   
\end{eqnarray}
In Graphene, $|\gamma| = 7.038(\sqrt{3}a)^{-2}$~\cite{Yu2024geomentryEPC} where $\bar{\gamma}=\gamma t>0$ and note that the lattice constant is defined differently from Ref.~\cite{Yu2024geomentryEPC}. Plugging Eq.~\eqref{eq:factor} into Eq.~\eqref{eq:epcoupling}, one obtains
\begin{eqnarray}\label{eq:EPHamiltonian}
H_{el-ph}&&=\frac{1}{\sqrt{N}}\sum_{\mathbf k_1,\mathbf k_2}\sum_{i=x,y} c^\dagger_{A\mathbf k_1}c_{B\mathbf k_2}\bar{\gamma}\Big(\delta_{\tau,A}i\partial_{k_{2,i}}\gamma_{\mathbf k_2}-\delta_{\tau,B}i\partial_{k_{1,i}}\gamma_{\mathbf k_1}\Big)u^i_{\tau,\mathbf k_1-\mathbf k_2}\nonumber\\
&&\qquad\qquad\qquad\quad +
c^\dagger_{B\mathbf k_1}c_{A\mathbf k_2}\bar{\gamma}\Big(\delta_{\tau,B}i\partial_{k_{2,i}}\gamma^\ast_{\mathbf k_2}-\delta_{\tau,A}i\partial_{k_{1,i}}\gamma^\ast_{\mathbf k_1}\Big)u^i_{\tau,\mathbf k_1-\mathbf k_2},  
\end{eqnarray}
where $\gamma_{\mathbf k}=\sum_{\mu}e^{-i\mathbf k\cdot\boldsymbol{\delta}_\mu}$.\\

\subsubsection{Phonon modes}

In the hexagonal lattice, the phonon Hamiltonian is given by
\begin{eqnarray}
H_{ph}=\sum_{\mathbf q}\sum_{\tau,\tau^\prime,i,i^\prime}\frac{P^i_{\mathbf q\tau}(P^i_{\mathbf q\tau})^\dagger}{2m_\tau} +\frac{1}{2}\sum_{\mathbf q}\sum_{\tau\tau^\prime,ii^\prime}D_{\tau\tau^\prime,ii^\prime}(u^i_{\mathbf q\tau})^\dagger u^{i^\prime}_{\mathbf q\tau^\prime}
\end{eqnarray}
where $\tau, \tau^\prime= A,B$ index the lattice, $P_{\mathbf q\tau}^i$ represents the momentum of phonon with $(P^i_{\mathbf q\tau})^\dagger=P_{-\mathbf q\tau}^i$, $m_\tau$ is the atom mass, $D_{\tau\tau^\prime,ii^\prime}$ is the force-constant matrix, and $(u^i_{\mathbf q\tau})^\dagger=u_{-\mathbf q\tau}^i$. For convenience, we make the following rescaling
\begin{eqnarray}
\Tilde{P}^i_{\mathbf q\tau}=\frac{P^i_{\mathbf q\tau}}{\sqrt{m_\tau}},\qquad 
\Tilde{u}^i_{\mathbf q\tau}=\sqrt{m_\tau}u^i_{\mathbf q\tau},\qquad
\Tilde{D}_{\tau\tau^\prime,ii^\prime}=\frac{D_{\tau\tau^\prime,ii^\prime}}{\sqrt{m_\tau m_\tau^\prime}},
\end{eqnarray}
by which the Hamiltonian is transformed to
\begin{eqnarray}\label{eq:phonon}
H_{ph}&&=\sum_{\mathbf q}\sum_{\tau,\tau^\prime,i,i^\prime}\frac{\Tilde{P}^i_{\mathbf q\tau}(\Tilde{P}^i_{\mathbf q\tau})^\dagger}{2} +\frac{1}{2}\sum_{q}\sum_{\tau\tau^\prime,ii^\prime}\Tilde{D}_{\tau\tau^\prime,ii^\prime}(\Tilde{u}^i_{\mathbf q\tau})^\dagger \Tilde{u}^{i^\prime}_{\mathbf q\tau^\prime}\nonumber\\
&&=\sum_{\mathbf q}\frac{\Tilde{P}^\dagger_{\mathbf q}\Tilde{P}_{\mathbf q}}{2}+\frac{1}{2}\sum_{\mathbf q}\Tilde{u}_{\mathbf q}^\dagger \Tilde{D}_{\mathbf q}\Tilde{u}_{\mathbf q},
\end{eqnarray}
where $\Tilde{u}_{\mathbf q}=\{\Tilde{u}^i_{\mathbf q\tau}\}$ with $i=x,y,z$, and $\Tilde{D}_{\mathbf q}$ is a $6\times 6$ matrix.  The matrix $\Tilde{D}_{\mathbf q}$ can be diagonalized by a set of eigenvector $\{v_\lambda(\mathbf q)\}$:
\begin{eqnarray}\label{eq:forceConstant}
\Tilde{D}_{\mathbf q}=\sum_\lambda\omega^2_\lambda(\mathbf q)v_\lambda(\mathbf q)v^\dagger_\lambda(\mathbf q),    
\end{eqnarray}
where $v_\lambda(\mathbf q)$ satisfies
\begin{eqnarray}\label{eq:completeness}
\sum_\lambda v_\lambda(\mathbf q)v^\dagger_\lambda(\mathbf q)=\mathbbm{1}.    
\end{eqnarray}
Inserting Eqs.~\eqref{eq:forceConstant} and~\eqref{eq:completeness} into Eq.~\eqref{eq:phonon}, we obtain 
\begin{eqnarray}
H_{ph}=\sum_{\mathbf q}\frac{\Tilde{P}^\dagger_{\mathbf q\lambda}\Tilde{P}_{\mathbf q\lambda}}{2}+\frac{1}{2}\sum_{\mathbf q}\omega^2_\lambda(\mathbf q)\Tilde{u}_{\mathbf q\lambda}^\dagger \Tilde{u}_{\mathbf q\lambda}   
\end{eqnarray}
where $\Tilde{P}^\dagger_{\mathbf q\lambda}=\Tilde{P}^\dagger_{\mathbf q}v_\lambda(\mathbf q)$ and $\Tilde{u}^\dagger_{\mathbf q\lambda}=\Tilde{u}^\dagger_{\mathbf q}v_\lambda(\mathbf q)$.\\ 

To proceed, we have to consider the time-reversal symmetry of the phonon Hamiltonian. The time reversal operation ($\mathcal T$) acts on field operators as below:
\begin{eqnarray}
\mathcal T \Tilde{P}_{\mathbf q}\mathcal T=\Tilde{P}_{-\mathbf q},\qquad
\mathcal T \Tilde{u}_{\mathbf q}\mathcal T= \Tilde{u}_{-\mathbf q}.   
\end{eqnarray}
The time-reversal symmetry of phonon Hamiltonian $\mathcal T H_{ph}\mathcal T^{-1}=H_{ph}$ leads to $\Tilde{D}_{\mathbf q}^\ast=\Tilde{D}_{-\mathbf q}$. Applying this relation to the eigen equation of $D_{\mathbf q}$ yields
\begin{eqnarray}
\omega_\lambda(\mathbf q)=\omega_\lambda(-\mathbf q),\qquad
v_\lambda(-\mathbf q)= v^\ast_\lambda(\mathbf q).
\end{eqnarray}
Now, we define the following phonon creation and annihilation operators for $\omega_\lambda(\mathbf q)\neq 0$:
\begin{eqnarray}\label{eq:phononOperator}
&&b^\dagger_{\mathbf q\lambda}=\sqrt{\frac{\omega_\lambda(\mathbf q)}{2}}\Big[\Tilde{u}_{\mathbf q \lambda}^\dagger-\frac{i}{\omega_\lambda(\mathbf q)}\Tilde{P}^\dagger_{\mathbf q \lambda}\Big],\nonumber\\
&&b_{\mathbf q\lambda}=\sqrt{\frac{\omega_\lambda(\mathbf q)}{2}}\Big[\Tilde{u}_{\mathbf q \lambda}+\frac{i}{\omega_\lambda(\mathbf q)}\Tilde{P}_{\mathbf q \lambda}\Big]. 
\end{eqnarray}
By this, one can show that
\begin{eqnarray}
H_{ph}=\sum_{\mathbf q,\lambda}\omega_\lambda(\mathbf q)\Big[b^\dagger_{\mathbf q \lambda}b_{\mathbf q,\lambda}+\frac{1}{2}\Big]    
\end{eqnarray}
where we neglected the contribution from the mode $\omega_{\lambda}(\mathbf q)=0$ by assuming its measure in momentum space is zero.

Note that from the relation: $\Tilde{u}_{\mathbf q\lambda}=v^\dagger_{\lambda}(\mathbf q)\Tilde{u}_{\mathbf q}$, we obtain
\begin{eqnarray}
\Tilde{u}_{\mathbf q}=\sum_{\lambda} v_\lambda(\mathbf q)v^\dagger_\lambda(\mathbf q)\Tilde{u}_{\mathbf q}=\sum_{\lambda}v_\lambda(\mathbf q)\Tilde{u}_{\mathbf q\lambda}.    
\end{eqnarray}
On the other hand, from Eq.~\eqref{eq:phononOperator}, we obtain
\begin{eqnarray}
\Tilde{u}_{\mathbf q\lambda}=\frac{1}{\sqrt{\omega_{2\lambda\mathbf q}}}(b_{\mathbf q\lambda}+b^\dagger_{-\mathbf q\lambda}).    
\end{eqnarray}
Therefore, we finally obtain
\begin{eqnarray}\label{eq:phononQuantize}
u^i_{\mathbf q\tau}=\frac{1}{\sqrt{m_\tau}}\Tilde{u}^i_{\mathbf q\tau}=\sum_\lambda\frac{[v_\lambda(\mathbf q)]_{(\tau i)}}{\sqrt{2m_\tau\omega_\lambda(\mathbf q)}}(b_{\mathbf q\lambda}+b^\dagger_{-\mathbf q\lambda}),   
\end{eqnarray}
where the label $(\tau i)$ in $[v_\lambda(\mathbf q)]_{(\tau i)}$ constitutes a single index containing the information of the sublattice and the position of the unit cell. \\

\subsubsection{Electron-phonon coupling matrix}

Substituting Eq.~\eqref{eq:phononQuantize} into Eq.~\eqref{eq:EPHamiltonian}, the Hamiltonian can be written as follows:
\begin{eqnarray}
H_{el-ph}&&=\frac{1}{\sqrt{N}}\sum_{\mathbf k_1,\mathbf k_2}\sum_{\lambda}c^\dagger_{\mathbf k_1}V^\lambda(\mathbf k_1,\mathbf k_2)c_{\mathbf k_2}(b_{\mathbf k_1-\mathbf k_2,\lambda}+b^\dagger_{\mathbf k_2-\mathbf k_1,\lambda}),   
\end{eqnarray}
where
\begin{eqnarray}
V^\lambda(\mathbf k_1,\mathbf k_2)=\left(
\begin{array}{cc}
  0  & F^\lambda_{AB}(\mathbf k_1,\mathbf k_2) \\
  F^\lambda_{BA}(\mathbf k_1,\mathbf k_2)  & 0
\end{array}
\right).   
\end{eqnarray}
Specifically, 
\begin{eqnarray}
F^\lambda_{AB}(\mathbf k_1,\mathbf k_2)&&=\sum_{i=x,y}\frac{\bar{\gamma}}{\sqrt{\omega_{\lambda}(\mathbf k_1-\mathbf k_2)}}\Big(
\frac{[v_{\lambda}(\mathbf k_1-\mathbf k_2)]_{(Ai)}i\partial_{k_{2,i}}\gamma_{\mathbf k_2}}{\sqrt{2m_A}}-
\frac{[v_{\lambda}(\mathbf k_1-\mathbf k_2)]_{(Bi)}i\partial_{k_{1,i}}\gamma_{\mathbf k_1}}{\sqrt{2m_B}}
\Big), \nonumber\\ 
&&=g_0\sqrt{\frac{\omega_0}{\omega_{\lambda}(\mathbf k_1-\mathbf k_2)}}\Big(\sum_{\mu}e^{-i\mathbf k_2\cdot\boldsymbol{\delta}_\mu}\boldsymbol{\delta}_\mu\cdot
\mathbf{v}_{\lambda,A}(\mathbf k_1-\mathbf k_2)-\sum_{\mu}e^{-i\mathbf k_1\cdot\boldsymbol{\delta}_\mu}\boldsymbol{\delta}_\mu\cdot
\mathbf{v}_{\lambda,B}(\mathbf k_1-\mathbf k_2)
\Big),\nonumber\\
F^\lambda_{BA}(\mathbf k_1,\mathbf k_2)&&=\sum_{i=x,y}\frac{\bar{\gamma}}{\sqrt{\omega_{\lambda}(\mathbf k_1-\mathbf k_2)}}\Big(
\frac{[v_{\lambda}(\mathbf k_1-\mathbf k_2)]_{(Bi)}i\partial_{k_{2,i}}\gamma^\ast_{\mathbf k_2}}{\sqrt{2m_B}}-
\frac{[v_{\lambda}(\mathbf k_1-\mathbf k_2)]_{(Ai)}i\partial_{k_{1,i}}\gamma_{\mathbf k^\ast_1}}{\sqrt{2m_A}}
\Big)\nonumber\\
&&=g_0\sqrt{\frac{\omega_0}{\omega_{\lambda}(\mathbf k_1-\mathbf k_2)}}\Big(\sum_{\mu}e^{i\mathbf k_1\cdot\boldsymbol{\delta}_\mu}\boldsymbol{\delta}_\mu\cdot
\mathbf{v}_{\lambda,A}(\mathbf k_1-\mathbf k_2)-\sum_{\mu}e^{i\mathbf k_2\cdot\boldsymbol{\delta}_\mu}\boldsymbol{\delta}_\mu\cdot
\mathbf{v}_{\lambda,B}(\mathbf k_1-\mathbf k_2)
\Big).
\end{eqnarray}
Note that $[F^\lambda_{AB}(\mathbf k_1,\mathbf k_2)]^\ast=F^\lambda_{BA}(\mathbf k_2,\mathbf k_1)$ due to the Hermiticity (the Hermitian conjugation on each term is equivalent to switching $\mathbf k_1,\mathbf k_2$). Here, we assumed $m_A=m_B=M$ and take
\begin{eqnarray}\label{eq:couplingconst}
g_0=\frac{\bar{\gamma}a}{\sqrt{2M\omega_0}}\approx 0.05 eV,  
\end{eqnarray}
where $\hbar$ is set to unit ($\hbar =1$), and $\omega_0$ is the phonon energy of optical mode at $\Gamma$ point.
This value is estimated as the following: in Graphene $g_{0,Graphene}=0.15 eV$ with hopping strength $t_{Graphene}=2.8 eV$~\cite{Thingstad2020phononSC}; provided with $g_0\propto t$, in our case $t=1 eV$, so we take $g_0\approx g_{0,Graphene}/3$.

\subsubsection{The phonon eigenmodes in a hexagonal lattice}.

In a 2D hexagonal lattice, due to the mirror symmetry with respect to the 2D plane, electrons only couple to $x,y$ components of phonons. In this part, we follow Ref.~\cite{FALKOVSKY2008GraphenPhonon,Thingstad2020phononSC} to review how to construct the phonon force-constant matrix by symmetry analysis. For the phonon modes, the $z$ component also decouples from the $x,y$ modes due to the mirror symmetry mentioned above. Here, we only concentrate on the $x,y$ components. The elastic potential from in-plane components reads
\begin{eqnarray}
 V_{ph,\perp}=\frac{1}{2}\sum_{\mathbf q}\Tilde{u}^\dagger_{\mathbf q,\perp} D_{\mathbf q,\perp}\Tilde{u}_{\mathbf q,\perp}  
\end{eqnarray}
where $\Tilde{u}_{\mathbf q,\perp}=(\Tilde{u}_A^x,\Tilde{u}_A^y,\Tilde{u}_B^x,\Tilde{u}_B^y)^T$ and $D_{\mathbf q,\perp}$ is the relevant $4\times 4$ block of $\Tilde{D}_{\mathbf q}$ in Eq.~\eqref{eq:phonon}. The hexagonal lattice under consideration has a threefold rotation symmetry $C_3$. It is convenient to transform the spatial variables to a pair of complex coordinates: $\xi= x+iy$, $\bar{\xi}=x-iy$. Specifically, the position displacement variables are transformed as
\begin{eqnarray}
w_{\mathbf q}=\Big(w_{A,\bar{\xi}}(\mathbf q),w_{A,\xi}(\mathbf q),w_{B,\bar{\xi}}(\mathbf q),w_{B,\xi}(\mathbf q)\Big)^T=T\Tilde{u}_{\mathbf q,\perp}    
\end{eqnarray}
where
\begin{eqnarray}
T=\frac{1}{\sqrt{2}}\left(
\begin{array}{cccc}
 1&-i&0&0 \\
 1&i&0&0\\
 0&0&1&-i\\
 0&0&1&i
\end{array}
\right).
\end{eqnarray}
By definition, $w_{\mathbf q}^\dagger=\Big(w_{A,\xi}(\mathbf q),w_{A,\bar{\xi}}(\mathbf q),w_{B,\xi}(\mathbf q),w_{B,\bar{\xi}}(\mathbf q)\Big)$. After the transformation, the potential energy becomes
\begin{eqnarray}
V_{ph,\perp}=\frac{1}{2}\sum_{\mathbf q}w^\dagger_{\mathbf q}\Phi_{\mathbf q}w_{\mathbf q}    
\end{eqnarray}
where $\Phi_{\mathbf q}=T D_{\mathbf q,\perp}T^\dagger$.
Explicitly, the $\Phi_{\mathbf q}$ matrix can be decomposed as follows
\begin{eqnarray}
\Phi_{\mathbf q}=\left(
\begin{array}{cc}
  \phi^{AA}(\mathbf q) & \phi^{AB}(\mathbf q)  \\
  \phi^{BA}(\mathbf q) & \phi^{BB}(\mathbf q)
\end{array}
\right),    
\end{eqnarray}
where
\begin{eqnarray}
\phi^{\tau\tau^\prime}(\mathbf q)=\left(
\begin{array}{cc}
  \phi_{\xi\bar{\xi}}^{\tau\tau^\prime}(\mathbf q) & \phi_{\xi\xi}^{\tau\tau^\prime}(\mathbf q)  \\
  \phi_{\bar{\xi}\bar{\xi}}^{\tau\tau^\prime}(\mathbf q) & \phi_{\bar{\xi}\xi}^{\tau\tau^\prime}(\mathbf q)
\end{array}
\right),\qquad \tau,\tau^\prime= A,B.  
\end{eqnarray}
The $\Phi_{\mathbf q}$ matrix can be fixed by symmetry analysis. Then, we can inversely obtain $D_{\mathbf q,\perp}=T^\dagger\Phi_{\mathbf q}T$ and its eigenstate.\\

In the following, we analyze the matrix element of $\Phi_{\mathbf q}$ by considering the symmetry constraint. First, due to the Hermiticity of the potential energy, we obtain
\begin{eqnarray}
 \phi^{AA}(\mathbf q)=[\phi^{AA}(\mathbf q)]^\dagger, \qquad \phi^{BB}(\mathbf q)=[\phi^{BB}(\mathbf q)]^\dagger , \qquad
 \phi^{BA}(\mathbf q)=[\phi^{AB}(\mathbf q)]^\dagger.
\end{eqnarray}
On the other hand, the $A$ and $B$ sublattices are equivalent (neglect the local spins in consideration of the phonon modes ), as they can be transformed to each other by a $C_2$ rotation; it is readily shown that 
\begin{eqnarray}
\phi^{AA}(\mathbf q)=\phi^{BB}(\mathbf q).    
\end{eqnarray}
Therefore, there are only two independent matrix blocks: $\phi^{AA}(\mathbf q)$ and $\phi^{AB}(\mathbf q)$.\\

\begin{figure}
    \centering
    \includegraphics[width=0.35\linewidth]{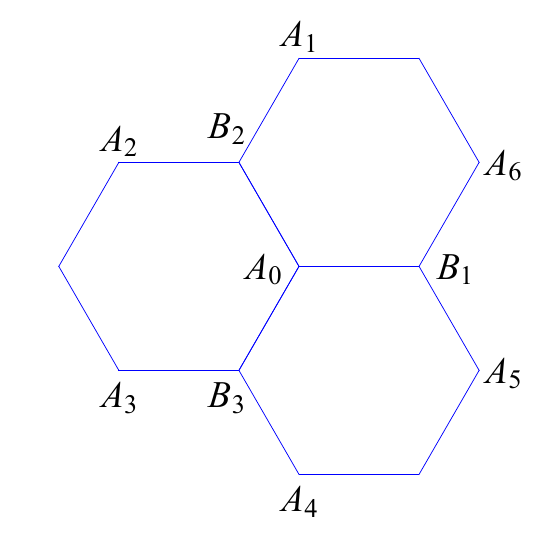}
    \caption{The hexagonal lattice where the first and second neighbors of the site $A_0$ are explicitly labeled.}
    \label{fig:hexagonal}
\end{figure}

\underline{\textbf{$\phi^{AA}(\mathbf q)$ block}}.
First, we figure out the form of $\phi^{AA}(\mathbf q)$. Because (i) the Hermiticity of $\phi^{AA}(\mathbf q)$ and (ii) $\xi\bar{\xi}=\bar{\xi}\xi$ (or $w_{\bar{\xi}}w_{\xi}=w_{\xi}w_{\bar{\xi}}$ ), the $\phi^{AA}(\mathbf q)$ block can be reduced to 
\begin{eqnarray}
\phi^{AA}(\mathbf q)=\left(
\begin{array}{cc}
\phi^{AA}_{\bar{\xi}\xi}(\mathbf q)  & \phi^{AA}_{\xi\xi}(\mathbf q) \\
(\phi^{AA}_{\xi\xi}(\mathbf q))^\ast    & \phi^{AA}_{\bar{\xi}\xi}(\mathbf q)
\end{array}
\right).    
\end{eqnarray}
The independent components are $\phi^{AA}_{\bar{\xi}\xi}(\mathbf q)$ and $\phi^{AA}_{\xi\xi}(\mathbf q)$. To count the elastic energy between the same sublattice, we only consider the second nearest-neighbor coupling, which gives the momentum-space force-constant matrix \begin{eqnarray}\label{eq:FourierA}
\phi^{AA}_{ij}(\mathbf q)=\phi^{AA}_{ij}(\mathbf A_0)+\sum_{\kappa=1}^6 e^{i\mathbf q\cdot\mathbf A_\kappa}\phi^{AA}_{ij}(\mathbf{A}_\kappa).  
\end{eqnarray}
For the case $(i,j)=(\bar{\xi},\xi)$, as $\bar{\xi}\xi$ is invariant under spatial operation, the $\phi^{AA}_{\bar{\xi}\xi}(\mathbf{A}_\kappa)$ components are equal for different $\mathbf A_\kappa$ that can be connected by symmetry. Note that $C_3$ symmetry leads to $\mathbf A_1\rightarrow \mathbf A_3\rightarrow \mathbf A_5\rightarrow\mathbf A_1$ and  $\mathbf A_2\rightarrow \mathbf A_4\rightarrow \mathbf A_6\rightarrow\mathbf A_2$; and $C_2$ around the axis along $\mathbf A_0-\mathbf B_1$ bond results in $\mathbf A_1\leftrightarrow\mathbf A_4$, $\mathbf A_2\leftrightarrow\mathbf A_3$, and $\mathbf A_5\leftrightarrow\mathbf A_6$. These symmetries yield
\begin{eqnarray}\label{eq:constraintAA}
\phi^{AA}_{\bar{\xi}\xi}(\mathbf{A}_1)=\phi^{AA}_{\bar{\xi}\xi}(\mathbf{A}_2)=\cdots =\phi^{AA}_{\bar{\xi}\xi}(\mathbf{A}_6)=\epsilon.    
\end{eqnarray}
To obtain the value of $\phi^{AA}_{\bar{\xi}\xi}(\mathbf A_0)$, we first investigate the constraint from a uniform translation. In real space, the equation of motion of phonon modes is given by~\cite{FALKOVSKY2008GraphenPhonon}
\begin{eqnarray}\label{eq:phononEOM}
\sum_{j,m,\tau^\prime} D^{\tau\tau^\prime}_{ij}(\mathbf r_n-\mathbf r_m)u^j_{\tau^\prime}(\mathbf r_m)-\omega^2 u^i_\tau(\mathbf r_n)=0,    
\end{eqnarray}
where $D^{\tau\tau^\prime}_{ij}(\mathbf r_n-\mathbf r_m)$ is the force-constant matrix in real space, $\mathbf r_{n,m}$ denotes the position of lattice sites.  Suppose all lattices displace from the original site by an in-plane vector $\mathbf u$, which costs no energy ($\omega=0$). Consequently, regarding the index $(i,j)$ as $(\bar{\xi},\xi)$, Eq.~\eqref{eq:phononEOM} is reduced to 
\begin{eqnarray}\label{eq:translation}
\sum_{m,\tau^\prime}\Phi^{\tau\tau^\prime}_{ij}(\mathbf r_n-\mathbf r_m)=0,    
\end{eqnarray}
where $\Phi^{\tau\tau^\prime}_{ij}(\mathbf r_n-\mathbf r_m)$ replaces $D^{\tau\tau^\prime}_{ij}(\mathbf r_n-\mathbf r_m)$ given that only in-plane components are considered, and the index $j$ in the summation is dropped because this relation applies to arbitrary $\mathbf u$. In the hexagonal model, let $\mathbf r_n=\mathbf A_0$ and abbreviate $\Phi^{\tau\tau^\prime}_{ij}(\mathbf r_n-\mathbf r_m)$ as $\Phi^{\tau\tau^\prime}_{ij}(\mathbf r_m)$. For the case $(i,j)=(\bar{\xi},\xi)$, Eq.~\eqref{eq:translation} becomes
\begin{eqnarray}\label{eq:constraintAsite}
\phi^{AA}_{\bar{\xi}\xi}(\mathbf A_0)+\sum_{\kappa=1}^6\phi^{AA}_{\bar{\xi}\xi}(\mathbf A_\kappa)+\sum_{\kappa=1}^3\phi^{AB}_{\bar{\xi}\xi}(\mathbf B_\kappa)=0.      
\end{eqnarray}
The $C_3$ symmetry leads to $\mathbf B_1\rightarrow \mathbf B_2\rightarrow \mathbf B_3\rightarrow \mathbf B_1$. Similar to Eq.~\eqref{eq:constraintAA}, the force-constant between $A$ and $B$ lattices satisfies
\begin{eqnarray}\label{eq:constraintAB}
\phi^{AB}_{\bar{\xi}\xi}(\mathbf{A}_1)=\phi^{AB}_{\bar{\xi}\xi}(\mathbf{A}_2)=\phi^{AB}_{\bar{\xi}\xi}(\mathbf{A}_3)=\alpha.    
\end{eqnarray}
Substituting Eqs.~\eqref{eq:constraintAA} and~\eqref{eq:constraintAB} into Eq.~\eqref{eq:constraintAsite}, we obtain 
\begin{eqnarray}\label{eq:phiA0}
\phi^{AA}_{\bar{\xi}\xi}(\mathbf A_0)=-6\epsilon-3\alpha.    
\end{eqnarray}
Finally, plugging Eq.~\eqref{eq:constraintAA} and~\eqref{eq:phiA0} into Eq.~\eqref{eq:FourierA} gives
\begin{eqnarray}\label{eq:FourierAAxixi1}
\boxed{\phi^{AA}_{\bar{\xi}\xi}(\mathbf q)= 2\epsilon[\cos(\sqrt{3}q_y)+2\cos(\frac{3q_x}{2})\cos(\frac{\sqrt{3}q_y}{2})-3]-3\alpha}.   
\end{eqnarray}

Now, we consider the form of $\phi^{AA}_{\xi\xi}(\mathbf q)$. The $C_3$ symmetry of the lattice leads to  $u^A_\xi(\mathbf A_3)\rightarrow e^{i2\pi/3}u^A_\xi(\mathbf A_1)$ and $C_3^{-1}$ leads to $u^A_\xi(\mathbf A_5)\rightarrow e^{-i2\pi/3}u^A_\xi(\mathbf A_1)$. The potential is invariant under these replacements, resulting in 
\begin{eqnarray}\label{eq:AAxixiValue}
&&\phi^{AA}_{\xi\xi}(\mathbf A_1)=\delta,\nonumber\\
&&\phi^{AA}_{\xi\xi}(\mathbf A_3)=\delta e^{-i2\pi/3},\nonumber\\
&&\phi^{AA}_{\xi\xi}(\mathbf A_5)=\delta e^{i2\pi/3}.
\end{eqnarray}
On the other hand, a reflection $\sigma_v$ about the plane crossing $\mathbf A_0-\mathbf B_1$ bond leads to $\mathbf A_1\leftrightarrow\mathbf A_2$ and $\xi\leftrightarrow\bar{\xi}$.  We consider an operation $\sigma_v\ast\mathcal C$ with $\mathcal C$ being complex conjugation, leading to 
\begin{eqnarray}\label{eq:reflcomplexSymm}
(\sigma_v\ast\mathcal C)\Big( u^A_{\xi}(\mathbf A_0)\Phi^{AA}_{\xi\xi}(\mathbf A_1) u^A_{\xi}(\mathbf A_1)\Big) (\sigma_v\ast\mathcal C)^{-1}= u^A_{\xi}(\mathbf A_0)[\Phi^{AA}_{\xi\xi}(\mathbf A_1)]^\ast u^A_{\xi}(\mathbf A_4).      
\end{eqnarray}
Because the system preserves the $\sigma_v$ symmetry and the potential takes a real value, this leads to 
\begin{eqnarray}
\phi^{AA}_{\xi\xi}(\mathbf A_4)=[\phi^{AA}_{\xi\xi}(\mathbf A_1)]^\ast=\delta^\ast.    
\end{eqnarray}
Similar to Eq.~\eqref{eq:AAxixiValue}, the $C_3$ symmetry cycle among $\mathbf A_2, \mathbf A_4, \mathbf A_6$ gives
\begin{eqnarray}\label{eq:AAxixiValue1}
&&\phi^{AA}_{\xi\xi}(\mathbf A_4)=\delta^\ast,\nonumber\\
&&\phi^{AA}_{\xi\xi}(\mathbf A_2)=\delta^\ast e^{i2\pi/3},\nonumber\\
&&\phi^{AA}_{\xi\xi}(\mathbf A_6)=\delta^\ast e^{-i2\pi/3}.
\end{eqnarray}
On the other hand, similar to the symmetry constraint on the $AA$ coupling Eqs.~\eqref{eq:AAxixiValue} and~\eqref{eq:AAxixiValue1}, the $C_3$ symmetry also imposes a symmetry constraint on the $AB$ coupling as below
\begin{eqnarray}\label{eq:ABxixiValue}
&&\phi^{AB}_{\xi\xi}(\mathbf B_1)=\beta,\nonumber\\
&&\phi^{AB}_{\xi\xi}(\mathbf B_2)=\beta e^{i2\pi/3},\nonumber\\
&&\phi^{AB}_{\xi\xi}(\mathbf B_3)=\beta e^{-i2\pi/3}.
\end{eqnarray}
Here, $\beta$ is a real number.  This can be seen by applying $\sigma_v\ast\mathcal C$ in Eq.~\eqref{eq:reflcomplexSymm} to $u^A_\xi(\mathbf A_0)\phi^{AB}_{\xi\xi}(\mathbf q)u^B_\xi(\mathbf B_1)$, yielding
\begin{eqnarray}
\phi^{AB}_{\xi\xi}(\mathbf q)=[\phi^{AB}_{\xi\xi}(\mathbf q)]^\ast=\beta \in \mathbbm R.
\end{eqnarray}
Now, combine Eqs.~\eqref{eq:AAxixiValue},\eqref{eq:AAxixiValue1} and~\eqref{eq:ABxixiValue} with the translation constraint Eq.~\eqref{eq:translation} [using $(i,j)=(\xi,\xi)$ and $\kappa=A$], we arrive at
\begin{eqnarray}\label{eq:constraintAsite1}
0&&=\phi^{AA}_{\xi\xi}(\mathbf A_0)+\sum_{\kappa=1}^6\phi^{AA}_{\xi\xi}(\mathbf A_\kappa)+\sum_{\kappa=1}^3\phi^{AB}_{\xi\xi}(\mathbf B_\kappa)\nonumber\\
&&=\phi^{AA}_{\xi\xi}(\mathbf A_0)+(\delta+\delta^\ast+\beta)(1+e^{i2\pi/3}+e^{i2\pi/3})\nonumber\\
&&=\phi^{AA}_{\xi\xi}(\mathbf A_0).
\end{eqnarray}
Applying Eqs.~\eqref{eq:AAxixiValue},\eqref{eq:AAxixiValue1},\eqref{eq:ABxixiValue} and~\eqref{eq:constraintAsite1} to the Fourier transformation Eq.~\eqref{eq:FourierA}, we get
\begin{eqnarray}\label{eq:FourierAAxixi2}
\boxed{\phi^{AA}_{\xi\xi}(\mathbf q)=\delta\Big(e^{i\sqrt{3}q_y}+2\cos(\frac{3q_x}{2}+\frac{2\pi}{3})e^{-i\sqrt{3}q_y/2}\Big)+\delta^\ast\Big(e^{-i\sqrt{3}q_y}+2\cos(\frac{3q_x}{2}-\frac{2\pi}{3})e^{i\sqrt{3}q_y/2}\Big)}.
\end{eqnarray}

\underline{\textbf{$\phi^{AB}(\mathbf q)$ block}}. Due to similar reason to the $\phi^{AA}$ block, the $\phi^{AB}(\mathbf q)$ block take the following form:
\begin{eqnarray}
\phi^{AB}(\mathbf q)=\left(
\begin{array}{cc}
\phi^{AB}_{\bar{\xi}\xi}(\mathbf q)  & \phi^{AB}_{\xi\xi}(\mathbf q) \\
\phi^{AB}_{\bar{\xi}\bar{\xi}}(\mathbf q)    & \phi^{AB}_{\bar{\xi}\xi}(\mathbf q)
\end{array}
\right),    
\end{eqnarray}
where $\phi^{AB}_{\bar{\xi}\bar{\xi}}(\mathbf q)$ and $\phi^{AB}_{\xi\xi}(\mathbf q)$ are related but not complex conjugate to each other. Under the reflection $\sigma_v$ about $\mathbf A_0-\mathbf B_1$ bond, $\xi\leftrightarrow\bar{\xi}$ and $(q_x,q_y)\leftrightarrow(q_x,-q_y)$, so we have the relation
\begin{eqnarray}
\phi^{AB}_{\bar{\xi}\bar{\xi}}(q_x,q_y)=\phi^{AB}_{\xi\xi}(q_x,-q_y).    
\end{eqnarray}
Therefore, there are only two independent elements: $\phi^{AB}_{\bar{\xi}\xi}(\mathbf q)$ and $\phi^{AB}_{\xi\xi}(\mathbf q)$.

First, from Eq.~\eqref{eq:constraintAB}, we obtain
\begin{eqnarray}\label{eq:FourierABxixi1}
\boxed{\phi^{AB}_{\bar{\xi}\xi}(\mathbf q)=\sum_{\kappa=1}^3 e^{i\mathbf q\cdot\mathbf A_\kappa}\phi^{AB}_{\bar{\xi}\xi}(\mathbf A_\kappa)=\alpha\Big(e^{iq_x}+2e^{-iq_x/2}\cos(\frac{\sqrt{3}q_y}{2})\Big)}.
\end{eqnarray}
To obtain $\phi^{AB}_{\xi\xi}(\mathbf q)$, we directly apply Eq.~\eqref{eq:ABxixiValue} to the Fourier transformation, leading to 
\begin{eqnarray}\label{eq:FourierABxixi2}
\boxed{\phi^{AB}_{\xi\xi}(\mathbf q)=\sum_{\kappa=1}^3 e^{i\mathbf q\cdot\mathbf A_\kappa}\phi^{AB}_{\xi\xi}(\mathbf A_\kappa)=\beta\Big(e^{iq_x}+2e^{-iq_x/2}\cos(\frac{\sqrt{3}q_y}{2}-\frac{2\pi}{3})\Big)}.   
\end{eqnarray}

We have obtained all the independent force-constant elements Eq.~\eqref{eq:FourierAAxixi1},\eqref{eq:FourierAAxixi2},\eqref{eq:FourierABxixi1} and \eqref{eq:FourierABxixi2}. Then, we can fully construct the matrix $\Phi_\mathbf q$ and subsequently obtain the matrix $D_{\mathbf q,\perp}=T^\dagger\Phi_{\mathbf q}T$.  According to Ref.~\cite{FALKOVSKY2008GraphenPhonon}, the involved force-constant parameters take the following values (in the unit of $10^5$ cm$^{-2}$)
\begin{eqnarray}
\alpha= -3.980,\qquad   \beta=-1.132,\qquad
\epsilon=-0.297\qquad
\delta=1.123.
\end{eqnarray}
Here, $\delta$ takes a real value.

\section{Spin current---the case of bosonic baths}

\subsection{Optical conductivity}

The optical conductivity in the Floquet system is given by~\cite{rudner2020floquetengineershandbook}
\begin{eqnarray}
\sigma_{\alpha\beta}(\Omega)=\frac{\chi^{(0)}_{\alpha\beta}(\Omega)+\mathcal K_{\alpha\beta}^{(0)}}{i\Omega}    
\end{eqnarray}
where
\begin{eqnarray}\label{eq:response}
\chi_{\alpha\beta}^{(0)}(\Omega)=\frac{1}{V}\sum_{\mathbf k}\sum_{\nu_1\nu_2}\sum_{m}\frac{(\rho_{\mathbf k\nu_1}-\rho_{\mathbf k\nu_2})j^{(m)}_{\alpha,\nu_1\nu_2}(\mathbf k)j^{(-m)}_{\beta,\nu_2\nu_1}(\mathbf k)}{\Omega-m\omega+(\varepsilon_{\mathbf k\nu_1}-\varepsilon_{\mathbf k\nu_2})+i\eta}   
\end{eqnarray}
is the paramagnetic current contribution
and
\begin{eqnarray}
\mathcal K_{\alpha\beta}^{(0)}=\frac{1}{T}\int dt \text{Tr}[\hat \rho_s\Big(-\frac{\partial^2H}{\partial k_{\alpha}\partial k_{\beta}}\Big)]    
\end{eqnarray}
denotes the diamagnetic contribution.
Here, 
\begin{eqnarray}
j_{\alpha,\nu_1\nu_2}(\mathbf k,t)=\langle\phi_{\mathbf k\nu_1}(t)|\frac{\partial H_0(t)}{\partial k_{\alpha}} |\phi_{\mathbf k\nu_2}(t)\rangle =\sum_m e^{-im\omega t}j^{(m)}_{\alpha,\nu_1\nu_2}(\mathbf k). 
\end{eqnarray}
Substituting the Fourier form of states, we obtain
\begin{eqnarray}
j^{(m)}_{\alpha,\nu_1\nu_2}(\mathbf k)=\sum_{n,l}\langle \phi_{\mathbf k\nu_1}^{(n)}|\partial_{k_\alpha}H^{(m+n-l)}|\phi^{(l)}_{\mathbf k\nu_2}\rangle    
\end{eqnarray}
which can be conveniently used to perform numerical calculations. Based on the general linear response results, we can derive the Hall response and longitudinal response formulas, respectively.\\

It is straightforward to show that for systems with non-degenerate band structure and large driving frequency
the paramagnetic and diamagnetic contributions satisfy the following relation~\cite{Kumar2020FloquetLinearResponse}: 
\begin{eqnarray}
\mathcal K^{(0)}_{\alpha\beta}=-\chi^{(0)}_{\alpha\beta}(0).    
\end{eqnarray}
More importantly, $\chi_{\alpha\beta}^{(0)}(0)$ takes a real value under above conditions. Therefore, the diamagnetic part only contribute to the imaginary part of the conductivity. In the following, we will focus on the real part of the conductivity, i.e., the paramagnetic part. \\

\textbf{Hall response}.
By using the equation $(\varepsilon_{\mathbf k\nu}+i\partial_t)|\phi_{\mathbf k\nu}(t)\rangle=H(t)|\phi_{\mathbf k\nu}(t)\rangle$, we can show that 
\begin{eqnarray}
j_{\alpha,\nu_1\nu_2}(\mathbf k,t)&&=\partial_{k_\alpha}[\langle\phi_{\mathbf k\nu_1}|H_0|\phi_{\mathbf k\nu_2}\rangle]-\langle\partial_{k_\alpha}\phi_{\mathbf k\nu_1}|H_0|\phi_{\mathbf k\nu_2}\rangle-\langle\phi_{\mathbf k\nu_1}|H_0|\partial_{k_\alpha}\phi_{\mathbf k\nu_2}\rangle
\nonumber\\
&&=\partial_{k_\alpha}[\langle\phi_{\mathbf k\nu_1}|(\varepsilon_{\mathbf k\nu_2}+i\overrightarrow{\partial}_t)|\phi_{\mathbf k\nu_2}\rangle]-\langle\partial_{k_\alpha}\phi_{\mathbf k\nu_1}|(\varepsilon_{\mathbf k\nu_2}+i\overrightarrow{\partial}_t)|\phi_{\mathbf k\nu_2}\rangle-\langle\phi_{\mathbf k\nu_1}|(\varepsilon_{\mathbf k\nu_1}-i\overleftarrow{\partial}_t)|\partial_{k_\alpha}\phi_{\mathbf k\nu_2}\rangle\nonumber\\
&&=\delta_{\nu_1\nu_2}\partial_{k_\alpha}\varepsilon_{\mathbf k\nu_1}+i\partial_t(\langle\phi_{\mathbf k\nu_1}|\partial_{k_\alpha}\phi_{\mathbf k\nu_2}\rangle)+(\varepsilon_{\mathbf k\nu_2}-\varepsilon_{\mathbf k\nu_1})\langle\phi_{\mathbf k\nu_1}|\partial_{k_\alpha}\phi_{\mathbf k\nu_2}\rangle.   
\end{eqnarray}
We further substitute the Fourier-transformed expression of states  into the equation above, then the Fourier component of the current operator elements is identified as below
\begin{eqnarray}
j^{(m)}_{\alpha,\nu_1\nu_2}(\mathbf k)=(m\omega+\varepsilon_{\nu_2}-\varepsilon_{\nu_1})\sum_{n}\langle\phi^{(n)}_{\mathbf k\nu_1}|\partial_{k_\alpha}\phi^{(n+m)}_{\mathbf k\nu_2}\rangle.    
\end{eqnarray}
Therefore, we have
\begin{eqnarray}
j^{(m)}_{\alpha,\nu_1\nu_2}(\mathbf k)j^{(-m)}_{\beta,\nu_2\nu_1}(\mathbf k)=-(\varepsilon_{\nu_1\mathbf k}-\varepsilon_{\nu_2\mathbf k}-m\omega)^2C^{(m)}_{\alpha,\nu_1\nu_2}C^{(-m)}_{\beta,\nu_2\nu_1}    
\end{eqnarray}
where $C^{(m)}_{\alpha,\nu_1\nu_2}=\sum_{n}\langle\phi^{(n)}_{\mathbf k\nu_1}|\partial_{k_\alpha}\phi^{(n+m)}_{\mathbf k\nu_2}\rangle$. In the DC limit, the response tensor in Eq.~\eqref{eq:response} is antisymmetric, hence the DC Hall response coefficient is given by
\begin{eqnarray}
\sigma_{\alpha\beta}(0)&&=\lim_{\Omega\rightarrow 0}\frac{\chi^{(0)}_{\alpha\beta}(\Omega)-\chi^{(0)}_{\beta\alpha}(\Omega)}{i2\Omega}\nonumber\\
&&=\lim_{\Omega\rightarrow 0}\frac{1}{i2\Omega}\frac{1}{V}\sum_{\mathbf k}\sum_{\nu_1\nu_2}\sum_{m}\frac{(\rho_{\mathbf k\nu_1}-\rho_{\mathbf k\nu_2})}{\Omega-m\omega+(\varepsilon_{\mathbf k\nu_1}-\varepsilon_{\mathbf k\nu_2})+i\eta}(\varepsilon_{\nu_1\mathbf k}-\varepsilon_{\nu_2\mathbf k}-m\omega)^2[C^{(m)}_{\beta,\nu_1\nu_2}C^{(-m)}_{\alpha,\nu_2\nu_1}-C^{(m)}_{\alpha,\nu_1\nu_2}C^{(-m)}_{\beta,\nu_2\nu_1}]\nonumber\\     
\end{eqnarray}
For a two-band model with upper and lower band being respectively labeled by $u,d$, the Hall coefficient is reduced to
\begin{eqnarray}\label{eq:hallcond}
\sigma_{\alpha\beta}(0)
&&=\lim_{\Omega\rightarrow 0}\frac{1}{i2\Omega}\frac{1}{V}\sum_{\mathbf k}\sum_{m}(\rho_{\mathbf k u}-\rho_{\mathbf kd})[\frac{1}{\Omega-m\omega+(\varepsilon_{\mathbf ku}-\varepsilon_{\mathbf kd})+i\eta}+\frac{1}{\Omega+m\omega+(\varepsilon_{\mathbf kd}-\varepsilon_{\mathbf ku})+i\eta}]\nonumber\\
&&\qquad\qquad\qquad\qquad\quad\times
(\varepsilon_{\mathbf ku}-\varepsilon_{\mathbf kd}-m\omega)^2[C^{(m)}_{\beta,ud}C^{(-m)}_{\alpha,du}-C^{(m)}_{\alpha,ud}C^{(-m)}_{\beta,du}]\nonumber\\
&&=\int \frac{d^2\mathbf{k}}{(2\pi)^2}\bar{\Omega}_{d}(\mathbf k)(\rho_{\mathbf k u}-\rho_{\mathbf k d}).     
\end{eqnarray}
Here, we used the relation~\cite{Dehghani2015FloquetHall}
\begin{eqnarray}
\bar{\Omega}_{d}(\mathbf k)=i\sum_m C^{(m)}_{\beta,ud}C^{(-m)}_{\alpha,du}-C^{(m)}_{\alpha,ud}C^{(-m)}_{\beta,du}=\frac{i}{T}\int_0^T dt \Big(\langle\partial_\alpha\phi_{\mathbf k d}(t)|\partial_\beta\phi_{\mathbf k d}(t)\rangle- \langle\partial_\beta\phi_{\mathbf k d}(t)|\partial_\alpha\phi_{\mathbf k d}(t)\rangle\Big).    
\end{eqnarray}
Moreover, from Eq.~\eqref{eq:hallcond} we obtain the optical Hall conductance expression
\begin{eqnarray}\label{eq:opticalHall}
\sigma_{\alpha\beta}(\Omega)
&&=\int\frac{d^2\mathbf k}{(2\pi)^2}\sum_{m}(\rho_{\mathbf k u}-\rho_{\mathbf kd})\frac{-i}{(\Omega+i\eta)^2-(\varepsilon_{\mathbf ku}-\varepsilon_{\mathbf kd}-m\omega)^2}
(\varepsilon_{\mathbf ku}-\varepsilon_{\mathbf kd}-m\omega)^2[C^{(m)}_{\beta,ud}C^{(-m)}_{\alpha,du}-C^{(m)}_{\alpha,ud}C^{(-m)}_{\beta,du}]\nonumber\\ 
&&=\int\frac{d^2\mathbf k}{(2\pi)^2}\sum_{m}(\rho_{\mathbf k u}-\rho_{\mathbf kd})\frac{2\text{Im}[j^{(m)}_{\beta,ud}j^{(-m)}_{\alpha,du}]}{(\Omega+i\eta)^2-(\varepsilon_{\mathbf ku}-\varepsilon_{\mathbf kd}-m\omega)^2}
.
\end{eqnarray}

\textbf{Longitudinal response}. For the longitudinal response, we consider the real part of the response function and two-band system is assumed, i.e. 
\begin{eqnarray}\label{eq:longitudinalcond}
\text{Re}[\sigma_{\alpha\alpha}]&&=\frac{\text{Im}[\chi^{(0)}_{\alpha\alpha}]}{\Omega} \nonumber\\
&&=\frac{1}{\Omega V}\sum_{\mathbf k}\sum_{\nu_1\nu_2}\sum_m(\rho_{\mathbf k\nu_1}-\rho_{\mathbf k\nu_2})|j_{\alpha,\nu_1\nu_2}^{(m)}|^2\frac{(-\eta)}{(\Omega-m\omega+\varepsilon_{\mathbf k\nu_1}-\varepsilon_{\mathbf k\nu_2})^2+\eta^2}\nonumber\\
&&=\int\frac{d^2\mathbf k}{(2\pi)^2}\sum_m(\rho_{\mathbf k u}-\rho_{\mathbf k d})|j_{\alpha,ud}^{(m)}|^2\frac{4\eta(\varepsilon_{\mathbf ku}-\varepsilon_{\mathbf k d}-m\omega)}{[\Omega^2-(m\omega+\varepsilon_{\mathbf kd}-\varepsilon_{\mathbf ku})^2]^2+2\eta^2[\Omega^2+(m\omega+\varepsilon_{\mathbf k d}-\varepsilon_{\mathbf k u})^2]}.
\end{eqnarray}
Here, we used the relation $[j^{(m)}_{\alpha,\nu_1\nu_2}]^\ast=j^{(-m)}_{\alpha,\nu_2\nu_1}$.

\subsection{Steady-state spin current (pumped spin current)} 

In the non-equilibrium steady state, a nonvanishing current is allowed to exist, which is the spin current pumped by the driving light radiation. If the system-bath coupling is much weaker than the driving frequency and Floquet level spacing, the steady state population varies slowly, and the off-diagonal component of the density matrix is highly suppressed, such that 
\begin{eqnarray}
\hat\rho_s(t)=\sum_n\rho_n|\psi_n(t)\rangle\langle \psi_n(t)|.   
\end{eqnarray}
The averaged steady-state current is calculated as follows:
\begin{eqnarray}
\bar{j}_{\mathbf k}&&=\frac{1}{T}\int_0^T dt \text{Tr}[\hat\rho_s(\mathbf k,t)\partial_{\mathbf k}H_{\mathbf k}]\nonumber\\
&&=\sum_n\rho_n(\mathbf k)\frac{1}{T}\int_0^T dt \langle\psi_n(\mathbf k,t)|\partial_{\mathbf k}H_{\mathbf k}|\psi_n(\mathbf k,t)\rangle\nonumber\\
&&=\sum_n\rho_n(\mathbf k)\frac{1}{T}\int_0^T dt[\langle \phi_n(\mathbf k,t)|\partial_{\mathbf k}\Big((\varepsilon_{n\mathbf k}+i\partial_t)|\phi_n(\mathbf k,t)\rangle\Big)-\langle\phi_n(\mathbf k,t)|H_\mathbf k\partial_{\mathbf k}|\phi_n(\mathbf k,t)\rangle]\nonumber\\
&&=\sum_n\rho_n(\mathbf k)\frac{1}{T}\int_0^T dt[\langle \phi_n(\mathbf k ,t)|\partial_{\mathbf k}\Big((\varepsilon_{n \mathbf k}+i\partial_t)|\phi_n(\mathbf k,t)\rangle\Big)-\langle\phi_n(\mathbf k,t)|(\varepsilon_{n\mathbf k}-i\overleftarrow{\partial}_t)\partial_{\mathbf k}|\phi_n(\mathbf k,t)\rangle]\nonumber\\
&&=\sum_n\rho_n(\mathbf k)\frac{1}{T}\int_0^T dt \partial_{\mathbf k}\varepsilon_{n\mathbf k}\langle\phi_n(\mathbf k,t)|\phi_n(\mathbf k,t)\rangle+i\partial_t\Big(\langle\phi_n(\mathbf k,t)|\partial_{\mathbf k}\phi_n(\mathbf k,t)\rangle\Big)\nonumber\\
&&=\sum_n\rho_n(\mathbf k)\partial_{\mathbf k}\varepsilon_{n \mathbf k},
\end{eqnarray}
where we used the equation $(\varepsilon_{n\mathbf k}+i\partial_t)|\phi_n(\mathbf k,t)\rangle=H_{\mathbf k}(t)|\phi_n(\mathbf k,t)\rangle$ in the derivation above, and we assumed that the system respects translational symmetry.

\begin{figure}
    \centering   \includegraphics[width=0.3\linewidth]{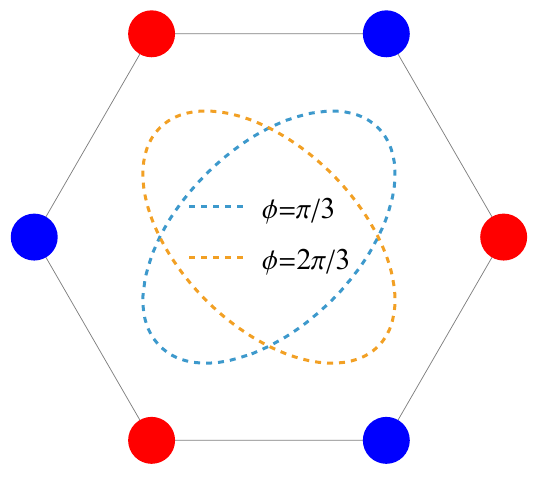}
    \caption{Light-dressed unit cell, where the dashed lines indicate the trajectory of the vector potential of the incident light.}
    \label{fig:dressed_lattice}
\end{figure}

\textbf{Symmetry of steady-state spin current}. In Fig.3(a) of the main text, the steady-state current exhibits an oscillatory dependence on the polarization angle $\varphi$. Two characteristic features can be identified:

(i) The current is antisymmetric with respect to $\varphi=\pi$, i.e.,
\begin{eqnarray}\label{eq:steadycurrent_symmetry1}
 \bar{j}^s_i(2\pi-\varphi)=-\bar{j}^s_i(\varphi).  
\end{eqnarray}

(ii) Within the interval $[0,\pi]$, $\bar{j}^s_x$ and $\bar{j}^s_y$ are symmetric and antisymmetric with respect to $\varphi=\frac{\pi}{2}$, respectively, i.e.,
\begin{eqnarray}\label{eq:steadycurrent_symmetry2}
&&\bar{j}^s_x(\frac{\pi}{2}-\varphi)=\bar{j}^s_x(\frac{\pi}{2}+\varphi),\nonumber\\
&&\bar{j}^s_y(\frac{\pi}{2}-\varphi)=-\bar{j}^s_y(\frac{\pi}{2}+\varphi). 
\end{eqnarray}
Specifically, the numerical data for $\varphi\in[0,\pi]$ is fitted to $\bar{j}^s_x(\varphi)=0.00338\sin(\varphi)+0.00618\sin(3\varphi)+0.00114\sin(5\varphi)+\cdots$ and  $\bar{j}^s_y(\varphi)=-0.0225\sin(2\varphi)+0.00194\sin(6\varphi)+\cdots$, where only the leading harmonics are retained; the omitted terms are at least one order of magnitude smaller.

These two features can be understood from symmetry considerations of the driven system. First, for characteristic (i), we note that
\begin{eqnarray}
\boldsymbol{\mathcal A}(2\pi-\varphi,t)=\boldsymbol{\mathcal A}(-\varphi,t)=A_0(\sin(\omega t),\sin(\omega t-\varphi),0)=-\boldsymbol{\mathcal A}(\varphi,-t),    
\end{eqnarray}
which implies 
\begin{eqnarray}\label{eq:phiantisymmetry}
H_{\sigma}[\mathbf k+\boldsymbol{\mathcal A}(2\pi-\varphi,t)]=H_{\sigma}[\mathbf k-\boldsymbol{\mathcal A}(\varphi,-t)]=\mathcal T H^\ast_{-\sigma}[-\mathbf k+\boldsymbol{\mathcal A}(\varphi,t)]\mathcal T^{-1}.  
\end{eqnarray}
Physically, replacing the polarization angle $\varphi$ by $2\pi-\varphi$ is equivalent to applying time-reversal symmetry to the vector potential, without reversing the spin. As a consequence,
\begin{eqnarray}
&&\varepsilon_{n\mathbf k}^{\sigma}(2\pi-\varphi)= \varepsilon_{n,-\mathbf k}^{-\sigma}(\varphi),\nonumber\\
&&\rho^{\sigma}_n(\mathbf k,2\pi-\varphi)=\rho^{-\sigma}_n(-\mathbf k,\varphi).
\end{eqnarray}
These relations further result in
\begin{eqnarray}
 \bar{j}^{\sigma}_{\mathbf k}(2\pi-\varphi)=\bar{j}^{-\sigma}_{-\mathbf k}(\varphi).    
\end{eqnarray}
Combining the equation above with $\bar{j}^s=\sum_{\sigma}\sigma\int\frac{d^2\mathbf k}{(2\pi)^2}j^\sigma_{\mathbf k}$, we recover the antisymmetry relation in feature (i) above.

Feature (ii) can be understood from the symmetry of the light-dressed system, which is determined by the combined configuration of the underlying lattice and the optical field trajectory. The latter is described by
\begin{eqnarray}
\Tilde{\mathcal A}_x^2 +\Tilde{\mathcal A}_y^2-2\Tilde{\mathcal A}_x\Tilde{\mathcal A}_y\cos\varphi=\sin^2\varphi   
\end{eqnarray}  
where $\Tilde{\mathcal A}_{x,y}=\mathcal A_{x,y}/A_0$. As shown in Fig.~\ref{fig:dressed_lattice}, the configurations of $\varphi=\frac{\pi}{2}\pm \Tilde{\varphi}$ can be related by a combined mirror reflection $\mathcal M_x$ about the $x$-axis and a time reversal $\mathcal T$. The spin current operator is defined as $\mathbf{j}^s=\frac{1}{2} \{\sigma^z,\mathbf{j}^e\}$ where $\mathbf j^e$ is charge current. Under the operation $\mathcal M_x\mathcal T$, the spin operator is invariant, i.e., $\sigma^z\rightarrow\sigma^z$, spatial variables transform as $(x,y)\rightarrow (x,-y)$, and the charge current as $(j^e_{x},j^e_y)\rightarrow (j^e_{x},-j^e_y)$. Consequently, the spin current transforms as
\begin{eqnarray}
&&j^s_x\rightarrow j^s_x,\nonumber\\
&&j^s_y\rightarrow -j^s_y,
\end{eqnarray}
which directly leads to the feature (ii). We note that in the combined operation $\mathcal M_x\mathcal T$, the reflection $\mathcal M_x$ maps the optical trajectory, while the time reversal $\mathcal T$ restores the handedness of the light, ensuring that the vector potential rotates in the same (clockwise or anticlockwise) direction.

\section{Couple to a fermionic bath}

We first consider a toy model, in which the system is coupled to a fermionic reservoir that is described by 
\begin{eqnarray}
H_{res}=\sum_l E_ld_l^\dagger d_l.    
\end{eqnarray}
The coupling is described by the tunneling between the system and the reservoir:
\begin{eqnarray}
H_{tun}=\sum_{l,\mathbf ka} J_{l,\mathbf ka}(c^\dagger_{\mathbf ka}d_l+h.c.)   
\end{eqnarray}
where $a$ is the sublattice index. 
According to the Floquet Fermi-golden rule, the rate for an electron to tunnel  from the reservoir to the state $|\psi_{\mathbf k\alpha}(t)\rangle$ via the harmonic component $|\phi^{n}_{\mathbf k\alpha}\rangle$ is given by~\cite{Seetharam2015population} 
\begin{eqnarray}\label{eq:Floquet_FG}
\Gamma_{\mathbf k\alpha}^n&&=\frac{2\pi}{\hbar}\sum_{l}|\langle \phi^n_{\mathbf k\alpha}|H_{tun}|l\rangle|^2\delta(\varepsilon_{\mathbf k\alpha}+n\omega-E_{l})\nonumber\\
&&=\frac{2\pi}{\hbar}\sum_{l,ab}J_{l,\mathbf ka}J_{l,\mathbf kb}\langle\phi^n_{\mathbf k\alpha}|\mathbf ka\rangle\langle\mathbf kb|\phi^n_{\mathbf k\alpha}\rangle\delta(\varepsilon_{\mathbf k\alpha}+n\omega-E_{l}).
\end{eqnarray}
If we assume the coupling matrix element $J_{l,\mathbf ka}$ is a constant, i.e, $J_{l,\mathbf ka}=J$, the rate above will be simplified to 
\begin{eqnarray}\label{eq:scatteringRate}
\Gamma_{\mathbf k\alpha}^n=\frac{2\pi}{\hbar}J^2\langle\phi^n_{\mathbf k\alpha}|\Xi|\phi^n_{\mathbf k\alpha}\rangle\sum_{l}\delta(\varepsilon_{\mathbf k\alpha}+n\omega-E_{l})=\frac{2\pi}{\hbar}J^2\langle\phi^n_{\mathbf k\alpha}|\Xi|\phi^n_{\mathbf k\alpha}\rangle,
\end{eqnarray}
where $\Xi=\sum_{a,b}|\mathbf ka\rangle\langle\mathbf kb|$ is a $N\times N$ matrix with all element filled by $1$, and $N$ is the number of sublattice in each unit cell. Here, we also assume that the spectrum of the reservoir is dense and broad enough such that $\sum_{l}\delta(\varepsilon_{\mathbf k\alpha}+n\omega-E_{l})=1$. On the other hand, the collision integral counting electrons tunneling into and out of the system is given by~\cite{Seetharam2015population}
\begin{eqnarray}
I_{\mathbf k\alpha}=\sum_n\Gamma^n_{\mathbf k\alpha}\Big[(1-\rho_{\mathbf k\alpha})f_0(\varepsilon_{\mathbf k\alpha}+n\omega)-\rho_{\mathbf k\alpha}\Big(1-f_0(\varepsilon_{\mathbf k\alpha}+n\omega)\Big)\Big].    
\end{eqnarray}
By setting $\partial_t\rho_{\mathbf k\alpha}=I_{\mathbf k\alpha}=0$, we obtain
\begin{eqnarray}\label{eq:fermionBathDensity}
\rho_{\mathbf k\alpha}=\frac{\sum_n\Gamma^n_{\mathbf k\alpha}f_0(\varepsilon_{\mathbf k\alpha}+n\omega)}{\sum_n\Gamma^n_{\mathbf k\alpha}}.    
\end{eqnarray}

\begin{figure}
    \centering   \includegraphics[width=0.7\linewidth]{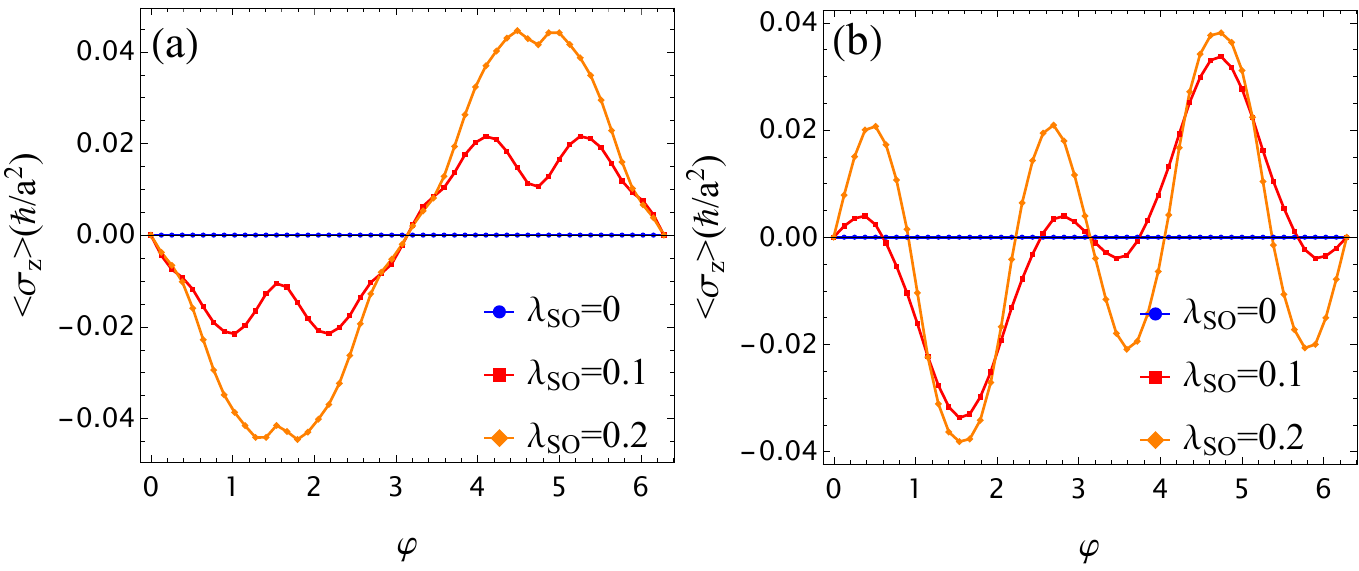}
    \caption{Net spin accumulation when the system is coupled to a fermionic bath. Parameters are $A_0a=1$, $t=1$, $\lambda=0.5$, and $T_{\text{bath}}=0.01t$. The driving frequencies are: (a) $\omega=1$ (b) $\omega=2$.}
    \label{fig:NetSpin}
\end{figure}

In our study, we expect the fast drive can induce a net spin accumulation in the steady state
\begin{eqnarray}\label{eq:spinpolarization}
s^z=\sum_{\alpha}\hbar\int \frac{d\mathbf k}{(2\pi)^2}[\rho_{\uparrow}(\varepsilon^{\uparrow}_{\mathbf k,\alpha})-\rho_{\downarrow}(\varepsilon^{\downarrow}_{\mathbf k,\alpha})].    
\end{eqnarray}
Given that two spin sectors are well separated, using the formula Eq.~\eqref{eq:fermionBathDensity} one can obtain the steady-state spin accumulation. As shown in Fig.~\ref{fig:NetSpin},we find the spin accumulation still requires a nonzero spin-orbit coupling to break the dual symmetry in the main text. Moreover, the magnitude and sign of the spin accumulation exhibit a high tunability with respect to the phase factor of the light field. 

\subsection{Couple to electrodes}

\begin{figure}
    \centering
    \includegraphics[width=0.8\linewidth]{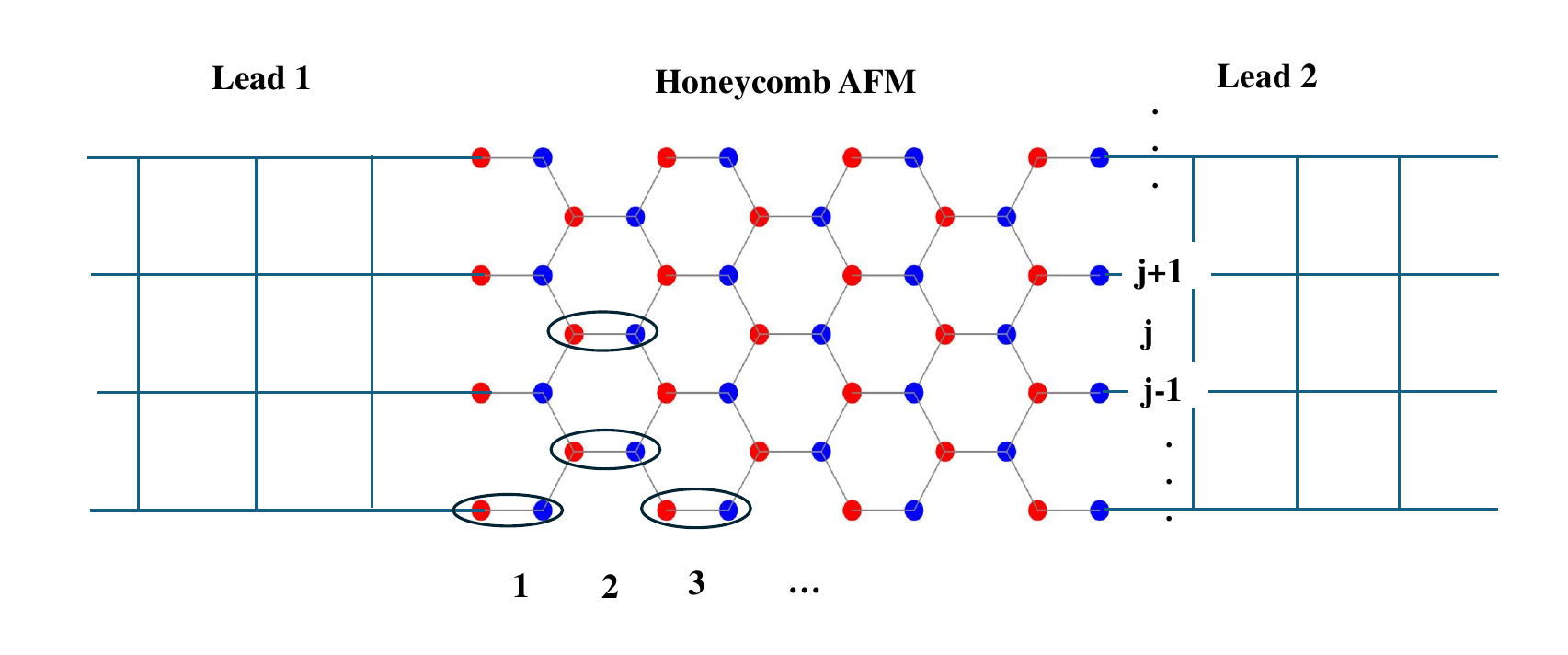}
    \caption{The honeycomb antiferromagnet is contacted to two metallic leads on its left and right. The two leads are characterized by chemical potentials $\mu_L$ and $\mu_R$, respectively.}
    \label{fig:junction}
\end{figure}

Above, we assumed uniform system-bath coupling, and the bath is undriven. This ideal case, however, is not easy to realize. In reality, the fermionic baths (leads) usually are coupled to the system at the interface, for instance, see the heterostructure in Fig.~\ref{fig:junction}. Here, we assume that the system takes a periodic boundary condition along the vertical direction ($y$ direction). This heterostructure can be described by 
\begin{eqnarray}
H= H_S + H_B+ H_{SB}.   
\end{eqnarray}

Here, 
\begin{eqnarray}
H_S&&=\frac{t}{2}\sum_{m,n}\Big[\psi^\dagger_{(m,n)}\tau^x\psi_{(m,n)}+\psi^\dagger_{(m,n)}\tau^-\psi_{(m+1,n+1)}+ \psi^\dagger_{(m,n)}\tau^-\psi_{(m+1,n-1)}+ \psi^\dagger_{(m,n)}\tau^+\psi_{(m-1,n+1)}\nonumber\\
&&\qquad\qquad+ \psi^\dagger_{(m,n)}\tau^+\psi_{(m-1,n-1)}+H.c.\Big]\nonumber\\
&&\quad+i\frac{\lambda_{\text{SO}}}{2}\sum_{m,n}\Big[\psi^\dagger_{(m,n)}(-\tau^z\otimes\sigma^z)\psi_{(m+1,n+1)}+ \psi^\dagger_{(m,n)}(-\tau^z\otimes\sigma^z)\psi_{(m-1,n+1)} +\psi^\dagger_{(m,n)}(-\tau^z\otimes\sigma^z)\psi_{(m,n-2)}\nonumber\\
&&\qquad\qquad\qquad+\psi^\dagger_{(m,n)}\tau^z\otimes\sigma^z\psi_{(m,n+2)}+\psi^\dagger_{(m,n)}\tau^z\otimes\sigma^z\psi_{(m-1,n-1)}+\psi^\dagger_{(m,n)}\tau^z\otimes\sigma^z\psi_{(m+1,n-1)}+H.c.\Big]\nonumber\\
&&\quad+\lambda\sum_{m,n}\psi^\dagger_{(m,n)}\tau^z\otimes\sigma^z\psi_{(m,n)}
\end{eqnarray}
where $\tau^{\pm}=\frac{1}{2}(\tau^x\pm i\tau^y)$ are the linear combinations of Pauli matrices in sublattice space; $H_B$ describes a semi-infinite lead of square lattice, and its momentum presentation reads $H_B=\sum_{k_x,k_y}2t_s[\cos(k_x c)+\cos (k_y c)]d_{\mathbf k}^\dagger d_{\mathbf k}$ with $c$ being the unit-cell length of the lattice; the coupling between the system and leads is given by
\begin{eqnarray}
H_{SB}=\sum_n J\psi^\dagger_{(1,2n),A}d_{1,n}+h.c.. 
\end{eqnarray}
Given the periodic boundary condition along $y$-direction, we perform the corresponding Fourier transformation:
\begin{eqnarray}\label{eq:system}
H_S&&= t\sum_m\sum_{k_y}\Big[\psi^\dagger_{(m,k_y)}\tau^x\psi_{(m,k_y)}+\psi^\dagger_{(m,k_y)}\tau^-\cos (k_y a_y) \psi_{(m+1,k_y)}+ \psi^\dagger_{(m,k_y)}\tau^+\cos (k_y a_y) \psi_{(m-1,k_y)}+H.c.\Big]\nonumber\\
&&\quad+\lambda_{\text{SO}}\sum_m\sum_{k_y}\Big[\psi^\dagger_{(m,k_y)}\tau^z\otimes\sigma^z\sin (k_y a_y) \psi_{(m+1,k_y)}
+\psi^\dagger_{(m,k_y)}\tau^z\otimes\sigma^z\sin (k_y a_y) \psi_{(m-1,k_y)}\nonumber\\
&&\qquad\qquad\qquad\quad
-\psi^\dagger_{(m,k_y)}\tau^z\otimes\sigma^z\sin (2k_y a_y) \psi_{(m,k_y)}+H.c.\Big]\nonumber\\
&&\quad+\lambda\sum_m\sum_{k_y}\psi^\dagger_{(m,k_y)}\tau^z\otimes\sigma^z \psi_{(m,k_y)},  
\end{eqnarray}
and 
\begin{eqnarray}
H_{SB}=\sum_{k_y} J\psi^\dagger_{(1,k_y),A}d_{1,k_y}+h.c..   
\end{eqnarray}

The system Hamiltonian can be represented in a matrix form under the basis $\Psi_{k_y}=(\psi_{(1,k_y)},\psi_{(2,k_y)},\cdots, \psi_{(N,k_y)})^T$, i.e., $H_S=\sum_{k_y}\Psi^\dagger_{k_y}H_s(k_y)\Psi_{k_y}$ with 
\begin{eqnarray}\label{eq:Hmatrix}
H_S(k_y)&&=\left(
\begin{array}{ccccc}
  u(k_y) & v(k_y)&0&0&\cdots  \\
  v^\dagger(k_y)&u(k_y)&v(k_y)&0 &\cdots\\
  0& v^\dagger(k_y)& u(k_y) & v(k_y) &\cdots\\
  0 & 0 & \ddots & u(k_y)&\vdots\\
  \vdots &\vdots &\vdots &\vdots &\vdots
\end{array}
\right)\nonumber\\ 
&&=t\mathbbm{1}_N\otimes\tau^x\otimes\sigma^0+ t\cos(k_ya_y)\Big(L^-\otimes\tau^-\otimes\sigma^0+ L^+\otimes\tau^+\otimes\sigma^0\Big)\nonumber\\
&&\quad +\lambda_{\text{SO}}\sin(k_ya_y)\Big(L^-\otimes\tau^z\otimes\sigma^z+ L^+\otimes\tau^z\otimes\sigma^z\Big)\nonumber\\
&&\quad +[\lambda-\lambda_{\text{SO}}\sin(2k_ya_y)]\mathbbm{1}_N\otimes\tau^z\otimes\sigma^z,
\end{eqnarray}
where $u(k_y) = [\lambda-\lambda_{\text{SO}}\sin(2k_ya_y)]\tau^z\otimes\sigma^z+t\tau^x\otimes\sigma^0$,  $v(k_y)= t\cos(k_ya_y)\tau^-\otimes\sigma^0+\lambda_{\text{SO}}\sin(k_ya_y)\tau^z\otimes\sigma^z$, $L^{+}_{mn}=\delta_{m,n+1}$ and $L^{-}_{mn}=\delta_{m,n-1}$.
We assume that only the system is driven by an optical field, which is implemented in Eq.~\eqref{eq:Hmatrix} as
\begin{eqnarray}
H_S(k_y,t)&&=t\mathbbm{1}_N\otimes(\tau^+e^{i\mathcal A _xa_x}+\tau^-e^{-i\mathcal A _xa_x})\otimes\sigma^0\nonumber\\
&&\quad+L^-\otimes\tau^-\otimes\sigma^0 e^{i\mathcal A_x \frac{a_x}{2}}t\cos[(k_y+\mathcal A_y) a_y]+ L^+\otimes\tau^+\otimes\sigma^0e^{-i\mathcal A_x \frac{a_x}{2}}t\cos[(k_y+\mathcal A_y) a_y]\nonumber\\
&&\quad+L^-\otimes\tau^z\otimes\sigma^z e^{i\mathcal A_x \frac{3a_x}{2}}\lambda_{\text{SO}}\sin[(k_y+\mathcal A_y)a_y]+ L^+\otimes\tau^z\otimes\sigma^ze^{-i\mathcal A_x \frac{3a_x}{2}}\lambda_{\text{SO}}\sin[(k_y+\mathcal A_y)a_y]\nonumber\\
&&\quad+\mathbbm{1}_N\otimes\tau^z\otimes\sigma^z\Big(\lambda-\lambda_{\text{SO}}\sin[2(k_y+\mathcal A_y)a_y]\Big).   
\end{eqnarray}
Here, the involved lattice lengths are given by $a_x=a, a_y=\frac{\sqrt{3}}{2}a$ and $c=2a_y=\sqrt{3}a$ with $a$ being the bond length of the honeycomb lattice. Note that the hopping distance projections on the $x$-direction are different for the nearest neighbor terms within the unit cell ($a_x$) and that between neighboring unit cells ($a_x/2$), and for the second nearest neighbor terms ($3a_x/2$) from intrinsic SOC. Performing a time-domain Fourier transformation, we obtain 
\begin{eqnarray}
H_S^{(m)}(k_y)&&=tJ_m(A_0a)\mathbbm{1}_N\otimes[(-1)^m\tau^++\tau^-]\otimes\sigma^0\nonumber\\
&&\quad+L^-\otimes\tau^-\otimes\sigma^0t(-1)^m[h_+^{(m)}(1/2)e^{ik_ya_y}+h_-^{(m)}(1/2)e^{-ik_ya_y}]\nonumber\\
&&\quad+L^+\otimes\tau^+\otimes\sigma^0t[h_-^{(m)}(1/2)e^{ik_ya_y}+h_+^{(m)}(1/2)e^{-ik_ya_y}]\nonumber\\
&&\quad+L^-\otimes\tau^z\otimes\sigma^z(-i\lambda_{\text{SO}})(-1)^m[h_+^{(m)}(3/2)e^{ik_ya_y}-h_-^{(m)}(3/2)e^{-ik_ya_y}]\nonumber\\
&&\quad+L^+\otimes\tau^z\otimes\sigma^z(-i\lambda_{\text{SO}})[h_-^{(m)}(3/2)e^{ik_ya_y}-h_+^{(m)}(3/2)e^{-ik_ya_y}]\nonumber\\
&&\quad+i\frac{\lambda_{\text{SO}}}{2}e^{-im\varphi}J_m(\sqrt{3}A_0a)[(-1)^me^{i2k_ya_y}-e^{-i2k_ya_y}]\mathbbm{1}_N\otimes\tau^z\otimes\sigma^z\nonumber\\
&&\quad+\delta_{m,0}\lambda\mathbbm{1}_N\otimes\tau^z\otimes\sigma^z
\end{eqnarray}
where 
\begin{eqnarray}
h^{(m)}_\pm(x)=\frac{1}{2} e^{-im\vartheta_\pm(x)}J_m\Big(A_0a\sqrt{\frac{3}{4}+x^2\pm\sqrt{3}x\cos\varphi}\Big).    
\end{eqnarray}
Here, it requires that $\cos\vartheta_\pm(x)=\frac{2x\pm\sqrt{3}\cos\varphi}{\sqrt{3+4x^2\pm 4\sqrt{3}x\cos\varphi}}$ and 
$\sin\vartheta_\pm(x)=\frac{\pm\sqrt{3}\sin\varphi}{\sqrt{3+4x^2\pm 4\sqrt{3}x\cos\varphi}}$, resulting in
\begin{eqnarray}
\vartheta_\pm(x)=\pm\text{sign}(\pi-\varphi)\arccos\frac{2x\pm\sqrt{3}\cos\varphi}{\sqrt{3+4x^2\pm 4\sqrt{3}x\cos\varphi}}.    
\end{eqnarray}

\textbf{Green's function and spin accumulation}. In the following, we calculate the steady-state spin polarization by using Green's function method.  For Floquet systems, we consider the Keldysh Green's function in both orbital and frequency space~\cite{Oka2009photoGraphen,Liu2025FloquetWeyl}. The corresponding Dyson's equation reads
\begin{eqnarray}
\left(\begin{array}{cc}
   G^R  & G^K \\
    0 &  G^A
\end{array}\right)^{-1}= \left(\begin{array}{cc}
   (G^R_0)^{-1}  & 0 \\
    0 &  (G^A_0)^{-1}
\end{array}\right)+
\left(\begin{array}{cc}
   \Sigma^R  & \Sigma^K \\
    0 &  \Sigma^A
\end{array}\right).
\end{eqnarray}
Here, $G^{\alpha}_{mn,ij}=G^{\alpha}_{mn,ij}(k_y)$ ($\alpha=R,K,A$), where $m,n$ refer to frequency index, and $i,j$ denote orbital freedoms; $(G^{R/A}_{0})^{-1}_{mn}=(\varepsilon+m\omega\pm i\delta)\delta_{mn}-H_{mn}$ with $H_{mn}=\frac{1}{T}\int_0^T dt e^{i(m-n)\omega t}H(t)$. Specifically, the frequency components of the Green function come from the Fourier transformation below:
\begin{eqnarray}
G(t,t^\prime)=\sum_{m,n}\int_0^\omega\frac{d\varepsilon}{2\pi} e^{-i(\varepsilon+m\omega)t}e^{i(\varepsilon+n\omega)t^\prime}G_{mn}(\varepsilon).    
\end{eqnarray}
In each spin section, the particle number is given by
\begin{eqnarray}\label{eq:particleNumber}
n_\sigma(t)=-i\sum_{k_y}\text{Tr}[G_\sigma^<(k_y;t,t)],    
\end{eqnarray}
where the trace acts on the spatial coordinate. Taking the time average over one period, the averaged particle number can be represented in the frequency presentation
\begin{eqnarray}\label{eq:averageddensity}
\overline{n}_\sigma=\frac{1}{T}\int_0^Tdt n_\sigma(t)=-i\sum_{k_y}\sum_n\int \frac{d\varepsilon}{2\pi}\text{Tr}[G^<_{\sigma,nn}(k_y,\varepsilon)],   
\end{eqnarray}
which is used to perform the numerical calculation.


Because the discussions can be performed separately in each spin space, we will omit the spin index in the following for simplicity; the resultant formulas apply well to each spin sector. Taking all freedoms into account (frequency, momentum, sites, and orbital), the lesser Green function is given by
\begin{eqnarray}\label{eq:GreenLess}
G^<= G^R\Sigma^< G^A    
\end{eqnarray}
where $\Sigma^<=(\Sigma^R+\Sigma^K-\Sigma^A)/2$. In the case of herterostructure, the self energy is given by
\begin{eqnarray}
\Sigma^R= -H_{SB}G^R_0(\varepsilon+m\omega) H_{SB}^\dagger    
\end{eqnarray}
where the coupling Hamiltonian $H_{SB}$ in matrix form reads $(H_{SB})_{i\alpha,k}=J\delta_{i1}\delta_{\alpha A}\delta_{k1}+J\delta_{iN}\delta_{\alpha B}\delta_{k1}$. Here, the $k$ index refers to the position of the leads, and we label both the sites in the two leads near the interface as $k=1$.  Consequently, the self-energy matrix element for the left lead is given by
\begin{eqnarray}
\Sigma^R_{L,(i\alpha,j\beta)}&&=-J^2\delta_{ij}\delta_{i1}\delta_{\alpha A}\delta_{\alpha\beta}G^R_{0,11}(\varepsilon+m\omega)\nonumber\\
&&=-J^2\delta_{ij}\delta_{i1}\delta_{\alpha A}\delta_{\alpha\beta}\sum_{n}\frac{|\psi_{B,n}(x=1)|^2}{\varepsilon+m\omega-E_n+i0^+}\nonumber\\
&&=-J^2\delta_{ij}\delta_{i1}\delta_{\alpha A}\delta_{\alpha\beta}\int dE\frac{\rho_B(E)|\psi_{B,\omega}(x=1)|^2}{\varepsilon+m\omega-E+i0^+}\nonumber\\
&&\approx i\pi J^2\delta_{ij}\delta_{i1}\delta_{\alpha A}\delta_{\alpha\beta}\rho_{B}(\varepsilon+m\omega)|\psi_{B,\varepsilon+m\omega}(x=1)|^2\nonumber\\
&&=i\frac{\Gamma_L}{2}\delta_{ij}\delta_{i1}\delta_{\alpha A}\delta_{\alpha\beta}
\end{eqnarray}
where $\Gamma_L=2\pi J^2 \rho_{B}(\varepsilon+m\omega)|\psi_{B,\varepsilon+m\omega}(x=1)|^2$. A similar result holds for the right leads. Here, the lead is assumed to have a broad band such that $|\psi_{B,\varepsilon+m\omega}(x=1)|^2\approx const.$ for $m$ within the frequency truncation range.  Moreover, the principal value of the integral is neglected in the calculation above. From the results above, the self-energy matrix is given by
\begin{eqnarray}
\left(\begin{array}{cc}
   \Sigma^R_{mn,ij}  & \Sigma^K_{mn,ij} \\
    0 &  \Sigma^A_{mn,ij}
\end{array}\right)=&&i\delta_{mn}\delta_{i1}\delta_{ij}
\left(
\begin{array}{cc}
  \Gamma_L/2 & -\Gamma_L(1-2f_L(\varepsilon+m\omega))  \\
   0  & -\Gamma_L/2 
\end{array}
\right)\otimes \tau^{A}\nonumber\\
&&+
i\delta_{mn}\delta_{iN}\delta_{ij}
\left(
\begin{array}{cc}
  \Gamma_R/2 & -\Gamma_R(1-2f_R(\varepsilon+m\omega))  \\
   0  & -\Gamma_R/2
\end{array}
\right)\otimes\tau^{B},\nonumber\\
\end{eqnarray}
where $\tau^{A/B}=\frac{1}{2}(\tau^1\pm \tau^3)$, and $f_a(x)=[e^{(x-\mu_a)/(k_BT)}+1]^{-1}$  ($a=L,R$) is the Fermi-Dirac distribution function for the leads. This leads to
\begin{eqnarray}
\Sigma^<_{mn,ij}=i2\delta_{mn}\delta_{i1}\delta_{ij}\Gamma_Lf_L(\varepsilon+m\omega)\tau^{A}+i2\delta_{mn}\delta_{iN}\delta_{ij}\Gamma_Rf_R(\varepsilon+m\omega)\tau^{B}.     
\end{eqnarray}
Applying this result in each spin sector and substituting it into Eq.~\eqref{eq:GreenLess}, we get the lesser Green function. By further applying the results to Eq.~\eqref{eq:particleNumber} and combining it with Eq.~\eqref{eq:spinpolarization}, we can finally obtain the spin accumulation Fig.4 (a) in the system, see the result in the main text. 
\begin{figure}
\centering
\includegraphics[width=0.8\linewidth]{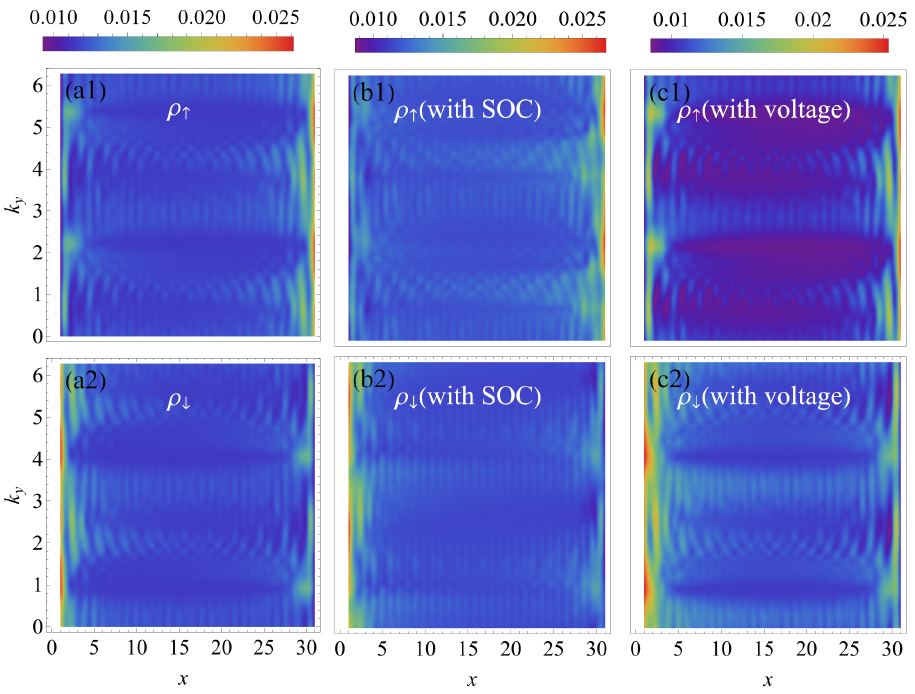}
\caption{Particle density  in the mixed $x-k_y$ representation for each spin sector. The parameters are  $\varphi=\pi/2, A_0a=1,\omega=1$ and $t=1,\lambda=0.5$, and there are $31$ unit cells along $x$-direction. In (a1,a2), the dual relation is preserved. In (b1,b2) and (c1,c2), the dual relation is explicitly broken due to the presence of spin–orbit coupling $\lambda_{\text{SO}}=0.2$ or an applied bias voltage $V=0.5$.}
\label{fig:mixed_density}
\end{figure}

\textbf{Symmetry of spin accumulation.} Here, we note that according to Fig.4 (a) in the main text, the spin accumulation is an antisymmetric function of $\varphi$ with respect to $\varphi=\pi$. As Eq.~\eqref{eq:phiantisymmetry}, the replacement $\varphi\rightarrow 2\pi-\varphi$ is equivalent to performing a time reversal on the vector potential. This imposes the following constraint on the time-averaged accumulated spins 
\begin{eqnarray}
\bar{\sigma}^z(2\pi-\varphi)&&=\sum_{\sigma}\sigma\frac{1}{T}\int_0^T dt n_{\sigma}[\boldsymbol{\mathcal A}(2\pi-\varphi,t)]\nonumber\\
&&=\sum_{\sigma}\sigma\frac{1}{T}\int_0^T dt n_{\sigma}[-\boldsymbol{\mathcal A}(\varphi,-t)]\nonumber\\
&&=\sum_{\sigma}\sigma\frac{1}{T}\int_0^T dt n_{-\sigma}[\boldsymbol{\mathcal A}(\varphi,t)]\nonumber\\
&&=-\bar{\sigma}^z(\varphi),
\end{eqnarray}
where we used the fact that the particle number in each spin sector is invariant under the time reversal operation, i.e., $n_{\sigma}[-\boldsymbol{\mathcal A}(\varphi,-t)]=\mathcal T n_{\sigma}[-\boldsymbol{\mathcal A}(\varphi,-t)]\mathcal T^{-1}=n_{-\sigma}[\boldsymbol{\mathcal A}(\varphi,t)]$. More rigorously, this relation can be derived as follows. The particle number density in a given spin sector is given by
\begin{eqnarray}
n_{\sigma}[-\boldsymbol{\mathcal A}(\varphi,-t)]&&= -i\sum_{k_y}\sum_j G^<_{\sigma,jj}(k_y,-\boldsymbol{\mathcal A}(\varphi,-t))\nonumber\\
&&=\sum_{k_y,j}\text{Tr}\Big[e^{-i\int_0^{-t}dt^\prime H_{\sigma}[k_y,-\boldsymbol{\mathcal A}(\varphi,t^\prime)]}\rho_{\sigma}(k_y,-\boldsymbol{\mathcal A}(\varphi,0))e^{i\int_0^{-t}dt^\prime H_{\sigma}[k_y,-\boldsymbol{\mathcal A}(\varphi,t^\prime)]}\psi^\dagger_{\sigma,j}(k_y,-t)\psi_{\sigma, j}(k_y,-t)\Big].\nonumber\\
\end{eqnarray}
Applying the Eq.~\eqref{eq:phiantisymmetry} to the expression above yields
\begin{eqnarray}
n_{\sigma}[-\boldsymbol{\mathcal A}(\varphi,-t)]
&&=\sum_{k_y,j}\text{Tr}\Big[\Big(e^{-i\int_0^{t}dt^\prime H_{-\sigma}[-k_y,\boldsymbol{\mathcal A}(\varphi,t^\prime)]}\rho_{-\sigma}(-k_y,\boldsymbol{\mathcal A}(\varphi,0))e^{i\int_0^{t}dt^\prime H_{-\sigma}[-k_y,\boldsymbol{\mathcal A}(\varphi,t^\prime)]}\Big)^\ast\nonumber\\
&&\qquad\qquad\times\psi^\dagger_{-\sigma,j}(-k_y,t)\psi_{-\sigma, j}(-k_y,t)\Big]\nonumber\\
&&=\sum_{k_y,j}[\rho_{-\sigma}(-k_y,\boldsymbol{\mathcal A}(\varphi,t))]_{jj}^\ast=\sum_{k_y,j}[\rho_{-\sigma}(-k_y,\boldsymbol{\mathcal A}(\varphi,t))]_{jj}\nonumber\\
&&=n_{-\sigma}[\boldsymbol{\mathcal A}(\varphi,t)].
\end{eqnarray}
Here, we used the following properties: $e^{-i\int_0^{-t}dt^\prime H_{\sigma}[k_y,-\boldsymbol{\mathcal A}(\varphi,t^\prime)]}=e^{-i\int_0^{-t}dt^\prime \mathcal TH^\ast_{-\sigma}[-k_y,\boldsymbol{\mathcal A}(\varphi,-t^\prime)]\mathcal T^{-1}}=\mathcal T\Big(e^{-i\int_0^{t}dt^\prime H_{-\sigma}[-k_y,\boldsymbol{\mathcal A}(\varphi,t^\prime)]}\Big)^\ast\mathcal T^{-1}$, $\mathcal T\rho_{\sigma}(k_y,-\boldsymbol{\mathcal A}(\varphi,0))\mathcal T^{-1}=\rho_{-\sigma}(-k_y,\boldsymbol{\mathcal A}(\varphi,0))$, and the diagonal components of the density matrix take real values.

On the other hand, the spin accumulation is symmetric about $\varphi=\pi/2$ in the interval $0\leq \varphi\leq \pi$: $\bar{\sigma}^z(\pi/2-\Tilde{\varphi})=\bar{\sigma}^z(\pi/2+\Tilde{\varphi})$. 
This result also follows from the fact that the two configurations at $\varphi=\pi/2\pm\Tilde{\varphi}$ are related by the combined operation of mirror reflection $\mathcal M_x$ and time reversal $\mathcal T$, and the spin is invariant under the combined operation of these two operators.\\

\textbf{The dual relation.}
We note that the dual relation given in Eq.(3) of the main text also applies to the heterostructure. The essential ingredient underlying this dual relation is that inversion symmetry interchanges the two sublattices, thereby mapping the Hamiltonians in the two spin sectors onto each other. Provided that inversion symmetry is preserved in the heterostructure, the dual relation can be expressed as
\begin{eqnarray}
\tau^xH_{\uparrow}[x, k_y,\boldsymbol{\mathcal A}(t)]\tau^x= H_{\uparrow}[-x, -k_y,-\boldsymbol{\mathcal A}(t)].   
\end{eqnarray}
Here, the Hamiltonian contains both the system and the bath. Similar to Eq.~\eqref{eq:averageddensity}, we can define a particle density in the $x-k_y$ mixed space:
\begin{eqnarray}\label{eq:localdensity}
\rho_\sigma(x=i,k_y)=-i\sum_n\int \frac{d\varepsilon}{2\pi}[G^<_{\sigma,nn}(\varepsilon,k_y)]_{ii},   
\end{eqnarray}
where $i$ indexes the lattice site in the $x$-direction. According to  the dual relation between Hamiltonian, we obtain a dual relation between local density in the mixed space:
\begin{eqnarray}
\rho_{\uparrow}(x,k_y)= \rho_{\downarrow}(-x,-k_y).   
\end{eqnarray}
As shown in Fig.~\ref{fig:mixed_density}(a1) and (a2), the dual relation is satisfied in the absence of spin–orbit coupling (SOC) and when the system is attached to inversion-symmetric leads. In contrast, as illustrated in Fig.~\ref{fig:mixed_density}(b1), (b2), and (c1), (c2), the dual symmetry is explicitly broken by the presence of SOC or by unequal chemical potentials in the two leads, i.e., upon applying a finite bias voltage.\\

\begin{figure}
    \centering    \includegraphics[width=0.7\linewidth]{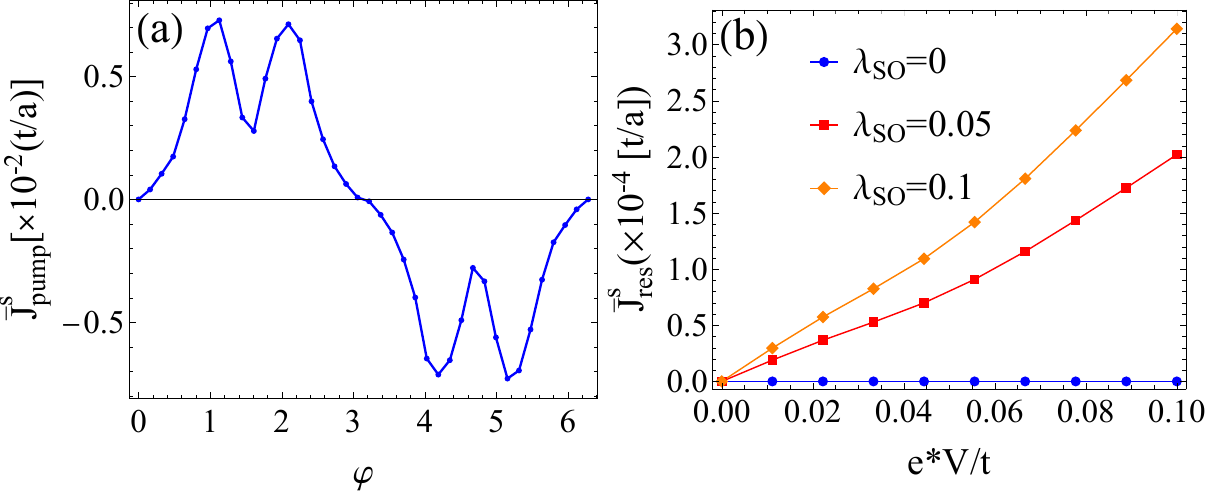}
    \caption{ (a) Steady spin current along longitudinal direction ($\lambda_{\text{SO}}=0$ and $\mu_L=\mu_R=0$). (b) Longitudinal spin current generated by an external bias with $\mu_{L/R}=\pm V/2$. Here, the central system under driven contains 11 unit cells along the longitudinal direction, and we used the following parameters: $\varphi=\pi/2, A_0a=1,\omega=1$, $t=1,\lambda=0.5$ and $\Gamma_L=\Gamma_R=0.1 t$.  }
    \label{fig:pump_response}
\end{figure}

\textbf{Spin pumping and spin conductance.} In the heterostructure, the spin current (per unit cell) measured in the right lead consists of a pumped contribution and a response contribution, $J^s_R= J^s_{pump}+J^s_{res}$~\cite{Kitagawa2011photoQuantumHall,Vahid2024FloquetGreen}, given by
\begin{eqnarray}
&&J^s_{pump}= \frac{1}{L_y}\sum_{\sigma}\sum_{mn}\sigma\int_{0}^\omega \frac{d\varepsilon}{2\pi} \text{Tr}[\Gamma_R G^R_{\sigma,mn}(\varepsilon)\Gamma_L G^A_{\sigma,nm}(\varepsilon)][f_R(\varepsilon+m
\omega)-f_R(\varepsilon+n\omega)],\nonumber\\
&&J^s_{res}= \frac{1}{L_y}\sum_{\sigma}\sum_{mn}\sigma\int_{0}^\omega \frac{d\varepsilon}{2\pi} \text{Tr}[\Gamma_R G^R_{\sigma,mn}(\varepsilon)\Gamma_L G^A_{\sigma,nm}(\varepsilon)][f_L(\varepsilon)-f_R(\varepsilon)],
\end{eqnarray}
where $L_y$ denotes the system length along the transverse direction. The pumped current corresponds to a steady-state current induced by the periodic light driving, whereas the response current arises from an external bias applied between the two leads. 

As shown in Fig.~\ref{fig:pump_response}, the steady-state current exhibits the same symmetry properties as in the case where the system is coupled to a phonon bath; see Eqs.~\eqref{eq:steadycurrent_symmetry1} and~\eqref{eq:steadycurrent_symmetry2}. Moreover, its magnitude is of the same order as that obtained in the phonon-coupled setup. The response current induced by the external bias is also consistent with linear-response expectations: (i) the insulating system becomes effectively metallic under periodic driving, and (ii) the longitudinal conductivity can be estimated as $\sigma_{xx}=\bar{J}^s_{res}/E_x\sim 10^{-2}e$ (where $E_x=V/L_x$), which is of the same order as that reported in the main text.\\



\section{Models}

In this section, we give the details of the calculation on the Floquet band structure of the non-symmorphic AFM and CuMnAs. The possibility to engineer the spin texture in CuMnAs is also demonstrated through a low-energy effective theory around the Dirac points.

\subsection{Nonsymmorphic AFM}

The Hamiltonian of the nonsymmorphic AFM is recapped below
\begin{eqnarray}
H_{\mathbf k}=-2t\cos\frac{k_x}{2}\cos\frac{k_y}{2}\tau^x-t^\prime(\cos k_x+\cos k_y)+w\sin\frac{k_y}{2}\cos\frac{k_x}{2}\tau^y+\lambda\tau^z\boldsymbol{\sigma}\cdot\mathbf n    
\end{eqnarray}
where $t=-(t_1+t_2)$ and $w=2(t_1-t_2)$. In terms of the Fourier components, the transformed expression is
\begin{eqnarray}
H^{(m)}=\varepsilon_0^{(m)}+V_{AB}^{(m)}\tau^x+S^{(m)}_{AB}\tau^y+\lambda\tau_z\mathbf{n}\cdot\boldsymbol{\sigma}\delta_{m,0}   
\end{eqnarray}
where $\tau_i$ is the Pauli matrix in sublattice space. Specifically,
\begin{eqnarray}\label{eq:Fourier}
\varepsilon_0^{(m)}&=&-\frac{t^\prime}{2}J_m(aA_0)\Big[(e^{ik_x}(-1)^m+e^{-ik_x})+(e^{ik_y}(-1)^m+e^{-ik_y})e^{-im\varphi}\Big],\nonumber\\
V_{AB}^{(m)}&=&-\frac{t}{2}J_{m}\Big(aA_0|\cos\frac{\varphi}{2}|\Big)e^{-im\theta_+}\Big[e^{i(k_x+k_y)/2}(-1)^m+e^{-i(k_x+k_y)/2}\Big]\nonumber\\
&&
-\frac{t}{2}J_{m}\Big(aA_0|\sin\frac{\varphi}{2}|\Big)e^{im\theta_-}\Big[e^{i(k_x-k_y)/2}(-1)^m+e^{-i(k_x-k_y)/2}\Big],\nonumber\\
S_{AB}^{(m)}&=&\frac{\Delta_{12}}{2}J_{m}\Big(aA_0|\cos\frac{\varphi}{2}|\Big)e^{-im\theta_+}\Big[e^{i(k_x+k_y)/2}(-1)^m-e^{-i(k_x+k_y)/2}\Big]\nonumber\\
&&
+\frac{\Delta_{12}}{2}J_{m}\Big(aA_0|\sin\frac{\varphi}{2}|\Big)e^{im\theta_-}\Big[e^{-i(k_x-k_y)/2}-(-1)^me^{i(k_x-k_y)/2}\Big].
\end{eqnarray}
Here, $\theta_+=\text{sign}(\pi-\varphi)\cos^{-1}\Big(|\cos\frac{\varphi}{2}|\Big)$ and $\theta_-=\text{sign}(\pi-\varphi)\cos^{-1}\Big(|\sin\frac{\varphi}{2}|\Big)$.
\\

\begin{figure}
    \centering    \includegraphics[width=0.5\linewidth]{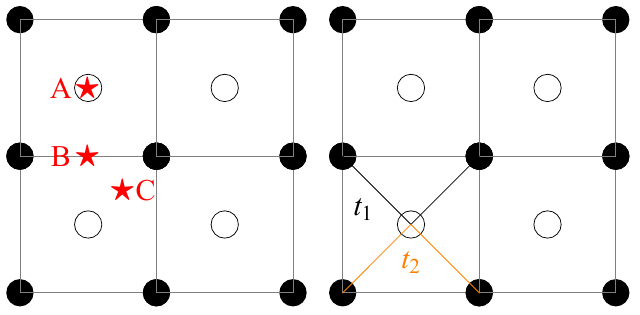}
    \caption{ Left: the unit cell of the square-lattice antiferromagnet, where the stars mark the possible inversion centers. Right: the unit cell of the nonsymmorphic antiferromagnet.  }
    \label{fig:nonsymorphic_unitCell}
\end{figure}

\textbf{Symmetry}. In the nonsymmorphic AFM model, the term proportional to $w$ plays an important role. When $w=0$, the Hamiltonian describes a square-lattice AFM, as shown in the left panel of Fig.~\ref{fig:nonsymorphic_unitCell}.  In this case, the system has both PT and inversion symmetry. The inversion center for the pure inversion symmetry is $A$ or $B$ point in the left panel of Fig.~\ref{fig:nonsymorphic_unitCell}; while the inversion center involved in the PT symmetry is the $C$ point. The inversion symmetry with respect to point $A$ or $B$ induces $H(\mathbf k+\boldsymbol{\mathcal A})=H(-\mathbf k-\boldsymbol{\mathcal A})$; the inversion operation with respect to point $C$ generates the dual symmetry $\tau^x H_{\uparrow}[\mathbf k+\boldsymbol{\mathcal A}(t)]\tau^x=H_{\downarrow}[-\mathbf k-\boldsymbol{\mathcal A}(t)]$. These two relations together enforce that
\begin{eqnarray}
\tau^xH_{\uparrow}[\mathbf k+\boldsymbol{\mathcal A}(t)]\tau^x=H_{\downarrow}[\mathbf k+\boldsymbol{\mathcal A}(t)],    
\end{eqnarray}
which leads to spin-degenerate band structure. However, the $w$ term breaks the inversion symmetry with respect to $A$ and $B$, thus making the spin-splitting possible.

\subsection{Tetragonal CuMnAs}
The tight-banding model reads 
\begin{eqnarray}\label{eq:CuMnAs}
H_{\mathbf k}=-2t\cos\frac{k_x}{2}\cos\frac{k_y}{2}\tau^x-t^\prime(\cos k_x+\cos k_y)+\alpha_R\tau^z(\sigma^y\sin k_x-\sigma^x\sin k_y)+\lambda\tau^z\boldsymbol{\sigma}\cdot\mathbf n
\end{eqnarray}
where $\alpha_R$ is Rashba spin orbit coupling coefficient. Performing Fourier transformation, one obtains
\begin{eqnarray}
H^{(m)}=\varepsilon_0^{(m)}+V_{AB}^{(m)}\tau_x+\tau_z\mathbf{h}^{(m)}\cdot\boldsymbol{\sigma}   
\end{eqnarray}
where $\tau_i$ is the Pauli matrix in sublattice space. Specifically, $\varepsilon_0^{(m)}$ and $V_{AB}^{(m)}$ are given by Eq.~\eqref{eq:Fourier}, and
\begin{eqnarray}
\mathbf h^{(m)}=\lambda\mathbf{n}\delta_{m,0}+\Big(
-\frac{\alpha_R}{2i}[e^{ik_y}(-1)^m-e^{-ik_y}]e^{-im\varphi}J_m(aA_0),
\frac{\alpha_R}{2i}[e^{ik_x}(-1)^m-e^{-ik_x}]J_m(aA_0),
0
\Big).
\end{eqnarray}
The eigenvalue of Eq.~\eqref{eq:CuMnAs} is
\begin{eqnarray}
E_{\mathbf k,\pm}=-t^\prime(\cos k_x+\cos k_y)\pm\sqrt{4t^2\cos^2(\frac{k_x}{2})\cos^2(\frac{k_y}{2})+(\lambda-\alpha_R\sin k_y)^2+\alpha_R^2\sin^2k_x}.    
\end{eqnarray}
There are two Dirac points in the spectrum at $\mathbf k_{1,\text{Dirac}}=(\pi, k_0)$ and $\mathbf k_{2,\text{Dirac}}=(\pi, \pi-k_0)$ where $k_0=\sin^{-1}(\frac{\lambda}{\alpha_R})$. To see how light engineer the spin texture, we take a close look at the case of weak field and high frequency, as did in the honeycomb model.
The long-wavelength expansion around the Dirac points reads
\begin{eqnarray}\label{eq:Dirac}
H(\mathbf k_{n,\text{Dirac}}+\mathbf q)=c_n+a q_y+b_n q_x\tau^x+\tau^z[(-1)^{n}d_1q_y\sigma^x-d_2q_x\sigma^y],\qquad n=1,2,   
\end{eqnarray}
where $c_n=t_0(1+(-1)^n \cos k_0)$, $a=t_0\lambda/\alpha_R$, $b_n=t_1\sin(\frac{n\pi-k_0}{2})$, $d_1=\alpha_R\cos k_0, d_2=\alpha_R$. The optical field is implemented via vector potential $\boldsymbol{\mathcal A}=A_0(\sin\omega t, \sin(\omega t+\varphi),0)$ and performing the Peierls substitution to the equation above. We further employ the Fourier transformation to obtain 
\begin{eqnarray}
\mathcal H_n^{(m)}=\mathcal H_n^{(+1)}\delta_{m,1}+\mathcal H_n^{(-1)}\delta_{m,-1},    
\end{eqnarray}
where 
\begin{eqnarray}
\mathcal H_n^{(\pm 1)}=\mp\frac{A_0}{2i}\Big[a e^{\mp i\varphi}+b_n\tau^x+\tau^z\Big((-1)^nd_1e^{\mp i\varphi}\sigma^x-d_2\sigma^y\Big)\Big].
\end{eqnarray}
From the Floquet-Magnus expansion, we get an optical field induced correction to Eq.~\eqref{eq:Dirac}
\begin{eqnarray}
\delta H_n&&=\frac{[\mathcal H^{-1},\mathcal H^{+1}]}{\omega}+ O(A_0^4)\nonumber\\
&&=A_0^2\sin\varphi d_1(-1)^n(d_2\tau^z\sigma^z-b_1\tau^y\sigma^x)+ O(A_0^4).
\end{eqnarray}
The correction term contains not only a staggered Zeeman term but also a spin-dependent hopping term, both of which can deeply affect the momentum-space spin texture of the band structure. These emergent terms can be certainly used to generate  N\'eel torque in AFM.

\section{Useful integrals}

In calculating Floquet band structures, many integrals containing Bessel functions are encountered. Here, we list useful properties and integrals relevant to the Bessel function for the convenience of the readers. The $n$-th order Bessel function can be defined via an integral 
\begin{eqnarray}
J_n(x)=\frac{1}{2\pi}\int_{-\pi}^\pi d\tau e^{i(n\tau-x\sin\tau)}.    
\end{eqnarray}
Moreover, we note the definition above is invariant upon shifting the integral interval by an arbitrary constant, i.e., $J_n(x,\eta)=\frac{1}{2\pi}\int_{-\pi+\eta}^{\pi+\eta} d\tau e^{i(n\tau-x\sin\tau)}=J_n(x)$. This can be verified as follow,
\begin{eqnarray}
\frac{d}{d\eta}J_n(x,\eta)=\frac{1}{2\pi}\{e^{i[m(\pi+\eta)+A\sin(\pi+\eta)]}-e^{i[m(-\pi+\eta)+A\sin(-\pi+\eta)]}\}=0.
\end{eqnarray}
Note the Bessel function respect the following relations
\begin{eqnarray}
J_n(-x)=J_{-n}(x)=(-1)^n J_n(x).    
\end{eqnarray}
With the assistance of above relations, we obtain two useful results 
\begin{eqnarray}
\frac{1}{T}\int_0^T dt e^{im\omega t}\cos[A_0\sin(\omega t+\varphi)]=\frac{(-1)^m+1}{2}e^{-im\varphi}J_m(A_0),\nonumber\\
\frac{1}{T}\int_0^T dt e^{im\omega t}\sin[A_0\sin(\omega t+\varphi)]=\frac{(-1)^m-1}{2i}e^{-im\varphi}J_m(A_0),
\end{eqnarray}
and 
\begin{eqnarray}
\frac{1}{T}\int_0^T dt e^{im\omega t}\exp[i A_0\sin(\omega t+\varphi)]=(-1)^me^{-im\varphi}J_m(A_0),\nonumber\\
\frac{1}{T}\int_0^T dt e^{im\omega t}\exp[-i A_0\sin(\omega t+\varphi)]=e^{-im\varphi}J_m(A_0).
\end{eqnarray}

\bibliographystyle{apsrev4-1-title}
\bibliography{splitting}